\documentclass[manuscript,screen,authordraft,nonacm,review=false,timestamfalse]{acmart}

\acmJournal{CSUR}

\newcommand{\parhl}[1]{\textbf{#1}}

\usepackage{todonotes}

\usepackage{longtable}
\usepackage{multirow}
\usepackage{changepage}
\usepackage{todonotes}
\usepackage{appendix}

\usepackage{pifont}
\usepackage{caption}
\usepackage{subcaption}
\usepackage{wrapfig}
\usepackage{amsmath}
\usepackage{tikz}
\usetikzlibrary{shapes.geometric}
\newcommand{\stars}[2][fill=black,draw=black]{
\begin{tikzpicture}[baseline=-0.34em,#1]
\foreach \X in {1,...,5}
{
\pgfmathsetmacro{\xfill}{min(1,max(1+#2-\X,0))}
\path (\X*1em,0) 
node[star,draw, ultra thin,star point height=0.23em,minimum size=0.75em,inner sep=0pt,
path picture={\fill (path picture bounding box.south west) 
rectangle  ([xshift=\xfill*0.722em]path picture bounding box.north west);}]{};
}
\end{tikzpicture}}

\AtBeginDocument{%
  }

\setcopyright{acmlicensed}
\copyrightyear{2026}
\acmYear{2026}
\acmDOI{XXXXXXX.XXXXXXX}

\acmISBN{978-1-4503-XXXX-X/18/06}

\begin{document}

\title{A Quarter of a Century of Neuromorphic Architectures on FPGAs - an Overview}

\author{Wiktor J. Szczerek}
\email{szczerek@kth.se}
\orcid{0009-0005-5561-440X}
\author{Artur Podobas}
\email{podobas@kth.se}
\orcid{0000-0001-5452-6794}
\affiliation{%
  \department{Department of Computing and Learning Systems}
  \institution{KTH Royal Institute of Technology}
  \city{Stockholm}
  \country{Sweden}
}


\begin{abstract}
  Neuromorphic computing is a relatively new discipline of computer science, where the principles of biological brain's computation and memory are used to create a new way of processing information, based on networks of spiking neurons. Those networks can be implemented as both analog and digital implementations, where for the latter, the Field Programmable Gate Arrays (FPGAs) are a frequent choice, due to their inherent flexibility, allowing the researchers to easily design hardware neuromorphic architecture (NMAs). Moreover, digital NMAs show good promise in simulating various spiking neural networks because of their inherent accuracy and resilience to noise, as opposed to analog implementations. This paper presents an overview of digital NMAs implemented on FPGAs, with a goal of providing useful references to various architectural design choices to the researchers interested in digital neuromorphic systems. We present a taxonomy of NMAs that highlights groups of distinct architectural features, their advantages and disadvantages and identify trends and predictions for the future of those architectures.
\end{abstract}

\ccsdesc[500]{General and reference~Surveys and overviews}

\keywords{neuromorphic systems, FPGA, computing architectures}

\received{XX XX XXXX}
\received[revised]{XX}
\received[accepted]{XX}

\maketitle

\section{Introduction}
Computers and improvements in their performance have historically relied on two pillars: Moore's law~\cite{schaller_moores_1997} and Dennard's scaling~\cite{bohr200730}. The former was formulated by Gordon Moore in the 1960s and states that the number of transistors per unit area doubles every 24 months, primarily due to improvements in lithography techniques. The latter was introduced by Richard Dennard, who found that as transistors grow smaller, the energy consumption reduces, resulting in constant power density. These two observations worked in tandem, allowing us to build more complex high-performing computers (Moore's law) without them consuming needlessly large amounts of energy (Dennard's scaling).
Unfortunately, Dennard's scaling ended in 2004-2005~\cite{dennard2018perspective}, and we are currently witnessing the slowdown of Moore's law~\cite{schaller_moores_1997} as well as the emergence of Dark Silicon~\cite{esmaeilzadeh2011dark} without any in-place replacement technology at hand~\cite{shalf2020future}. Instead, researchers and industry are looking for alternative ways of computing in the hope of continuing the trend in computer performance we have seen so far. These technologies are collectively called \textit{post-Moore technologies}~\cite{leiserson2020there,shalf2020future} and include quantum computers~\cite{gyongyosi2019survey}, adiabatic reversible logic~\cite{bommi2018survey}, and analog computing~\cite{maclennan2007review}. Two post-Moore technologies, however, are perhaps the most salient alternatives: Neuromorphic--~\cite{schuman2017survey} and Reconfigurable--computing~\cite{podobas2020survey,kuon_fpga_2008}. Introduced in the 1980s by Carver Mead, neuromorphic systems~\cite{mead1990neuromorphic} are computers designed to replicate -- to a varying extent -- the impressive computing capabilities of the biological brain. Initially designed to replicate brain circuitry using analog electronic components, neuromorphic systems today are implemented in different technologies, ranging from digitally synchronous (e.g., Intel Loihi~\cite{davies2018loihi} or IBM TrueNorth~\cite{akopyan2015truenorth}), mixed analog/digital (e.g., BrainDrop~\cite{neckar2018braindrop}), fully analog (e.g. \cite{indiveri2011neuromorphic}), to even memristor- or photonics-based systems~\cite{li2018review,shastri2021photonics}. Regardless of how they are implemented, they share two fundamental traits: \textbf{(i) }they communicate through discrete events called spikes and \textbf{(ii)} they are programmed in a non-imperative way, generally by encoding the problem to be solved through a spiking neural network (SNN)~\cite{maass1997networks}. Solving a problem using a neuromorphic system often results in a significantly more energy-efficient solution in several application domains~\cite{davies_advancing_2021} as opposed to a traditional von Neumann system~\cite{nikhil1989can}. Numerous functionalities of neuromorphic systems are driven by discoveries and innovation in neuroscience (e.g.,  Spike-Timing Dependent Plasticity (STDP)~\cite{song2000competitive}, Bayesian Confidence Propagation Neural Network (BCPNN)~\cite{ravichandran2024unsupervised}, or Eligibility Propagation (E-prop)~\cite{bellec_solution_2020}), which has motivated many researchers to consider reconfigurable platforms, such as Field Programmable Gate Arrays (FPGAs), for their designs to keep up with the fast pace of neuroscientific research. FPGAs~\cite{gandhare2019survey} are a group of reconfigurable devices facilitating the implementation of a variety of digital circuits without the need for fabricating a bespoke ASIC. Initially designed to prototype ASICs prior to tape-out, nowadays the FPGAs are used in a plethora of different fields from low-volume electronics~\cite{andina2017fpgas}, through space exploration~\cite{jacobs2012reconfigurable}, to large data centers~\cite{weerasinghe2015enabling}. A hardware designer can program an FPGA to act as a central processing unit (CPU)~\cite{calderon2005soft}, a Graphics Processing Unit (GPU)~\cite{andryc2013flexgrip}, or a neuromorphic device. FPGAs have shown a lot of promise in accelerating neuromorphic workloads, resulting in significant performance gains over CPU- and GPU-based solutions~\cite{podobas2017designing,ju_fpga_2020}, with examples of even outperforming ASIC-based neuromorphic systems (e.g., for brain simulation~\cite{lindqvist2024fast}).

This paper presents the outcome of a systematic review of neuromorphic hardware architectures integrated using FPGAs. We included the majority of the literature on the subject starting from 1998, discarded unrelated works based on their abstract and/or topic, and read a total of 135 papers in detail, which we then summarized and classified according to the extended taxonomy we propose. Furthermore, we identified and collected performance metrics across the surveyed literature and condensed this information into trends and observations on the direction of the field's development. Finally, we conclude the survey with a discussion on future opportunities. There are many excellent surveys on neuromorphic systems and spiking neural networks (SNNs) from different perspectives, tied to hardware~\cite{schuman2017survey}, specific devices~\cite{davies_advancing_2021} (e.g., Intel Loihi), different models~\cite{shrestha2022survey}, and more. However, as opposed to the prior work surveying FPGAs and SNNs~\cite{karamimanesh2025spiking,mehrabi2024fpga,maguire_challenges_2007}, we introduced a hardware design choice-driven Taxonomy, which we then applied to the surveyed architectures using the Taxonomy as a framework. We then acquired metrics and performed an in-depth analysis to make predictions for the future of the NMAs on FPGAs. Our taxonomy-centric approach contribute to a more structured and systematic survey, bringing unique insights and predictions to NMAs on FPGAs that existing taxonomy-agnostic surveys are unable to unveil.

\vspace{-0.25cm}
\section{Background}
In the following section, we present the key concepts related to this survey: \textbf{(i)} the Field-Programmable Gate Arrays (FPGAs), \textbf{(ii)} Neuromorphic Architectures (NMAs) and \textbf{(iii)} Spiking Neural Networks (SNNs). Furthermore, we propose a Taxonomy for the NMAs, which we argue is necessary, considering the current lack of agreement on how to understand the term \textit{neuromorphic hardware architecture}\cite{indiveri2025neuromorphic}.

\vspace{-0.25cm}
\subsection{Field-Programmable Gate Arrays}
Field-Programmable Gate Arrays (FPGAs) \cite{kuon_fpga_2008} belong to a family of reconfigurable architectures, whose underlying silicon can be reprogrammed to yield different functionalities. Historically used to prototype ASICs before tape-out, the FPGAs are used in various fields from low-volume electronics, through embedded systems and telecommunications to high-performance computing (HPC)\cite{podobas2017evaluating, zohouri2016evaluating}. The performance of FPGA-based systems can generally be better than that of systems based on CPUs or GPUs, but worse than that of custom ASICs. Another strength of using FPGAs is that they can be iteratively tailored to the application's needs without incurring the cost of developing a new ASIC each time. To put this matter into perspective, an FPGA capable of executing tens of billions of floating-point operations per second (tens of TFLOPs) and typically costs thousands of dollars, while creating a similarly capable ASIC is between two and three orders of magnitude more expensive.

\begin{wrapfigure}[16]{o}{0.5\textwidth}
    \centering
    \includegraphics[width=0.9\linewidth, trim={0 30pt 0 40pt}]{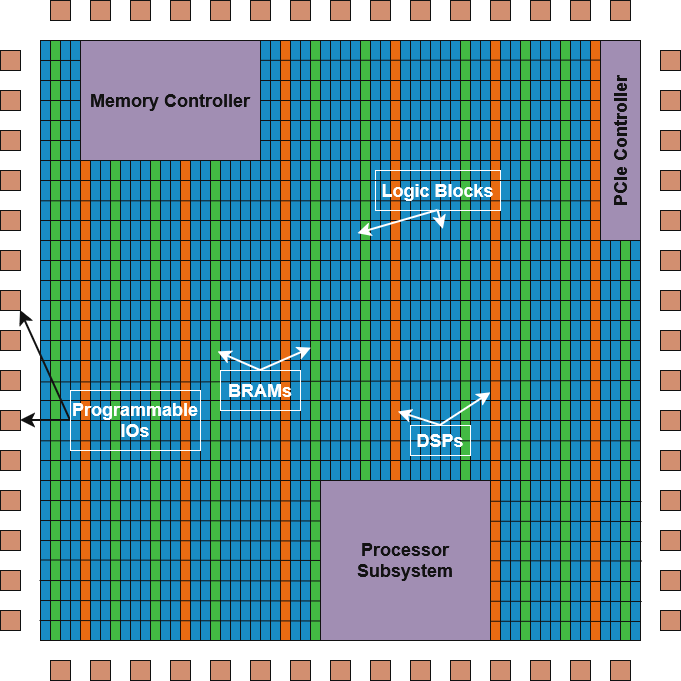}
    \caption{Abstract view of an FPGA, showing reconfigurable logic (blue), DSP-blocks (orange), BRAM (green), and hardened memory controller and processor blocks. Illustration based on~\cite{kuon_fpga_2008}.}
    \label{fig:fpga_arch}
\end{wrapfigure}
\noindent
Figure~\ref{fig:fpga_arch} shows the abstract view of a typical FPGA. Fundamentally, every FPGA consists of a number of \textit{logic blocks} (also called \textit{configurable logic blocks} or \textit{logic elements}), a system of programmable interconnects (called \textit{fabric}) that routes signals between the logic blocks and input/output (I/O) blocks for communication with external devices. Logic blocks are usually made up of a look-up table (LUT) that stores a predefined list of logic outputs for any combination of logic input vectors and standard logic circuitry, such as multiplexers, full adders, and flip-flops. The content of these LUTs can be programmed, and by connecting several LUTs together, we can achieve any digital functionality. Originally, FPGAs comprised only the aforementioned logic blocks and fabric, but it was soon realized that certain digital circuits (e.g., multipliers) consumed much of the FPGA's logic resources. Thus, dedicated (called hardened) functionality was added to the subsequent iterations of the FPGA chips, e.g., bespoke on-chip memory called block random access memory (BRAM), digital signal processing (DSP) blocks (which, in some FPGAs, include floating-point units), and many others (phase-locked loops, PCIe controllers, etc.).

\subsection{Spiking Neural Networks}
\label{sec:bg_snn}
Computations in the biological brains are mainly performed using \textit{neurons}, which are connected to each other through \textit{synapses}. Biological neurons communicate via \textit{action potentials} - voltage pulses traveling from one neuron to another through the synapse, caused by rapid changes in neurons' membrane potentials - commonly referred to as \textit{spikes} or \textit{spike events}. Networks built using neurons that mimic this behavior are called \textit{Spiking Neural Networks} (SNNs), often called \textit{third-generation networks}\cite{maass1997networks} that add a notion of time to the Artificial Neural Networks (ANNs). SNNs have been used as tools to understand the biological brain in the neuroscience community\cite{potjans_cell-type_2014} through simulators such as Neuron\cite{hines1997neuron}, NEST\cite{gewaltig2007nest}, or Brian\cite{stimberg2019brian}. Lately, these models have also been considered for solving machine learning (ML) problems. Moreover, SNNs are believed to reduce the energy consumption needed to solve those problems and/or solve them faster, primarily by leveraging the sparseness of spike communication. Spiking neurons can be arranged in various topologies that fit different scenarios, from classification tasks to neuroscientific experiments. Those include, but are not limited to: Feed-Forward (FF) and its fully-connected subtype (FF-FC) -- equivalent of the multilevel perceptron (MLP), Spiking Convolutional Neural Networks (SCNNs) - spiking equivalents of Convolutional Neural Networks (CNNs), Liquid State Machines (LSM)\cite{maass2011liquid} - randomly-connected reservoirs of neurons connected to an output layer, randomly-connected (RAND) - reservoirs of recurrently-connected neurons, all-to-all-connected (A2A) and biological SNNs (BIO)\footnote{Presented abbreviations are \textbf{not necessarily} used unanimously. Still, they commonly appear in the literature.} - e.g., Central Pattern Generators (CPGs)\cite{wilson1961central,jankowska1972electrophysiological}, able to produce rhythmic motor patterns. Recently, examples of spike-based implementations of transformer networks\cite{gao_advancing_2025} were presented. As for communication, the most popular way of exchanging spike events is the Address Event Representation (AER)\cite{mahowald1994analog} protocol, representing the event as the address of a destination neuron and the timestep when it occurred.

\begin{table}[h]
    \caption{The most commonly used spiking neuron models in our survey (see Section \ref{sec:trends} for analysis). The biological plausibility and complexity of the model metrics are based on \cite{izhikevich_which_2004} and our own analysis. The complexity metric relates to the estimated number of floating-point operations (addition, multiplication, etc.) necessary to advance to the next timestep (1 ms) with a selected neuron model ($v$ -- membrane potential, $I$ -- synaptic current, $R$ -- leakage resistance, $\tau_m$ -- membrane time constant, $g_L$ -- leakage conductance, $C$ -- membrane capacitance, $g_K$ -- conductance of the potassium ion channel, $g_{NA}$ -- conductance of the sodium ion channel; other parameters are model-specific, and interested readers are referred to the related literature for detailed explanations).}
    \label{tab:neuron_models}
    \resizebox{\textwidth}{!}{
        \begin{tabular}{c | l c c }
            \toprule
            \multicolumn{4}{c}{\textbf{Neuron Models}}                                                                                                                                                      \\
            Name                    & Formula                                                                                                                             & Bio. plausibility & Complexity  \\
            \midrule  \midrule
            Integrate-and-Fire (IF) & $\begin{array}{ll}
                                               \frac{dv}{dt} = RI
                                           \end{array}$                                                                                                                  & V. Low            & $\approx$5 FLOPs \\ 
            Leaky IF (LIF)          & $\begin{array}{ll}
                                               \tau_m\frac{dv}{dt} = -(v - V_{reset}) + RI
                                           \end{array}$                                                                                         & Low               & $\approx$10 FLOPs                         \\ 
            Izhikevich (IZH)        & $\begin{cases}
                                               \frac{dv}{dt} = 0.04v^2 + 5v + 140 - u + I \\
                                               \frac{du}{dt} = a(bv - u)
                                           \end{cases}$                                                                                      & Low               & $\approx$60 FLOPs                            \\ 

            FitzHugh-Nagumo (FHN)   & $\begin{cases}
                                               \frac{dv}{dt} = v - \frac{v^3}{3} - w + I \\
                                               \frac{dw}{dt} = 0.08 (v + 0.7 - 0.8w)
                                           \end{cases}$                                                                                      & Medium            & $\approx$100 FLOPs                           \\ 

            Adaptive IF (AdEX)      & $\begin{cases}
                                               \frac{dv}{dt} = \frac{-g_L(v-E_L) + g_L \Delta_T exp(\frac{v-v_T}{\Delta_T}) - w + I}{C} \\
                                               \frac{dw}{dt} = a(v-E_L) - w
                                           \end{cases}$                                          & Medium            & $\approx$200 FLOPs                                                                       \\ 

            Hodgkin-Huxley (HH)     & $\begin{cases}
                                               \frac{du}{dt} =                              \\ \frac{-g_{K}n^4(U-V_{K})-g_{Na}m^3h(U - V_{Na})-g_{leak}(U - u_{leak}) + I}{C_m} \\
                                               \frac{dn}{dt} = \alpha_n(U)(1-n)-\beta_n(U)n \\
                                               \frac{dm}{dt} = \alpha_m(U)(1-m)-\beta_m(U)m \\
                                               \frac{dh}{dt} = \alpha_h(U)(1-h)-\beta_h(U)h
                                           \end{cases}$ & High              & $\approx$1200 FLOPs                                     \\

            \bottomrule
        \end{tabular}
    }
\end{table}

An SNN uses a \textit{spiking neuron model} to describe the functionality of generating outgoing spikes. Those models are formulated using a set of ordinary differential equations (ODEs) that describe how a neuron's cell membrane potential interacts with an injected current or modified conductance over time. In addition to the ODEs, models have a condition for emitting a spike, typically when the membrane voltage potential reaches a certain value (\textit{threshold voltage}). Neuron models vary in biological plausibility and complexity, with different models being suitable for various use cases. Table \ref{tab:neuron_models} outlines six important neuron models, their mathematical descriptions, biological plausibility, and an estimate of the number of floating-point operations (FLOPs) needed to advance their simulation time by one millisecond. The biologically implausible Integrate-and-Fire (IF) neuron model is the least computationally expensive and requires only a handful of operations, whereas the Hodgkin-Huxley (HH) model is complex, describes in great detail the behavior of biological neurons, and requires thousands of operations to solve the ODEs. Hence, the choice of neuron model comes primarily from the use case: an SNN used for detailed simulation of a biological system might need HH-like neurons (as in OpenWorm\cite{sarma2018openworm}), whereas an ML system performing image inference could use a less expensive model (e.g., the Leaky Integrate-and-Fire (LIF) model, which is a popular choice for image inference\cite{diehl2015unsupervised}). The neurons in Table \ref{tab:neuron_models} were selected based on popularity in surveyed neuromorphic hardware, but there are various other models positioned somewhere in the span of complexity between IF and HH\cite{hodgkin_quantitative_1952}. These include the Spike Response Model (SRM)\cite{gerstner_neuronal_2014}, pulsed neurons\cite{waldemark_pulse_1998}, Message-Based Event-Driven (MBED) neurons\cite{claverol_event-driven_2000}, Digital Spiking Silicon Neuron (DSSN)\cite{nanami_fpga-based_2016} and Traub's model\cite{traub1991model}, to name a few. When implemented in hardware, the solutions for spiking neuron models ODEs (and other ODEs, related for example to synaptic current decay) are often approximated via \textit{Runge-Kutta methods}, of which the most commonly found are \textit{forward Euler method} and \textit{RK4} (or simply \textit{Runge-Kutta method}). Those methods allow for numerical integration of the ODEs with a known step $h$, which models well the SNN simulation task. The \textit{forward Euler} method is the simplest first-order Runge-Kutta method and is primarily used for simpler neuron models, such as the LIF and IZH models. A more complex RK4 method was commonly used for HH ODEs, due to the more intricate dynamics of those equations. Neurons connect through synapses, which accept incoming \textit{spikes} from other neurons. Those spikes cause a change in current or conductance of the targeting neuron's extensions (called \textit{dendrites}). This modifies the neuron's membrane potential and can cause it to spike. Interestingly, the choice of a synapse model has potentially a far greater impact on the complexity of the hardware than the choice of neuron model, mainly because there are many orders of magnitude more synapses than there are neurons\cite{azevedo2009equal}.
Synapses, which from the biological point of view include gamma-aminobutyric acid (GABA), N-methyl-D-aspartate (NMDA), and $\alpha$-amino-3-hydroxy-5-methyl-4-isoxazolepropionic acid (AMPA) ion channels\cite{ben-ari_gabaa_1997}, can be modeled as current-based, where they modify the injected current into the target membrane or conductance-based (CUBA), where they modify the conductance (COBA). As with neurons, synaptic dynamics are typically described using ODEs.
Synapses are extremely important and are thought to play a significantly larger role than neurons in an SNN. This is because synapses facilitate \textit{learning} in the change of their \textit{strength} (modeling the change in conductance). This process is known as \textit{synaptic plasticity} \cite{citri2008synaptic}, and numerous models of this phenomenon exist. Most models are Hebbian\cite{gerstner2011hebbian} in nature, meaning that they relate the increase or decrease in strength to the simultaneous activity of pre- and post-synaptic neurons.

\begin{wrapfigure}[14]{o}{0.4\textwidth}
    \centering
    \includegraphics[width=0.95\linewidth, trim={0 0 0 30pt}]{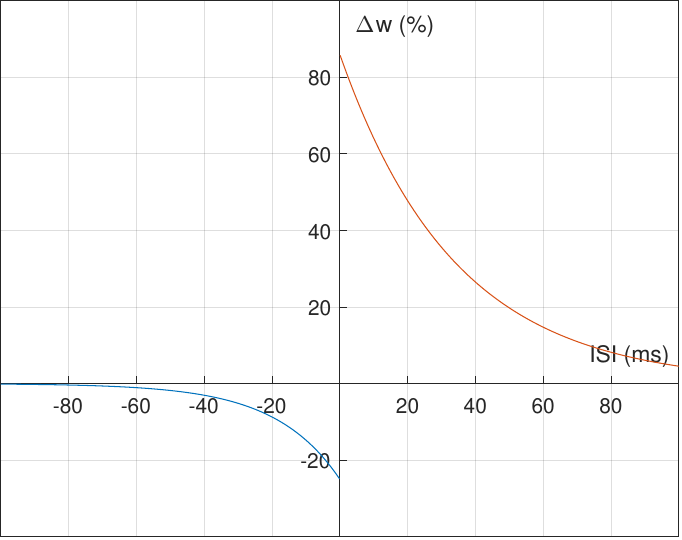}
    \caption{STDP learning window adapted from and based on the data points presented by Bi and Poo\cite{bi1998synaptic} ($\Delta w$ -- change of synaptic weight in \%, ISI - inter-spike interval in ms, blue line corresponds to weakening and red line to strengthening the synaptic connection).}
    \label{fig:bg_stdp}
\end{wrapfigure}

For an SNN to perform any meaningful operation, it must first be trained, which typically involves adjusting the synaptic connections between neurons. Moreover, the designers may choose to adjust the threshold voltage of the neurons or perform a structural adjustment, such as pruning unnecessary connections.
In this survey, we were primarily interested in whether the architecture supports online learning, as we firmly believe that supporting some kind of online reconfigurability brings these systems closer to brain-like operation. We did not, however, focus on the intricacies of these methods, and we refer interested readers to the relevant literature\cite {purves_neuroscience_2018}. We do, however, inform the reader about the most common approaches to the topic.
Overall, we can distinguish two main training types that the NMAs supported based on the process occurring during operation -- \textit{online} -- or before the architecture is deployed on the platform -- \textit{offline}. Online training assumes supporting some form of plasticity, be it synaptic -- where the synaptic weights are adjusted according to a predefined rule, structural -- where the connections between the neurons can be severed or created, or some special rules, like adjusting the synaptic delay on the fly.
The most well-known type of online learning related to the synaptic plasticity is the Hebbian learning\cite{hebb2005organization}, proposed in 1949 by Donald O. Hebb, also called the \textit{correlation learning}. In this context, \textit{correlation} refers to the dependency of the synaptic weights on the intervals between pre- and postsynaptic spike times.
A particular and rather popular implementation of Hebbian learning is Spiking-Time-Dependent Plasticity (STDP), which in its most recognizable form was proposed by Bi and Poo\cite{bi1998synaptic}. The process relies on the concept of long-term strengthening the connections -- Long-Term Potentiation (LTP) -- where pre-synaptic neurons directly correspond to the generation of postsynaptic spikes and weakening those connections -- Long-Term Depression (LTD) -- if presynaptic spikes appear after the postsynaptic spikes. Those processes are defined by the \textit{learning window} -- curve that relates synaptic weight change to the timing difference between consecutive spikes. Currently, there are many learning windows used in research, as their shape allows mimicking the behavior of different brain parts and types of cells. The one used by Bi and Poo\cite{bi1998synaptic} can be thought of as a \textit{foundation} for the other subtypes and is shown in Figure \ref{fig:bg_stdp}. This learning method has its own subtypes, such as Pair-wise STDP (PSTDP) and Triplet-wise STDP (TSTDP), as well as more unique ones, including Self-Organizing-Map-based STDP (SOM-STDP).
Apart from the STDP and its modifications, another Hebbian learning rule is the Bayesian Confidence Propagation Neural Network (BCPNN)\cite{lansner1989one}, which focuses on the relations between hyper- and minicolumns, whose hardware implementation is also under investigation in current research, as shown by Al Hafiz et al.\cite{al_hafiz_reconfigurable_2025}. Moreover, there exist other important plas--

\noindent
--ticity models, e.g., Bienenstock–Cooper–Munro (BCM) rule\cite{bienenstock_theory_1982} based on observations in homeostatic plasticity, an entire range of backpropagation-related methods (e.g., SpikeProp\cite{bohte_spikeprop_2000}, Backpropagation Through Time (BPTT)\cite{werbos_generalization_1988}), and direct-feedback-related methods (e-prop\cite{bellec_solution_2020}, DECOLLE\cite{kaiser_synaptic_2020}). Offline learning usually means that the implemented SNN is either the result of converting an equivalent ANN to SNN or an SNN that was pre-trained using Stochastic Gradient Descent (SGD) or Backpropagation (BP) and transferring the resulting weights to an SNN. This can be achieved by training an ANN with a framework like \textbf{PyTorch}\cite{paszke2019pytorch} or setting appropriate network parameters in \textbf{NEST}\cite{gewaltig2007nest}. The trained networks can then be converted to their FPGA-compliant SNN counterparts, i.e., supporting fixed-point arithmetic and other features, making them focused on available resources on those kinds of platforms, via specialized tools such as \textbf{Brevitas}\cite{brevitas}. However, there are also software-based frameworks, such as \textbf{snnTorch}\cite{eshraghian2021training} that apply training methods directly to SNNs, which can then be implemented in hardware.

\vspace{-0.35cm}
\section{Proposed Taxonomy for Neuromorphic Architectures}
\label{sec:taxonomy}

\begin{wrapfigure}[14]{o}{0.5\textwidth}
  \centering
  \includegraphics[width=0.9\linewidth, trim ={0 0 0 60pt}]{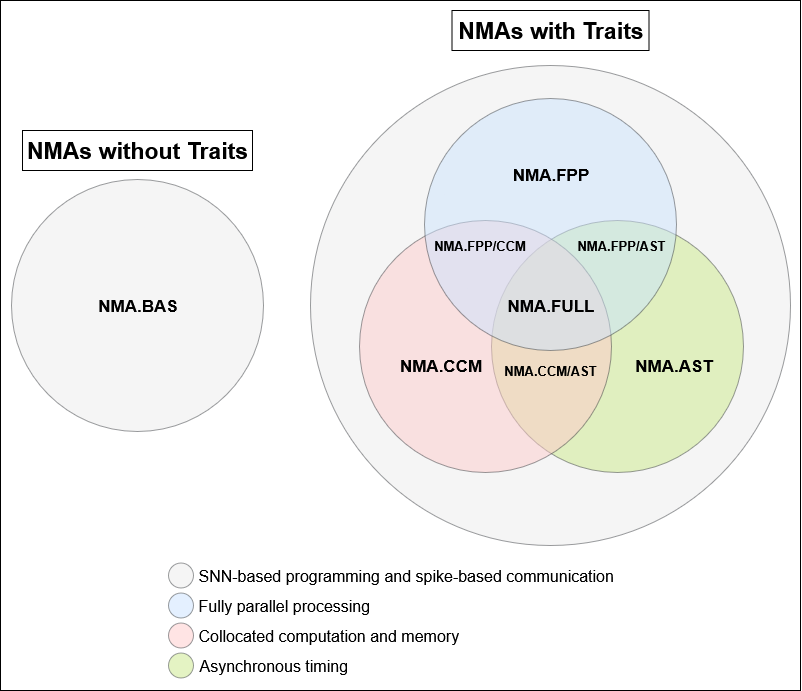}
  \caption{Proposed taxonomy - classes.}
  \label{fig:taxonomy_venn}
\end{wrapfigure}
Schuman et al.\cite{schuman_opportunities_2022} suggest that for a system to be called \textit{neuromorphic}, it must fulfill the following requirements: \textbf{(i)} it must compute using neurons and synapses (as opposed to instructions of a classical computer), \textbf{(ii)} it must communicate with spikes (as opposed to multi-bit values in a classical system), \textbf{(iii)}
it is massively parallel (that is, it can compute all neurons/synapses fully in parallel), \textbf{(iv)} it co-locates computation and memory (unlike von Neumann-based classical system), \textbf{(v)} it operates in an event-driven manner (the network state can be updated asynchronously, i.e., the neurons are not updated in sequence). Table~\ref{tab:schuman} contrasts these properties with those of a classical (von Neumann) computer system. We propose a Taxonomy for hardware neuromorphic architectures (NMAs) that builds upon that of Schuman et al., but relaxes its constraints. We argue that while a neuromorphic system must be programmed using neurons and synapses and communicate with spikes \textbf{(i--ii)}, the remaining requirements \textbf{(iii-v)} are, in fact, architectural design decisions that define the properties of the implementation of a hardware neuromorphic architecture (NMA). We assume that an NMA \textbf{must} perform computations using neurons and synapses and communicate with spikes, but it \textbf{may} also be \textit{massively parallel}, \textit{asynchronous}, and/or \textit{co-locate computations and memory}. Ultimately, this results in a Taxonomy consisting of eight classes, shown in Figure~\ref{fig:taxonomy_venn}.
\begin{table}
  \centering
  \caption{Comparison between von Neumann architectures and NMAs according to Schuman et al. Adapted from the Figure in \cite{schuman_opportunities_2022}.}
  \label{tab:schuman}
  \begin{tabular}{c|cc}
    \toprule
    \textbf{Property} & \textbf{von Neumann}             & \textbf{Neuromorphic}             \\
    \midrule
    Operation         & Sequential processing            & Massively parallel processing     \\
    Organization      & Separated computation and memory & Co-located computation and memory \\
    Programming       & Code as binary instructions      & Spiking Neural Network            \\
    Communication     & Binary data                      & Spikes                            \\
    Network update    & Synchronous (clock-driven)       & Asynchronous (event-driven)       \\
    \bottomrule
  \end{tabular}
\end{table}

The latter three optional characteristics will be referred to as \textbf{Traits} for the remainder of this paper. It is important to mention that the introduction of classes was not done to grade the architectures, i.e., we do not believe that there are objectively \textit{better} or \textit{worse} classes. Instead, the classification is intended to compare NMAs and implementations that are similar in terms of hardware and relate them to each other, similar to Flynn's taxonomy~\cite{skillicorn1988taxonomy}. Every class comes with its own advantages and disadvantages. The full names of the classes contain the base prefix \textbf{NMA.} symbolizing the \textit{neuromorphic architecture}, followed by a three-letter abbreviation stating which Traits are supported -- apart from the class \textbf{NMA.FULL} which supports all characteristics mentioned above \textbf{(i--v)}, hence the suggestive, four-letter suffix \textit{FULL}. The abbreviations for the Traits are constructed as follows: \textit{Fully Parallel Processing} - \textbf{FPP}, \textit{Co-located Computation and Memory} - \textbf{CCM}, and \textit{ASynchronous Timing} - \textbf{AST}. Moreover, if the NMA does not support any of the Traits, it belongs to the class \textbf{NMA.BAS}, stemming from its support for only the \textit{BASic} characteristics \textbf{(i--ii)}. If an NMA supports two Traits, it is also reflected in the suffix, which takes the form of \textbf{SUFFIX\textsubscript{1}/SUFFIX\textsubscript{2}}, resulting in classes \textbf{NMA.CCM/AST}, \textbf{NMA.FPP/AST} and \textbf{NMA.FPP/CCM}. For convenience, we can also refer to the classes with their suffixes, leaving out the \textbf{NMA.} prefix -- we will use this method in the main text for improved readability.

Every class is a viable choice for implementing a neuromorphic system on an FPGA. However, because of the design choices, implementing architectures that follow any class specification can pose several challenges and limitations typical of this choice. Thus, it is crucial for designers to estimate the key implementation properties that need to be supported and choose a class for their architecture accordingly. In our Taxonomy, this choice should be prefaced with selecting the appropriate Traits that an architecture should support, which will result in a specific set of design options. For example, if one wishes to create an architecture that can support millions of HH neurons, they should probably refrain from building an architecture that supports the CCM Trait, as even the largest available platforms today may struggle to support the amount of memory and logic required for such networks. On the other hand, if energy saving is of the highest importance, they should focus on implementing an architecture that supports the AST Trait, as asynchronous (or event-driven) operation allows for potentially reducing power consumption by updating the network state only when a spiking event occurs. In Table \ref{tab:adv_disadv}, we present the potential advantages and disadvantages associated with supporting different sets of Traits in the NMAs.
\begin{table}
  \caption{Priority of different \textit{features} (performance, energy-efficiency, brain-likeness, and capacity) for the proposed classes and summaries of advantages and disadvantages of using them.}
  \label{tab:adv_disadv}
  \resizebox{\textwidth}{!}{
    \begin{tabular}{c||c|c|c|c|p{0.5\linewidth}}
      \toprule
                     & \multicolumn{4}{|c|}{\textit{Features}} &                                                                                                                                                                                                                                                                                  \\
      \textbf{Class} & \textbf{Performance}                    & \textbf{Energy-efficency} & \textbf{Brain-like} & \textbf{Capacity} & \textbf{Comments}                                                                                                                                                                                          \\ \hline \hline

      BAS            & \stars{1}                               & \stars{1}                 & \stars{1}           & \stars{5}         & Least brain-like NMAs with the highest capacity, but heavily bottlenecked by most of the state held in the auxiliary memory.                                                                               \\
      \hline
      FPP            & \stars{2.5}                             & \stars{2}                 & \stars{2.5}         & \stars{3}         & Reduces execution time by having each neuron execute in parallel with others. Memory/latency-bound subject to neuron spike activity. Capacity bound by on-chip memory for neurons.                         \\
      \hline
      CCM            & \stars{2}                               & \stars{2.5}               & \stars{2.5}         & \stars{1}         & Co-located memory and computation improve execution time through increased memory bandwidth and reduced latency by keeping state on-chip, but heavily limited in capacity due to restricted on-chip space. \\
      \hline
      AST            & \stars{1.5}                             & \stars{2.5}               & \stars{1}           & \stars{5}         & Improved energy efficiency by only computing when spikes are sent to neurons, but increased complexity and cost of the design due to inclusion of a spike distribution algorithm.                          \\
      \hline
      FPP/CCM        & \stars{4.5}                             & \stars{3.5}               & \stars{4.5}         & \stars{1.5}       & Reduced execution time due to inherently parallel operation and state held on-chip, but high resource utilization because of it.                                                                           \\
      \hline
      FPP/AST        & \stars{2.5}                             & \stars{3}                 & \stars{2.5}         & \stars{3}         & Fully-parallel processing of sparse events may greatly reduce the power consumption and increase performance, but can still be latency- and memory-bound.                                                  \\
      \hline

      CCM/AST        & \stars{2.5}                             & \stars{3.5}               & \stars{2.5}         & \stars{1}         & Most state kept on-chip increases bandwidth and reduces access latency. Event-driven nature improves energy-efficiency without fully exploiting parallelism.                                               \\
      \hline
      FULL           & \stars{5}                               & \stars{5}                 & \stars{5}           & \stars{0.5}       & Most brain-like approach that combines fully parallel execution with keeping most state on-chip with asynchronous execution, but with potentially the smallest capacity.
      \\
      \hline
      \bottomrule
    \end{tabular}
  }
\end{table}

\section{Overview of the existing NMAs}
\label{sec:overview}
\begin{wrapfigure}[11]{o}{0.4\textwidth}
  \centering
  \includegraphics[width=0.95\linewidth, trim = {30pt 30pt 30pt 60pt}]{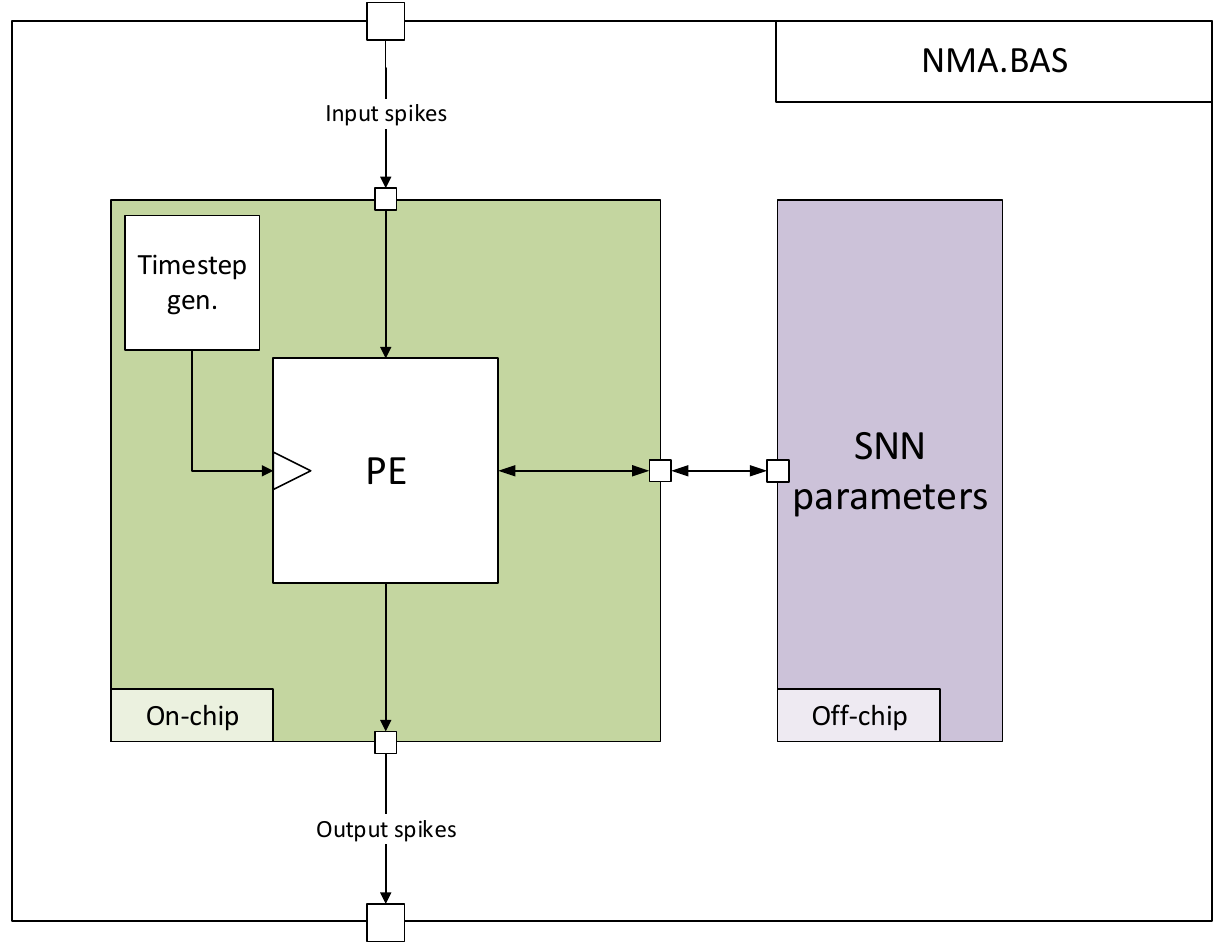}
  \caption{High-level diagram of an BAS system.}
  \label{fig:class0}
\end{wrapfigure}

\begin{wrapfigure}[11]{o}{0.4\textwidth}
    \centering
    \includegraphics[width=0.95\linewidth, trim = {30pt 30pt 30pt 50pt}]{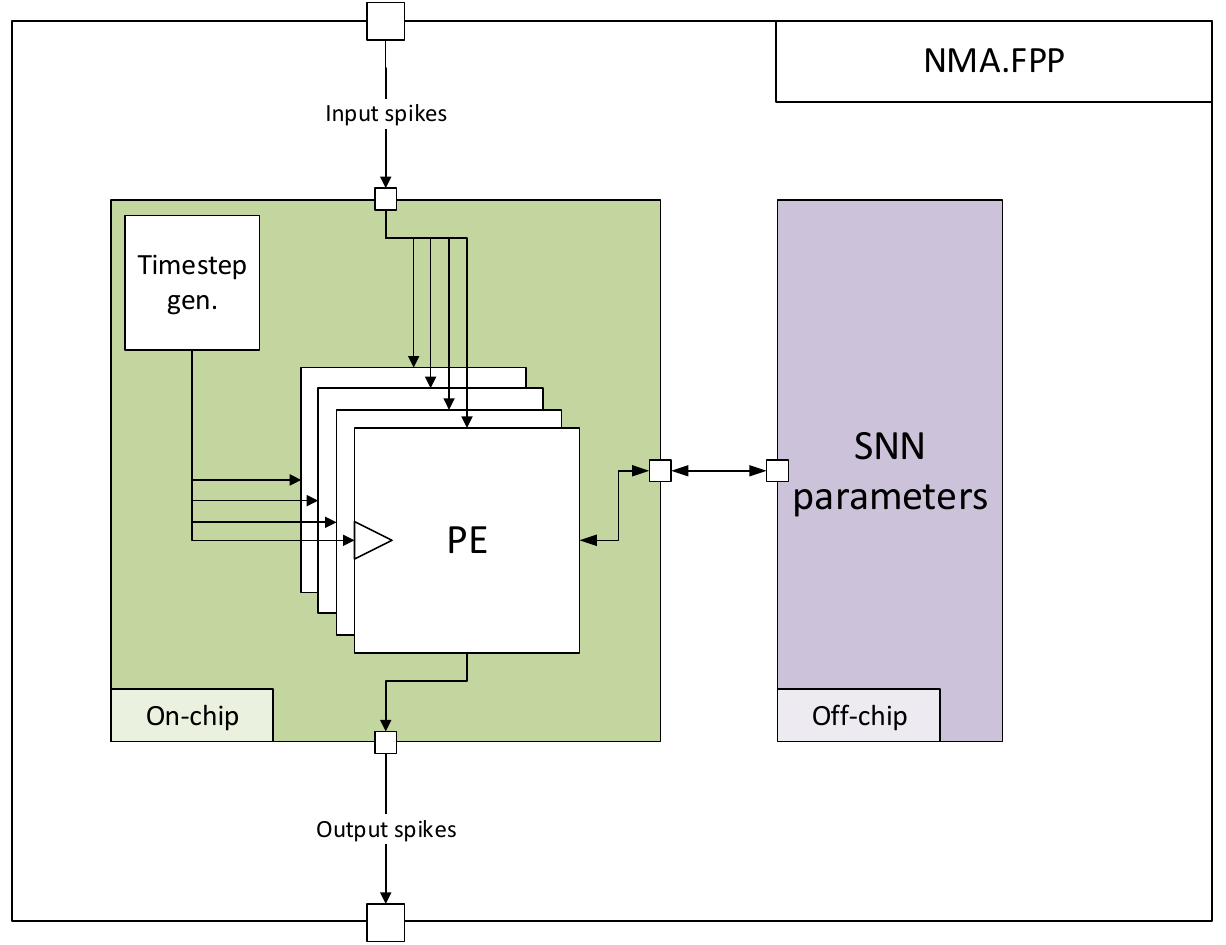}
    \caption{High-level diagram of an FPP system.}
    \label{fig:class1}
\end{wrapfigure}

\begin{wrapfigure}[11]{o}{0.4\textwidth}
    \centering
    \includegraphics[width=0.95\linewidth, trim = {30pt 30pt 30pt 60pt }]{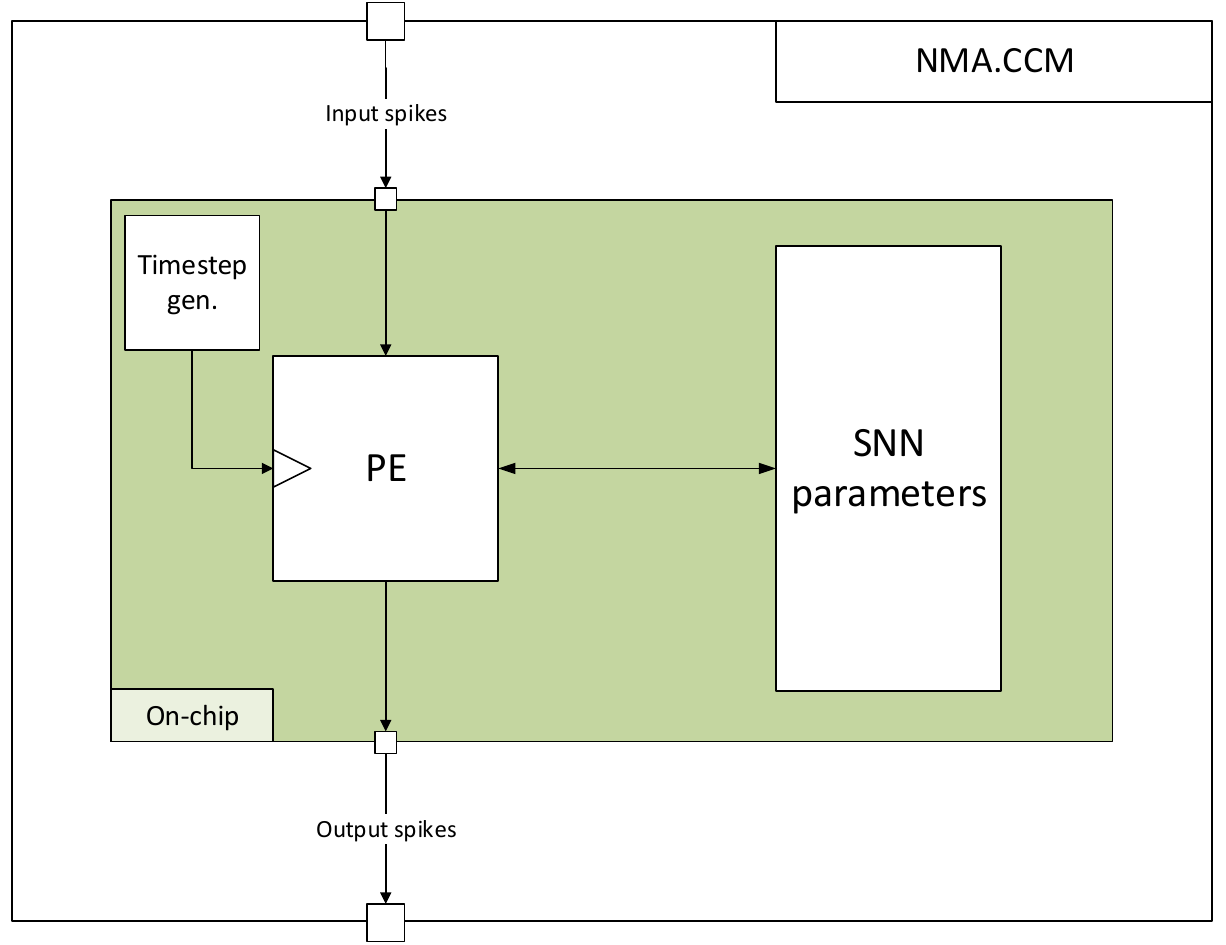}
    \caption{High-level diagram of a CCM system.}
    \label{fig:class2}
\end{wrapfigure}

\begin{wrapfigure}[11]{o}{0.4\textwidth}
  \centering
  \includegraphics[width=0.95\linewidth, trim = {30pt 30pt 30pt 50pt}]{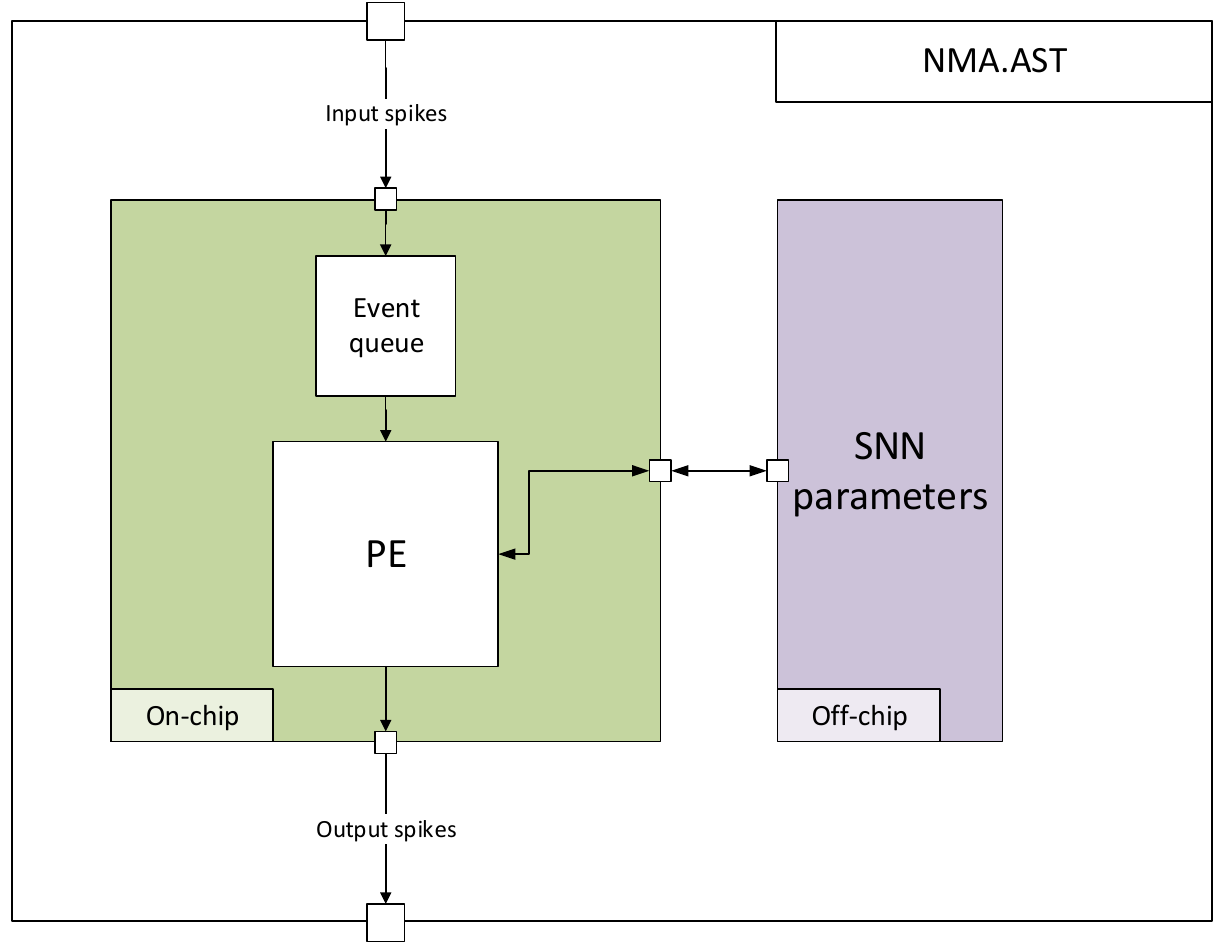}
  \caption{High-level diagram of an AST system.}
  \label{fig:class3}
\end{wrapfigure}

\begin{wrapfigure}[11]{o}{0.4\textwidth}
    \centering
    \includegraphics[width=0.95\linewidth, trim = {30pt 30pt 30pt 50pt}]{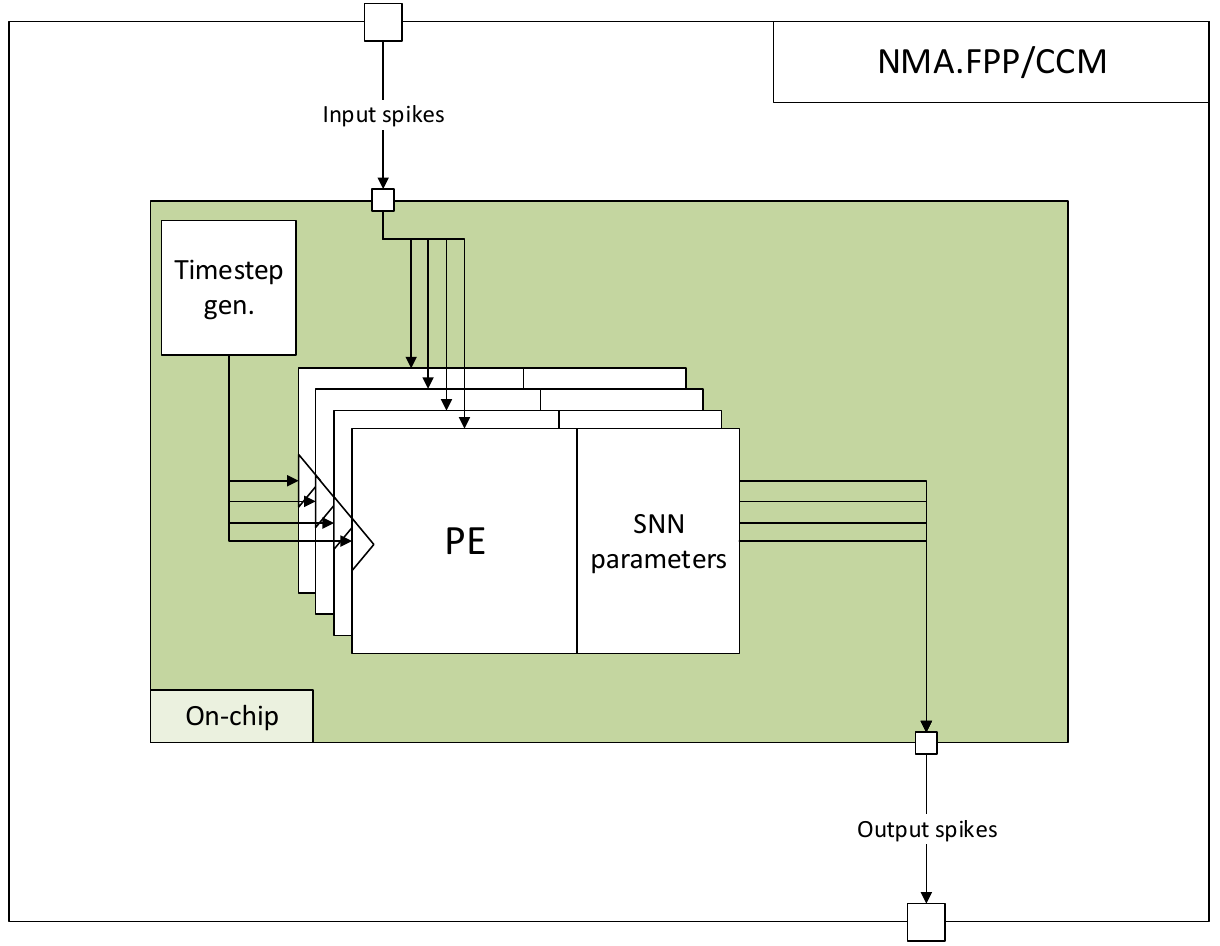}
    \caption{High-level diagram of FPP/CCM system.}
    \label{fig:class4}
\end{wrapfigure}

\begin{wrapfigure}[10]{o}{0.4\textwidth}
  \centering
  \includegraphics[width=0.95\linewidth, trim = {30pt 30pt 30pt 80pt}]{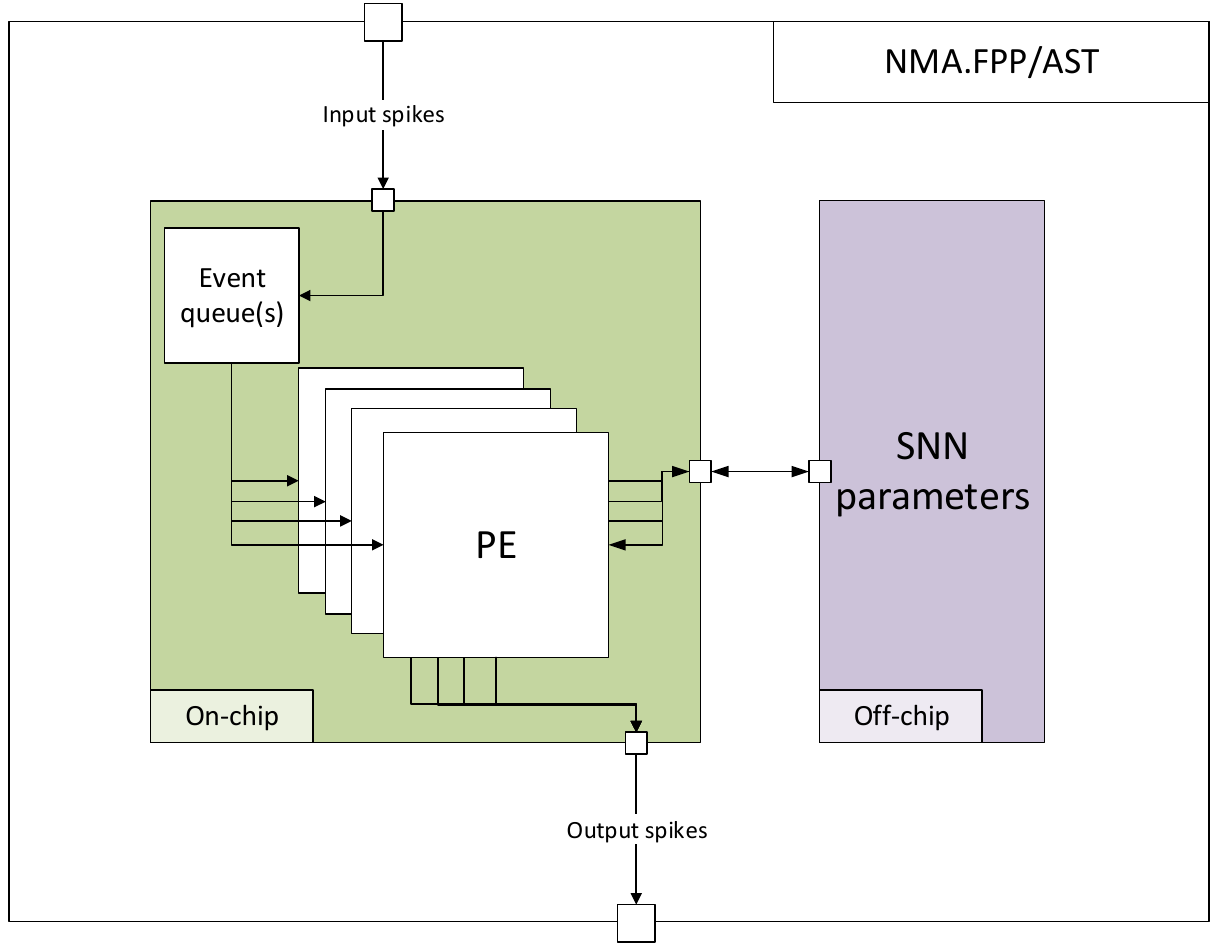}
  \caption{High-level diagram of FPP/AST system.}
  \label{fig:class5}
\end{wrapfigure}

\begin{wrapfigure}[11]{o}{0.4\textwidth}
  \centering
  \includegraphics[width=0.95\linewidth, trim = {30pt 30pt 30pt 50pt}]{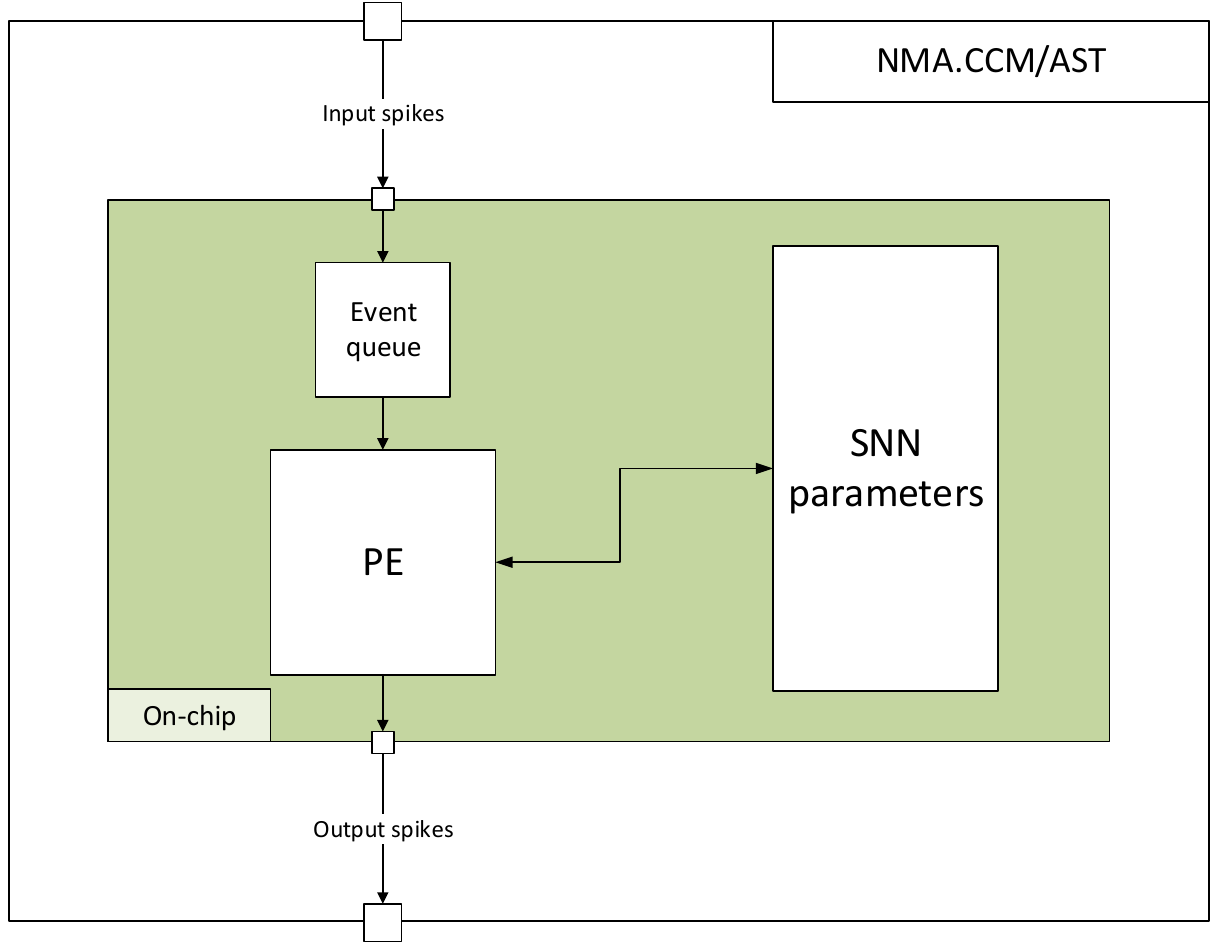}
  \caption{High-level diagram of CCM/AST system.}
  \label{fig:class6}
\end{wrapfigure}

\begin{wrapfigure}[11]{o}{0.4\textwidth}
    \centering
    \includegraphics[width=0.95\linewidth, trim = {30pt 30pt 30pt 60pt}]{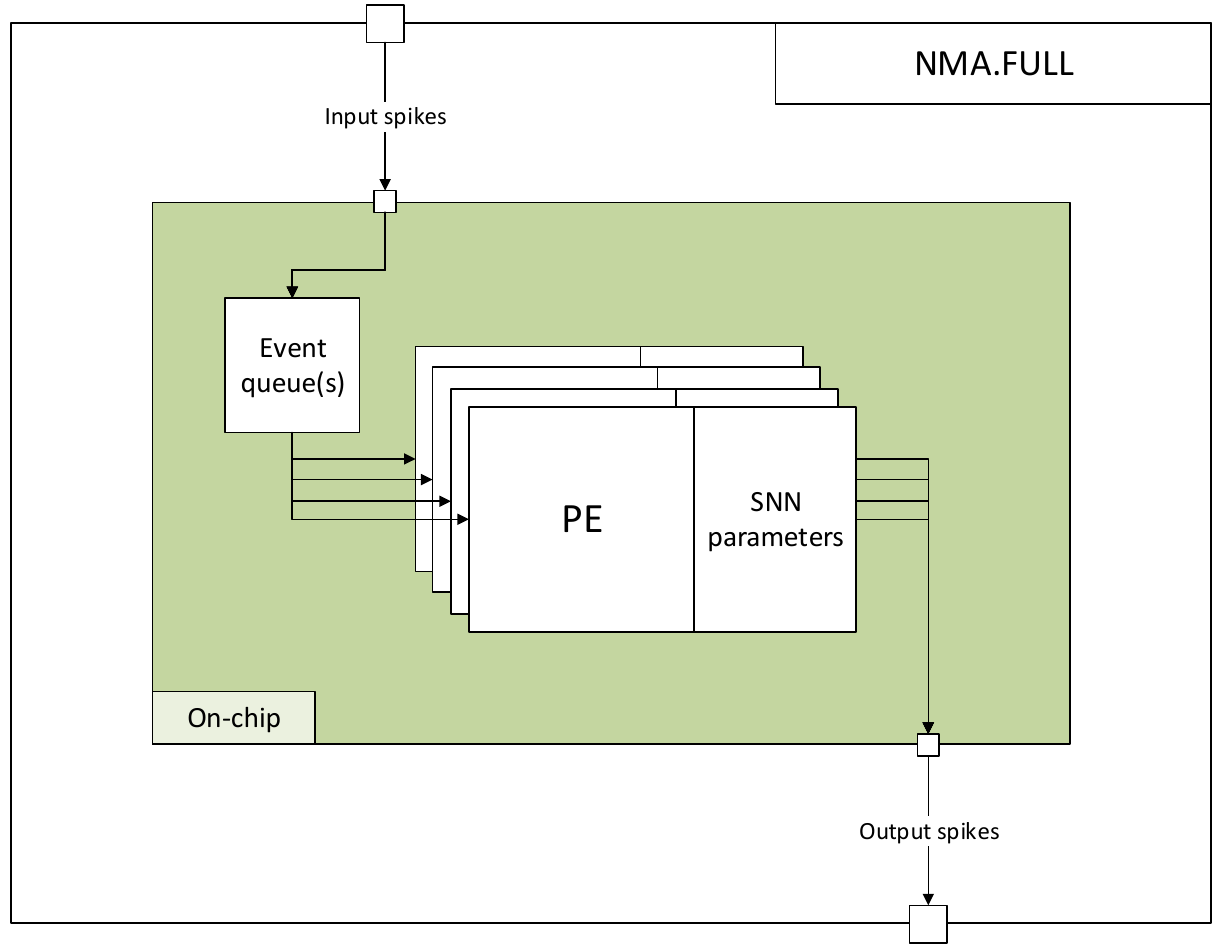}
    \caption{High-level diagram of an FULL system.}
    \label{fig:class7}
\end{wrapfigure}

In this section, we present the NMAs arranged by the classes of the Taxonomy. Every following subsection consists of the following parts. Firstly, we provide a more elaborate description of typical choices present in architectures belonging to a given class. We then present a set of \textit{landmark architectures} and describe their implementation in greater detail, with focus on the interesting design choices and compliance with the Taxonomy. Those systems can be treated as a quick reference for what architectural design choices one can expect from a specific class. We include also abstract visualizations of the architectures in figures \ref{fig:class0}, \ref{fig:class1}, \ref{fig:class2}, \ref{fig:class3}, \ref{fig:class4}, \ref{fig:class5}, \ref{fig:class6} and \ref{fig:class7}. In those figures, if a single PE unit is shown, it represents a situation where $\#PE < \#VN$, whereas if multiple PEs are shown -- $\#PE = \#VN$. Moreover, tables \ref{tab:class0}, \ref{tab:class1}, \ref{tab:class2}, \ref{fig:class3}, \ref{fig:class4}, \ref{tab:class5}, \ref{tab:class6} and \ref{tab:class7} list the information that allow for quick comparison between the architectures within the same class based on their most basic statistics. Due to the scope of this survey and the varying nature of the provided details about the architectural choices, we were unable to fit every piece of information into those tables. Instead, we focused on presenting the features that allowed us to group them into classes of our Taxonomy, as well as showcase their capabilities in realizing the selected (presented) use cases. To identify and select which publications to review in our survey, we adapted systematic and reproducible methods, which are outlined in the steps below:
\vspace{-0.25cm}
\begin{enumerate}
  \item With the help of the search experts from KTH Royal Institute of Technology library, we crafted a search string aimed at capturing all publications that might be relevant for the study, which we used to search in the Web of Science, Scopus, and IEEE Xplore databases. The search string we used was:
        \begin{quote}
          \texttt{( TITLE-ABS-KEY ( "spiking model*" OR "spiking neur* model*" OR "spiking network*" OR "Spiking Neur* Network*" OR neuromorphic OR ( ( biological* OR  realistic* ) W/3 \\( "Neur* Network*" OR "neur* model*" ) ) )AND TITLE-ABS-KEY ( "coarse grained\\reconfigurable architecture*" OR "coarse grained reconfigurable array*" OR cgra\\OR fpga OR "field programmable gate array*" OR "reconfigurable architecture*" ))}
        \end{quote}
  \item The initial search query yielded more than 2128 results. We filtered and removed any duplicates and also pruned the results by reading all remaining abstracts and discarding publications irrelevant to the study (e.g., those targeting Deep Neural Networks (DNNs), architectures based on CPUs, or GPUs).
  \item The remaining 135 publications were read, analyzed in detail, classified according to our proposed taxonomy (see Section \ref{sec:taxonomy}), and included in the survey.
\end{enumerate}
\vspace{-0.1cm}

The data gathered from the articles relates to the implementations that were physically implemented in hardware. Moreover, the metrics presented in the tables -- related to e.g., number of synapses, topologies -- were gathered based on the following descriptions. \textbf{Supported (Supp.) models} relate to implemented spiking neuron models. \textbf{Number (Num.) representation (repres.)} relate to chosen fixed-point or floating-point (half, float, double) number representation for weights (\textit{w}), membrane (\textit{m}) or both (if not specified) -- more on this topic in Appendix \ref{sec:app_ext_repr}. We describe this further in the paragraph below. \textbf{Online learning} shows what kind of on-chip learning algorithm is supported (if only \textit{yes} is stated, check footnote). \textbf{Max. \#Neurons/\#Synapses (\#Neu./\#Syn.)} is the maximum \textit{reported} number of neurons and synapses supported by the architecture. \textbf{Use case} shows selected test case/use case for the architecture and \textbf{Use case \#Neurons/\#Synapses (\#Neu./\#Syn.)} -- number of \textit{implemented} neurons/synapses for the provided test/use case. \textbf{Topology (Topo.)} relates to selected SNN topology for the provided test/use case. \textbf{Device} is the family of the FPGA device used as a platform for the architecture, \textbf{Utilization} is the percentage of Logic/Memory/Digital Signal Processing (DSP) units used to implement the system and \textbf{Frequency (Freq.)} is the clock speed of the system.

\subsection{NMA.BAS}
\label{sec:overview_class0}
\begin{table}[h]
  \caption{NMA.BAS implementations. Hardware implementation comparison with quantity capabilities. (in bold -- a holotype specimen described in detail in the paragraph, * -- Flip-Flops as on-chip memory reported, \textsuperscript{$\dagger$} -- \textit{multi-device-by-design} system).}
  \label{tab:class0}
  \resizebox{\textwidth}{!}{
    \begin{tabular}{p{0.15\linewidth}|c|cccccp{0.15\textwidth}cp{0.08\textwidth}p{0.13\textwidth}p{0.15\textwidth}c}
      \toprule
      \multirow{2}{*}{\textbf{Source}}                                   & \multirow{2}{*}{\textbf{Year}} & \textbf{Supp.}               & \textbf{\#PE} & \textbf{Num.}                   & \textbf{Online}                    & \textbf{Max.}          & \textbf{Use case}                      & \textbf{Use case}      & \textbf{Topo.}                 & \textbf{Device}             & \textbf{Utilization}    & \textbf{Freq.}                \\
                                                                         &                                & \textbf{models}              &               & \textbf{repres.}                & \textbf{learning}                  & \textbf{\#Neu./\#Syn.} &                                        & \textbf{\#Neu./\#Syn.} &                                &                             & \textbf{Logic/Mem/DSP}  & (MHz)                         \\
      \midrule  \midrule
      Waldemark et al.\cite{waldemark_pulse_1998}                        & 1998                           & PULSED                       & 1             & --                              & --                                 & 1/2                    & image processing                       & 1/2                    & SINGLE                         & Flex 10K                    & 19\%/--/--              & 21.5                          \\
      \hline
      Jung et al.\cite{jung_advanced_2005}                               & 2005                           & IF                           & 1             & --                              & STDP                               & 1/8                    & neuroscience                           & 1/8                    & SINGLE                         & Virtex-4                    & --                      & --                            \\
      \hline
      Glackin et al.\cite{glackin_novel_2005}                            & 2005                           & LIF                          & 16            & 18b                             & STDP                               & 4.2k/1.96M             & 1-D coord. translation                 & 4.2k/1.96M             & FF-FC                          & Virtex-II Pro               & 100\%/100\%*/--         & 100                           \\
      \hline
      Rice et al.\cite{rice_fpga_2009}                                   & 2009                           & IZH                          & 25            & 16b                             & --                                 & 9.2k/44.6k             & classification task (96x96 binary map) & 9.2k/44.6k             & FF-FC                          & Virtex-4                    & 79\%/53\%/--            & 199                           \\
      \hline
      Nuño-Maganda et al.\cite{nuno-maganda_hardware_2009}               & 2009                           & SRM                          & 32            & w: 16b                          & SpikeProp                          & 768/131k               & NN research                            & 768/131k               & FF-FC                          & Virtex-II Pro               & 62\%/37\%/96\%          & --                            \\
      \hline
      Kousanakis et al.\cite{kousanakis_architecture_2017}               & 2017                           & LIF                          & 40            & w: 32b                          & LTP/LTD                            & 15.3k/7.86M            & NN research                            & 240/810k               & RAND                           & Virtex-6                    & 52\%/99\%/28\%          & 150                           \\
      \hline
      \textbf{Podobas \& Matsuoka\cite{podobas2017designing}}            & \textbf{2017}                  & \textbf{IZH/HH}              & \textbf{1}    & \textbf{--}                     & \textbf{--}                        & \textbf{14k/196M}      & \textbf{NN research}                   & \textbf{14k/196M}      & \textbf{A2A}                   & \textbf{Stratix V}          & \textbf{33\%/43\%/75\%} & \textbf{247.4}                \\
      \hline
      Galindo Sanchez \& Nunez-Yanez\cite{galindo_sanchez_energy_2017}   & 2017                           & IZH                          & 32            & float or 32/16/8b               & --                                 & 28.9k/4.91M            & NN research                            & 28.9k/4.91M            & FF-FC                          & Zynq-7020                   & 18\%/81\%/28\%          & --                            \\
      \hline
      Mostafa et al.\cite{mostafa_fast_2017}                             & 2017                           & IF                           & 8             & m: 16b, w: 8b                   & --                                 & 2k/--                  & classification task (MNIST)            & 610/476k               & FF-FC                          & Spartan-6                   & 4\%/1\%/--              & --                            \\
      \hline
      Humaidi \& Kadhim\cite{humaidi_spiking_2018}                       & 2018                           & --                           & 1             & --                              & STDP                               & 48/720                 & letter recognition (5x3, 4 letters)    & 48/720                 & FF-FC                          & Cyclone II                  & 14\%/0\%/28\%           & --                            \\
      \hline
      \textbf{Ahn\cite{ahn_implementation_2020}}                         & \textbf{2020}                  & \multirow{2}{*}{\textbf{HH}} & \textbf{8}    & \multirow{2}{*}{\textbf{float}} & \multirow{2}{*}{\textbf{STDP/LTP}} & \textbf{12M/600M}      & \multirow{2}{*}{\textbf{NN research}}  & \textbf{12M/600M}      & \multirow{2}{*}{\textbf{RAND}} & \textbf{Virtex}             & \textbf{12\%/50\%/15\%} & \multirow{2}{*}{\textbf{300}} \\
                                                                         &                                &                              & \textbf{32}   &                                 &                                    & \textbf{1M/50M}        &                                        & \textbf{1M/50M}        &                                & \textbf{UltraScale+}        & \textbf{38\%/87\%/50\%} &                               \\
      \hline
      Zhou \& Hu\cite{zhou_neuromorphic_2021}\textsuperscript{$\dagger$} & 2021                           & LIF                          & 64            & w: 16b                          & --                                 & 65.5k/65.5M            & classification task (MNIST)            & 2k/1.79M               & FF-FC                          & Virtex-6\textsuperscript{1} & 15.6\%/--/--            & 105                           \\
      \hline
      \textbf{Ogaki \& Sato\cite{ogaki_hodgkin-huxley-based_2021}}       & \textbf{2021}                  & \textbf{HH}                  & \textbf{2}    & \textbf{float}                  & \textbf{--}                        & \textbf{20.7k/82.1k}   & \textbf{NN optimization}               & \textbf{10k/20.2k}     & \textbf{2D array}              & \textbf{Alveo U200}         & \textbf{25\%/27\%/23\%} & \textbf{200}                  \\
      \bottomrule
      \multicolumn{13}{l}{\textsuperscript{1}The system used four Virtex-6 devices, with the neuromorphic processing part implemented on one of them.}
    \end{tabular}
  }
\end{table}

BAS architectures support only the SNN-based programming model and spike-based communication from the Traits listed in Section \ref{sec:taxonomy}. It is common for them to include fewer Processing Elements (PEs) than the neurons they simulate -- they incorporate some form of time-domain multiplexing of resources (TDM) to realize larger networks. The PEs process neuron data stored in external memory and update appropriate neurons in sequence, making them well-suited for e.g., neuroscientific simulations of larger SNNs. Those systems have the potential to simulate the largest networks on a single device due to the arbitrary size of the attached off-chip memory, but they are also primarily memory-bound, and their performance is dictated by the memory transfers, which can be costly\cite{upadhyay_emerging_2019}. They are the closest to standard von Neumann-like accelerator systems and draw the least inspiration from biology among the architectures listed in this survey. \textbf{Landmark architectures for this class include three systems.} In the article by \textbf{Ahn\cite{ahn_implementation_2020}, a Neuron Machine (NM)} was presented -- a physical pipeline for HH neurons with the possibility of interconnecting eight such

\noindent
pipelines in a ring topology, with a central control unit connected in a star pattern. The architecture implemented data duplication per router, which allowed for reduced latency in updating the network state and its availability to the rest of the NMs in the network. Specifically, the spike information was available immediately, on the next network timestep. Moreover, the dual-port memory units located at corresponding addresses in all the routers were tied together to form an axon bus, allowing for common axon delay modeling for appropriate synapses. The network data was stored in HBM memory and/or on-chip, depending on the network conditions -- if the network was too large, the entire network state and connectivity information was stored in HBM, while with smaller networks, the neuron states were stored on-chip to increase the computation speed. Another \textit{landmark} architecture can be found in the article by \textbf{Ogaki \& Sato\cite{ogaki_hodgkin-huxley-based_2021}}, featuring multiple parallel computation pipelines for HH neurons, fed by a common streaming unit connected to the host machine with the network parameters stored in the auxiliary memory. Another example can be found in a work by \textbf{Podobas \& Matsuoka\cite{podobas2017designing}}, where a single PE was used to process a network of HH or IZH neurons (depending on the configuration), with parameters stored in auxiliary memory.

\vspace{-0.35cm}
\subsection{NMA.FPP}
\label{sec:overview_class1}
FPP architectures expand the basic implementations of BAS by providing a bespoke physical PE for every neuron of the programmed SNN, potentially increasing the computation speed by the cost of higher logic utilization (in comparison to a TDM-ed set of PEs or a single datapath). As is the case with BAS implementations, these systems are also limited

\noindent
by the memory bottleneck, as they store the majority of the necessary network states in auxiliary memory. They also process the neurons in a synchronous manner, which means that there is a potential power consumption increase due to multiple physical PEs processing neuron states in every occurring timestep. Networks simulated by those architectures are expected to be smaller than it is the case with BAS, as every neuron occupies its own portion of on-chip resources. \textbf{Landmark architectures for this class include two systems.} The first selected \textit{landmark} system for this class is \textbf{Spiker presented by Carpegna et al.\cite{carpegna_spiker_2022}}, however, we acknowledge the fact that only four systems were assigned to this class in this survey, and thus, more fitting architectures could potentially be selected had more examples been found. However, this system fulfills all the necessary requirements stated for the assignment as a class FPP. In the article, the authors proposed a hierarchical architecture, organized into network, layer, and neuron subunits, connected to their bespoke control units. Due to the selected type of the WTA network, the architecture consisted of two-layer units with excitatory connections between the layers and lateral inhibition. Every layer consisted of a bespoke neuron processing datapath, and thus, the entire architecture was rather fixed for a specific SNN. This also seems to be the case with this kind of systems, where they often follow the SNN topology they aim at simulating. Other than Spiker, a system presented by \textbf{Moctezuma et al.\cite{moctezuma_numerically_2013}} can be viewed as another potential \textit{landmark}, with fully parallel PEs responsible for computing elaborate Traub/HH neuron models, connected to a common external memory and central control unit.
\begin{table}[h]
    \caption{NMA.FPP implementations. Hardware implementation comparison with quantity capabilities.}
    \label{tab:class1}
    \resizebox{\textwidth}{!}{
        \begin{tabular}{p{0.15\linewidth}|c|cccccp{0.15\textwidth}cp{0.08\textwidth}p{0.13\textwidth}p{0.15\textwidth}c}
            \toprule
            \multirow{2}{*}{\textbf{Source}}                           & \multirow{2}{*}{\textbf{Year}} & \textbf{Supp.}    & \textbf{\#PE} & \textbf{Num.}          & \textbf{Online}   & \textbf{Max.}                 & \textbf{Use case}                    & \textbf{\#Neu./\#Syn.} & \textbf{Topo.} & \textbf{Device}   & \textbf{Utilization}   & \textbf{Freq.} \\
                                                                       &                                & \textbf{models}   &               & \textbf{repres.}       & \textbf{learning} & \textbf{\#Neu./\#Syn.}        &                                      &                        &                &                   & \textbf{Logic/Mem/DSP} & (MHz)          \\
            \midrule  \midrule
            Maya et al.\cite{maya_compact_2000}                        & 2000                           & PULSED            & 3             & --                     & --                & "1000s"\textsuperscript{1}/-- & XOR                                  & 3/2                    & FF-FC          & Virtex 2.5V       & 28\%/--/--             & 90             \\
            \hline
            Xicotencatl \& Arias-Estrada\cite{xicotencatl_fpga_2003}   & 2003                           & PULSED            & 1.1k          & --                     & --                & 1.1k/1k                       & classification task (not specified)  & 1.1k/1k                & 2D array       & Virtex-II         & 97\%/--/--             & 35.6           \\
            \hline
            \textbf{Moctezuma et al.\cite{moctezuma_numerically_2013}} & \textbf{2013}                  & \textbf{HH/Traub} & \textbf{24}   & \textbf{float}         & \textbf{--}       & \textbf{24/576}               & \textbf{neuroscience}                & \textbf{24/576}        & \textbf{RAND}  & \textbf{Virtex-6} & \textbf{97\%/--/--}    & \textbf{100}   \\
            \hline
            \textbf{Carpegna et al.\cite{carpegna_spiker_2022}}        & \textbf{2022}                  & \textbf{LIF}      & \textbf{400}  & \textbf{m: 16b, w: 5b} & \textbf{--}       & \textbf{400/40.8k}            & \textbf{classification task (MNIST)} & \textbf{400/40.8k}     & \textbf{WTA}   & \textbf{Artix-7}  & \textbf{55\%/32\%/--}  & \textbf{100}   \\
            \bottomrule
            \multicolumn{13}{l}{\textsuperscript{1}The authors claimed it is possible to implement 1000s of neurons, but did not provide a specific number or reference.}
        \end{tabular}
    }
\end{table}

\vspace{-0.5cm}
\subsection{NMA.CCM}
\label{sec:overview_class2}
\begin{table}[htp]
    \caption{NMA.CCM implementations. Hardware implementation comparison with quantity capabilities (* -- Flip-Flops as on-chip memory reported, $^{\dagger}$ -- \textit{multi-device-by-design} system).}
    \label{tab:class2}
    \resizebox{\textwidth}{!}{
        \begin{tabular}{p{0.15\linewidth}|c|cccccp{0.15\textwidth}cp{0.08\textwidth}p{0.13\textwidth}p{0.15\textwidth}c}
            \toprule
            \multirow{2}{*}{\textbf{Source}}                                         & \multirow{2}{*}{\textbf{Year}} & \textbf{Supp.}               & \textbf{\#PE}                & \textbf{Num.}                & \textbf{Online}              & \textbf{Max.}          & \textbf{Use case}                               & \textbf{\#Neu./\#Syn.} & \textbf{Topo.}                 & \textbf{Device}                    & \textbf{Utilization}                     & \textbf{Freq.}                \\
                                                                                     &                                & \textbf{models}              &                              & \textbf{repres.}             & \textbf{learning}            & \textbf{\#Neu./\#Syn.} &                                                 &                        &                                &                                    & \textbf{Logic/Mem/DSP}                   & (MHz)                         \\
            \midrule  \midrule
            Schrauwen et al.\cite{schrauwen_compact_2008}                            & 2008                           & SRM\textsubscript{0}         & 5                            & 10b                          & --                           & 1.6k/19.2k             & classification task (spoken digit)              & 200/2.4k               & LSM                            & Virtex-4                           & 4\%/5\%/0\%                              & 115                           \\
            \hline
            Thomas \& Luk\cite{thomas_fpga_2009}                                     & 2009                           & IZH                          & 1                            & double/8b                    & --                           & 1k/1.05M               & neuroscience                                    & 1k/1.05M               & A2A                            & Virtex-5                           & 21\%/84\%/50\%                           & 133                           \\
            \hline
            Wildie et al.\cite{wildie_reconfigurable_2009}                           & 2009                           & IZH                          & 8                            & 24b                          & --                           & 1k/--                  & neuroscience                                    & 1k/--                  & RAND                           & Virtex-5                           & 18\%/43\%/--                             & 20                            \\
            \hline
            \textbf{Ambroise et al.\cite{ambroise_biorealistic_2013}}                & \textbf{2013}                  & \textbf{IZH}                 & \textbf{1}                   & \textbf{m: 18b}              & \textbf{--}                  & \textbf{117/13.7k}     & \textbf{NN research}                            & \textbf{117/13.7k}     & \textbf{A2A}                   & \textbf{Virtex-4}                  & \textbf{4\%/3\%/1\%}                     & \textbf{--}                   \\
            \hline
            Wang et al.\cite{wang_fpga_2013}                                         & 2013                           & IF                           & 129\textsuperscript{1}       & --\textsuperscript{2}        & STDP                         & 4.1k/1.15M             & NN-based memory                                 & 4.1/1.15M              & RAND                           & Virtex-6                           & 89\%/37\%/--                             & 66                            \\
            \hline
            \textbf{Ambroise et al.\cite{ambroise_real-time_2013}}                   & \textbf{2013}                  & \textbf{IZH}                 & \textbf{1}                   & \textbf{w: 26b}              & \textbf{LTD}                 & \textbf{1.92k/3.36k}   & \textbf{neuroscience (240 CPGs)}                & \textbf{1.92k/3.36k}   & \textbf{BIO}                   & \textbf{Spartan-6}                 & \textbf{2\%/16\%/6\%}                    & \textbf{--}                   \\
            \hline
            Yang et al.\cite{yang_cost-efficient_2015}                               & 2015                           & IZH                          & 1                            & 28b                          & --                           & 3k/--                  & neuroscience (basal ganglia)                    & 256/2k                 & BIO                            & Cyclone IV                         & 3\%/1\%/20\%                             & 12.5                          \\
            \hline
            Wang et al.\cite{wang_multi-fpga_2015}$^{\dagger}$                       & 2015                           & FHN                          & 12                           & --                           & --                           & 288/82.9k              & NN research                                     & 288/82.9k              & FF-FC                          & Cyclone IV                         & 8\%/2\%/--                               & 48                            \\
            \hline
            Ahn\cite{ahn_neuron-like_2015}                                           & 2015                           & HH                           & 1                            & float                        & STDP                         & 1k/1M                  & neuroscience                                    & 1k/1M                  & RAND                           & Kintex-7                           & 31\%/83\%/33\%                           & --                            \\
            \hline
            Molin et al.\cite{molin_fpga_2015}                                       & 2015                           & IF                           & 49                           & 4/6/8/10/12b                 & --                           & 19.2k/14.2k            & image processing (dewarping)                    & 19.2k/14.2k            & SCNN                           & Spartan-6                          & --                                       & 100                           \\
            \hline
            \textbf{Lei et al.\cite{lei_efficient_2016}   }                          & \textbf{2016}                  & \textbf{LIF}                 & \textbf{22}                  & \textbf{32b}                 & \textbf{--}                  & \textbf{3.98k/398k}    & \textbf{NN research}                            & \textbf{398/398k}      & \textbf{RAND}                  & \textbf{Zynq-7020}                 & \textbf{14\%/1\%/8\%}                    & \textbf{125}                  \\
            \hline
            Pani et al.\cite{pani_fpga_2017}                                         & 2017                           & IZH                          & 8                            & --                           & --                           & 1.44k/2.07M            & neuroscience                                    & 1k/1.05M               & A2A                            & Virtex-6                           & 37\%/94\%/53\%                           & 100                           \\
            \hline
            Thakur et al.\cite{thakur_real-time_2017}                                & 2017                           & IF                           & 1                            & --                           & --                           & 22.5k/44.7k            & image processing                                & 22.5k/44.7k            & 2D array                       & --                                 & 3\%/26\%/1\%                             & --                            \\
            \hline
            Akbarzadeh-Sherbaf et al.\cite{akbarzadeh-sherbaf_scalable_2018}         & 2018                           & HH\textsuperscript{3}        & 4                            & 33b                          & --                           & 5.12k/26.2M            & NN research                                     & 4.1k/16.8M             & RAND                           & Artix-7                            & 21\%/6\%/1\%                             & 71.4                          \\
            \hline
            Sripad et al.\cite{sripad_snavareal-time_2018}                           & 2018                           & LIF/IZH/HH                   & 100                          & 16b                          & STDP                         & 200/7.5k               & NN research (synfire)                           & 200/7.5k               & FF-FC                          & Kintex-7                           & 46\%/48\%12\%                            & 125                           \\
            \hline
            Zhang et al.\cite{zhang_versatile_2019}                                  & 2019                           & LIF                          & 4                            & 9b\textsuperscript{4}        & --                           & 1k/134k                & classification task (MNIST [16x16])             & 1K/134k                & FF-FC                          & Cyclone IV                         & 62\%/--/--                               & 100                           \\
            \hline
            Guo et al.\cite{guo_systolic_2019}                                       & 2019                           & IF                           & 512                          & 32b                          & --                           & 2.41k/2.39M            & classification task (MNIST)                     & 2.41k/2.39M            & FF-FC\textsuperscript{5}       & Virtex-7                           & 13\%/5\%/--                              & 100                           \\
            \hline
            Huang et al.\cite{huang_spiking_2020}                                    & 2020                           & IF                           & 2                            & w: 8b                        & --                           & 48/4.32k               & classification task (radioscope identification) & 48/4.32k               & FF-FC                          & --                                 & 5\%/2\%*/--                              & 100                           \\
            \hline
            Abdoli \& Safari \cite{abdoli_reconfigurable_2020}                       & 2020                           & IZH                          & 4                            & w: 30b                       & --                           & 18k/--                 & neuroscience (center-annular surround network)  & 4.5k/4M                & WTA                            & Artix-7                            & 23\%/6\%/5\%                             & 52.75                         \\
            \hline
            Guo et al.\cite{guo_towards_2020}                                        & 2020                           & LIF                          & 25                           & w: 16b                       & T-STDP                       & 500/392k               & classification task (MNIST)                     & 500/392k               & WTA                            & Virtex-7                           & --/7\%/--                                & --                            \\
            \hline
            Ju et al.\cite{ju_fpga_2020}                                             & 2020                           & IF                           & 3                            & w: 8b                        & --                           & 78.5k/518k             & classification task (MNIST)                     & 78.5k/518k             & SCNN                           & Zynq-UltraScale+                   & 52\%/29\%/--                             & 150                           \\
            \hline
            Gholami et al.\cite{gholami_reconfigurable_2021}                         & 2020                           & IZH                          & 1                            & 32b                          & yes\textsuperscript{6}       & 18/81                  & neuroscience (spiking behaviour)                & 18/81                  & FF-FC                          & Virtex-6                           & 14\%/1\%/0\%                             & 83.209                        \\
            \hline
            Wu et al.\cite{wu_efficient_2021}                                        & 2021                           & LIF                          & 24                           & m: 16b, w: 8b                & batch learning               & 100/78.4k              & classification task (MNIST)                     & 100/78.4k              & WTA                            & Zynq-7030                          & 100\%/14.8\%*/107.5\%\textsuperscript{7} & 301.8                         \\
            \hline
            Zheng et al.\cite{zheng_balancing_2021}                                  & 2021                           & LIF                          & 24                           & m: 16b, w: 2b                & T-STDP                       & 2.3k/--                & classification task (MNIST)                     & 2.3k/--                & VFA                            & Zynq-UltraScale+                   & 4\%/100\%/0\%                            & 200                           \\
            \hline
            \multirow{2}{*}{Aung et al.\cite{aung_deepfire_2021}}                    & \multirow{2}{*}{2021}          & \multirow{2}{*}{IF}          & \multirow{2}{*}{271}         & \multirow{2}{*}{w: 8b}       & \multirow{2}{*}{--}          & 395/252k               & classification task (MNIST)                     & 395/252k               & SCNN                           & \multirow{2}{*}{Zynq-UltraScale+}  & 5\%/16\%/4\%                             & 500                           \\
                                                                                     &                                &                              &                              &                              &                              & 5.51k/12M              & (CIFAR10)                                       & 5.51k/12M              & SCNN                           & 33\%/100\%/44\%                    & 425                                                                      \\
            \hline
            Ali et al.\cite{ali_energy_2022}                                         & 2022                           & IF                           & 2                            & 8b                           & --                           & 7/135                  & classification task (5x5 binary map)            & 7/135                  & FF-FC                          & Artix-7                            & 1\%/1\%*/--                              & 25                            \\
            \hline
            Sommer et al.\cite{sommer_efficient_2022}                                & 2022                           & IF                           & 1                            & 8b                           & --                           & 685/--                 & classification task (MNIST)                     & 685/--                 & SCNN                           & Zynq-UltraScale+                   & 9\%/20\%/--                              & 333                           \\
            \hline
            Hwang et al.\cite{hwang_replacenet_2023}                                 & 2023                           & LIF                          & 9                            & --                           & --                           & 60/3.6k                & neuroscience                                    & 60/3.6k                & A2A                            & Zynq-7020                          & 62\%/65\%/--                             & --                            \\
            \hline
            Kauth et al.\cite{kauth_neuroaix-framework_2023}$^{\dagger}$             & 2023                           & LIF/IZH                      & 350                          & m: float, w: 32b             & --                           & 89.6k/--               & neuroscience (cortical microcircuit)            & 77.2k/300M             & BIO                            & Virtex-7\textsuperscript{8}        & --                                       & --                            \\
            \hline
            \multirow{2}{*}{\textbf{Wang et al.\cite{wang_resource-efficient_2023}}} & \multirow{2}{*}{\textbf{2023}} & \multirow{2}{*}{\textbf{--}} & \multirow{2}{*}{\textbf{25}} & \multirow{2}{*}{\textbf{--}} & \multirow{2}{*}{\textbf{--}} & \textbf{464/--}        & \textbf{classification task (MNIST)}            & \textbf{464/--}        & \multirow{2}{*}{\textbf{SCNN}} & \multirow{2}{*}{\textbf{Kintex-7}} & \textbf{2\%/5\%/0\%}                     & \multirow{2}{*}{\textbf{100}} \\
                                                                                     &                                &                              &                              &                              &                              & \textbf{--/--}         & \textbf{(CIFAR10)}                              & \textbf{--/--}         &                                &                                    & \textbf{8\%/30\%/0\%}                    &                               \\
            \hline
            Wang et al.\cite{wang_large-scale_2023}                                  & 2023                           & IZH                          & 36                           & m: 15b, w: 11b               & STDP                         & 20.7k/41.2k            & path planning                                   & 20.7k/41.2k            & 2D array                       & Cyclone IV                         & 70\%/69\%/0\%                            & 134.36                        \\
            \hline
            Shi et al.\cite{shi_ghost_2024}                                          & 2024                           & IF                           & 4                            & --                           & BP/STDP                      & 1.38k/--               & classification task (MNIST)                     & 1.08k/1.11M            & FF-FC                          & Zynq-7010                          & 54\%/44\%/0\%                            & 100                           \\
            \hline
            Chen et al.\cite{chen_hardware_2024}                                     & 2024                           & IF                           & --                           & --                           & STDP/S-STDP                  & --/--                  & classification task (MNIST)                     & --/--                  & SCNN                           & Zynq-UltraScale+                   & --                                       & 100                           \\
            \hline
            Liu et al.\cite{liu_sc-plr_2024}                                         & 2024                           & LIF\textsuperscript{9}       & 64                           & --                           & SC-PLR                       & 256/201k               & classification task (MNIST)                     & 256/201k               & FF-FC                          & Zynq-7035                          & 4\%/12\%/0\%                             & --                            \\
            \hline
            \textbf{Mohammad-hassani et al.\cite{mohammadhassani_digital_2025}}      & \textbf{2025}                  & \textbf{IF}                  & \textbf{350}                 & \textbf{--}                  & \textbf{E-prop}              & \textbf{400k/6M}       & \textbf{graph traversal (shortest path)}        & \textbf{400k/6M}       & \textbf{RAND}                  & \textbf{Versal Premium}            & \textbf{76\%/100\%/--}                   & \textbf{100}                  \\
            \bottomrule
            \multicolumn{13}{l}{\textsuperscript{1}128 for neurons and one for axons.}                                                                                                                                                                                                                                                                                                                                                                                                                 \\
            \multicolumn{13}{l}{\textsuperscript{2}The authors provided only information about the delay, which was 9b.}                                                                                                                                                                                                                                                                                                                                                                               \\
            \multicolumn{13}{l}{\textsuperscript{3}The authors used COBA HH neuron, which is an even more elaborate than a standard HH neuron model, due to two additional terms in the differential equations defining its dynamics.}                                                                                                                                                                                                                                                                 \\
            \multicolumn{13}{l}{\textsuperscript{4}Only four values were available.}                                                                                                                                                                                                                                                                                                                                                                                                                   \\
            \multicolumn{13}{l}{\textsuperscript{5}The authors also implemented an SCNN for MNIST.}                                                                                                                                                                                                                                                                                                                                                                                                    \\
            \multicolumn{13}{l}{\textsuperscript{6}Online learning type not explicitly stated.}                                                                                                                                                                                                                                                                                                                                                                                                        \\
            \multicolumn{13}{l}{\textsuperscript{7}We evaluated this value based on the data provided by the authors and provided platform. However, the excess in DSP resource usage was not addressed in the paper.}                                                                                                                                                                                                                                                                                 \\
            \multicolumn{13}{l}{\textsuperscript{8}This system used 35 NetFPGA SUME boards.}                                                                                                                                                                                                                                                                                                                                                                                                           \\
            \multicolumn{13}{l}{\textsuperscript{9}The neuron model used here was a stochastic LIF.}
        \end{tabular}
    }
\end{table}

CCM architectures assume storing all the necessary circuitry and memory on-chip, which alleviates the issue with memory bottleneck and expensive auxiliary memory transfers, by the cost of increased on-chip resource utilization, where not only the neuron processing circuitry needs to \textit{fit} on-chip, but also logic related to memory movements,

\noindent
as well as the actual memory elements utilization -- which is often limited. As such, this kind of implementation is more dependent on the used platform than the previous two classes, as not only the available logic resources are the boundary for the possible size of the network to be implemented, but also the memory capacity. However, this class is also the broadest of the surveyed architectures, as the requirements for co-located memory and computation are rather easy to satisfy. Thus, selecting an appropriate exemplary implementation required a bit more thought, as multiple systems were equally good picks. It is important to keep in mind, though, that those were different types of systems, with a specific focus on different implementation details, often stemming from different SNNs that they aim to support. \textbf{Landmark architectures for this class include four systems, which represent four main organizational schemes,} namely: \textbf{(i)}several parallel units, which consist of at least one PE supporting TDM connected through a shared bus to on-chip memory, as well as some form of centralized control circuitry that monitors and manages the operation of the system, \textbf{(ii)}single PE, \textbf{(iii)}NoC-like structures and \textbf{(iv)}direct implementation of the simulated network. We selected \textbf{ENA presented by Lei et al.\cite{lei_efficient_2016}}, \textbf{SENTIENCE presented by Mohammadhassani et al.\cite{mohammadhassani_digital_2025}}, system proposed by \textbf{Ambroise et al.\cite{ambroise_biorealistic_2013, ambroise_real-time_2013}}, and architecture presented by \textbf{Wang et al.\cite{wang_resource-efficient_2023}} for the aforementioned groups, respectively. Firstly, \textbf{ENA} itself represented a unit for simulating a particular layer of neurons, thus making it suitable primarily for FF-like networks. Within those layers, a number of physical PEs were connected to a common communication interface, through which the global communication scheme was accessed -- in the form of a shared bus, which was responsible for delivering the spikes to the appropriate layer units. What is interesting is that the authors suggested a rather high precision of the fixed-point representation -- 32 bits -- which is not that common for this particular neuron model, as satisfactory performance is usually achieved with lower precision. \textbf{SENTIENCE}, which was a NoC-like structure, was used for graph traversal and finding the shortest path by realizing the nodes of the graph as IF neurons and the vertices as trainable synapses. The shortest path is found through the \textit{spiking wavefront} reaching the target from the start node, which is obtained in the amount of time it takes said wavefront to reach the destination. \textbf{Ambroise et al.}\cite{ambroise_biorealistic_2013, ambroise_real-time_2013} used a single PE to process the neurons, and included a bespoke exponential decay computation core for the synaptic current. Finally, \textbf{Wang et al. (2023) \cite{wang_resource-efficient_2023}} focused on creating an architecture for SCNNs with module reusability (not storing the intermediate states of neurons in memory but propagating it directly through the network).

\subsection{NMA.AST}
\label{sec:overview_class3}
AST architectures follow the principle of event-driven operation, i.e., updating the network state if and only if a spiking event occurs, which is then propagated to the postsynaptic neurons. This approach, due to inherent sparsity of occurring spiking events in SNNs, has the potential of reducing the power consumption of the system, however the access to the external memory is less organized, making benefiting from optimization in memory accesses such as burst reads/writes less probable, which could be of high importance considering the necessity to fetch the data from auxiliary memory. \textbf{Landmark architectures for this class include three systems.} \textbf{Minitaur introduced by Neil et al.\cite{neil_minitaur_2014}} contained a number of \textit{Minitaur cores} (primarily consisting of neuronal and synaptic datapaths), which were connected to auxiliary memory storing the network parameters, as well as in parallel to a common spike distributor. Every core had a 2kB state cache and 8kB weight cache, which exploited the potential spiking patterns occurring between a subpopulation of closely located neurons. This mechanism allowed the cores to hold information about the recently active and/or fre--

\noindent
--quently spiking neurons to alleviate the aforementioned problem with sparse and unorganized access to the auxiliary memory. For completeness, we also mention the \textbf{n-Minitaur} system\cite{kiselev_event-driven_2016}, which was heavily based on Minitaur, adapted to support the processing of data from up to three spike-based sensors. As for the other two \textit{landmark} systems, we nominate \textbf{SiBrain presented by Chen et al.\cite{chen_sibrain_2024} and NeuroFlow presented by Cheung et al.\cite{cheung_neuroflow_2016}}. The former was a spatio-temporal parallel NMA targeting SCNNs, with a square array of interconnected PEs and a bespoke core with a parallel set of PEs for the neurons in the fully-connected output layer. Those PEs were connected to a common spike and parameter buffers, which fetched the information from the off-chip memory. As for the latter, the proposed system relied on the \textit{dataflow computing} approach, where spiking data was streamed through a set of neuron PEs, with neural states of currently computed neurons being held on-chip after being fetched from auxiliary memory, and then through a set of parallel synaptic integration kernels, resulting in spikes being fed back to the auxiliary memory.
\begin{table}[h]
  \caption{NMA.AST implementations. Hardware implementation comparison with quantity capabilities (* -- Flip-Flops as on-chip memory reported, \textsuperscript{$\dagger$} -- \textit{multi-device-by-design} system).}
  \label{tab:class3}
  \resizebox{\textwidth}{!}{
    \begin{tabular}{p{0.15\linewidth}|c|cccccp{0.15\textwidth}cp{0.08\textwidth}p{0.13\textwidth}p{0.15\textwidth}c}
      \toprule
      \multirow{2}{*}{\textbf{Source}}                                  & \multirow{2}{*}{\textbf{Year}} & \textbf{Supp.}  & \textbf{\#PE} & \textbf{Num.}    & \textbf{Online}   & \textbf{Max.}          & \textbf{Use case}                    & \textbf{\#Neu./\#Syn.} & \textbf{Topo.}    & \textbf{Device}             & \textbf{Utilization}   & \textbf{Freq.} \\
                                                                        &                                & \textbf{models} &               & \textbf{repres.} & \textbf{learning} & \textbf{\#Neu./\#Syn.} &                                      &                        &                   &                             & \textbf{Logic/Mem/DSP} & (MHz)          \\
      \midrule  \midrule
      Hellmich et al.\cite{hellmich_fpga_2004, hellmich_emulation_2005} & 2004                           & IF              & 1             & --               & --                & 524k/800M              & NN research                          & 900/26.1k              & 2D array          & Virtex-II                   & --                     & 50             \\
      \hline
      Ros et al.\cite{ros_real-time_2006}                               & 2006                           & SRM             & 3             & 14b              & --                & 1k/--                  & robotics                             & 256/--                 & BIO               & Virtex-II                   & 24\%/50\%/--           & 25             \\
      \hline
      Glackin et al.\cite{glackin_emulating_2009}                       & 2009                           & LIF             & 4             & --               & STDP              & 1.05M/52.4M            & image processing (edge detection)    & 1.05M/52.4M            & SCNN              & Virtex-4                    & --                     & --             \\
      \hline
      Yang et al.\cite{yang_case_2011}\textsuperscript{$\dagger$}       & 2011                           & LIF             & 64            & w: 16b           & --                & 1M/2M                  & image processing                     & 256k/--                & FF                & Virtex-4\textsuperscript{1} & 11\%/80\%/24\%         & 200            \\
      \hline
      \textbf{Neil \& Liu\cite{neil_minitaur_2014}}                     & \textbf{2011}                  & \textbf{LIF}    & \textbf{32}   & \textbf{16b}     & \textbf{--}       & \textbf{65.5k/16.8M}   & \textbf{classification task (MNIST)} & \textbf{1.79k/647k}    & \textbf{FF-FC}    & \textbf{Spartan-6}          & \textbf{23\%/--/36\%}  & \textbf{75}    \\
      \hline
      \textbf{Cheung et al.\cite{cheung_neuroflow_2016}}                & \textbf{2016}                  & \textbf{LIF}    & \textbf{32}   & \textbf{--}      & \textbf{STDP}     & \textbf{98.3k/492M}    & \textbf{NN research/ neuroscience}   & \textbf{98.3k/492M}    & \textbf{Toroidal} & \textbf{Stratix-V}          & \textbf{--}            & \textbf{--}    \\
      \hline
      Wang et al.\cite{wang_fpga-based_2018}                            & 2018                           & LIF             & 128           & m: 8b, w: 4b     & --                & 20M/4T                 & neuroscience (100 hypercolumns)      & 100M/--                & BIO               & Stratix V                   & 67\%/79\%/100\%        & --             \\
      \hline
      Han et al.\cite{han_hardware_2020}                                & 2020                           & LIF             & 1             & w: 16b           & --                & 16.4k/16.8M            & classification task (MNIST)          & 2.06k/1.86M            & FF-FC             & Kintex-7                    & 3\%/8\%/--             & 200            \\
      \hline
      Li et al.\cite{li_fast_2021}                                      & 2021                           & LIF             & 12            & half             & STDP              & 310/178k               & classification task (MNIST)          & 310/178k               & FF-FC             & Virtex-7                    & --                     & 100            \\
      \hline
      Liu et al.\cite{liu_fpga-nhap_2022}                               & 2022                           & LIF/IZH         & 16            & w: 16b           & --                & 16.4k/16.8M            & classification task (MNIST)          & 2.06k/1.06M            & FF-FC             & Kintex-7                    & 6\%/19\%/8\%           & 200            \\
      \hline
      \textbf{Chen et al.\cite{chen_sibrain_2024}}                      & \textbf{2024}                  & \textbf{LIF}    & \textbf{64}   & \textbf{w: 8b}   & \textbf{--}       & \textbf{469k/606M}     & \textbf{classification task (MNIST)} & \textbf{51.5k/536K}    & \textbf{SCNN}     & \textbf{Virtex-7}           & \textbf{9\%/41\%1\%}   & \textbf{200}   \\
      \bottomrule
      \multicolumn{13}{l}{\textsuperscript{1}Implementation used four Virtex-4 devices.}
    \end{tabular}
  }
\end{table}

\subsection{NMA.FPP/CCM}
\label{sec:overview_class4}
\begin{table}[h]
    \caption{NMA.FPP/CCM implementations. Hardware implementation comparison with quantity capabilities (* -- Flip-Flops as on-chip memory reported, \textsuperscript{$\dagger$} -- \textit{multi-device-by-design} system).}
    \label{tab:class4}
    \resizebox{\textwidth}{!}{
        \begin{tabular}{p{0.15\linewidth}|c|cccccp{0.15\textwidth}cp{0.08\textwidth}p{0.13\textwidth}p{0.15\textwidth}c}
            \toprule
            \multirow{2}{*}{\textbf{Source}}                                            & \multirow{2}{*}{\textbf{Year}} & \textbf{Supp.}                                   & \textbf{\#PE} & \textbf{Num.}                          & \textbf{Online}                 & \textbf{Max.}          & \textbf{Use case}                                            & \textbf{\#Neu./\#Syn.} & \textbf{Topo.} & \textbf{Device}                   & \textbf{Utilization}                   & \textbf{Freq.}                \\
                                                                                        &                                & \textbf{models}                                  &               & \textbf{repres.}                       & \textbf{learning}               & \textbf{\#Neu./\#Syn.} &                                                              &                        &                &                                   & \textbf{Logic/Mem/DSP}                 & (MHz)                         \\
            \midrule  \midrule
            Roggen et al.\cite{roggen_hardware_2003}                                    & 2003                           & 64                                               & LIF           & --                                     & --                              & 64/112                 & robotics (object avoidance)                                  & 64/112                 & 2D array       & APEX 20K                          & 71.38\%/--/--                          & 33                            \\
            \hline
            Bellis et al.\cite{bellis_fpga_2004}                                        & 2004                           & LIF                                              & 4             & m: 7b, w: 5b                           & --                              & 40/--                  & robotics (wall-following, object avoidance)                  & 4/4                    & FF-FC          & Spartan-2                         & 10\%/--/--                             & 118.189                       \\
            \hline
            Upegui et al.\cite{upegui_fpga_2005}                                        & 2005                           & LIF                                              & 30            & w: 9b                                  & STDP                            & 30/90                  & mathematics (freq. differentiation)                          & 30/90                  & FF-R           & Virtex-II                         & 63.78\%/--/--                          & 54.4                          \\
            \hline
            Guerrero-Rivera et al.\cite{guerrero-rivera_programmable_2006}              & 2006                           & LIF                                              & 100           & m: 32b, w: 16b                         & STDP                            & --                     & neuroscience (olfactory bulb)                                & 100/675                & BIO            & Virtex-II Pro                     & --                                     & 33                            \\
            \hline
            Shayani et al.\cite{shayani_fpga-based_2008}                                & 2008                           & LIF                                              & 161           & m: 16b                                 & yes\textsuperscript{1}          & 161/1.61k              & NN research                                                  & 161/1.61k              & RAND           & Virtex-5                          & 85\%/--/--                             & 160                           \\
            \hline
            Johnston et al.\cite{johnston_fpga_2010}                                    & 2010                           & LIF                                              & 10            & half                                   & STDP                            & --                     & robotics (Khepera robot)                                     & 10/16                  & FF-FC          & Virtex-II                         & 18\%/--/--                             & 100                           \\
            \hline
            Caron et al.\cite{caron_fpga_2011}                                          & 2011                           & LIF                                              & 648           & m: 16b, w: 11b                         & --                              & 648/420k               & neuroscience (ODLM\textsuperscript{2})                       & 648/420k               & A2A            & Virtex-5                          & 69\%/95\%/--                           & 100                           \\
            \hline
            Rossello et al.\cite{rossello_hardware_2012}                                & 2012                           & IF                                               & 100           & 8b                                     & --                              & 63/972                 & image processing (Gabor fitlers)                             & 100/10k                & FF-FC          & Cyclone III                       & --                                     & 25                            \\
            \hline
            Iakymchuk et al.\cite{iakymchuk_fast_2012}                                  & 2012                           & SRM                                              & 22            & 16b                                    & --                              & 100/10k                & classification task (3x3 binary map)                         & 22/96                  & WTA            & Spartan-3                         & 15\%/6\%*/41\%                         & 69                            \\
            \hline
            Deng et al.\cite{deng_implementation_2014}                                  & 2014                           & LIF                                              & 24            & --                                     & --                              & 24/128                 & NN research (synfire chain)                                  & 24/128                 & FF-FC          & Stratix III                       & 7\%/1\%*/55\%                          & 50                            \\
            \hline
            Farsa et al.\cite{farsa_function_2015}                                      & 2015                           & LIF                                              & 81            & --                                     & --                              & 81/162                 & mathematics (approx. Gaussian force fields)                  & 81/162                 & WTA            & Zynq-7000                         & 11\%/1\%/--                            & 21.207                        \\
            \hline
            \textbf{Wang et al.\cite{wang_general-purpose_2015}}                        & \textbf{2015}                  & \textbf{LIF}                                     & \textbf{161}  & --                                     & \textbf{yes\textsuperscript{3}} & \textbf{161/--}        & \textbf{classification task ( 4 custom vision/speech tasks)} & \textbf{161/--}        & \textbf{LSM}   & \textbf{Virtex-6}                 & \textbf{37\%/7\%/--}                   & \textbf{390}                  \\
            \hline
            Gomar et al.\cite{gomar_digital_2016}                                       & 2016                           & AdEx                                             & 2             & --                                     & --                              & 2/1                    & neuroscience                                                 & 2/1                    & A2A            & Virtex-II                         & --                                     & --                            \\
            \hline
            Lammie et al.\cite{lammie_unsupervised_2018}                                & 2018                           & IZH                                              & 10            & 18b                                    & --                              & 10/160                 & classification task (MNIST)                                  & 10/160                 & WTA            & --                                & --                                     & --                            \\
            \hline
            Heidarpur et al.\cite{heidarpur_cordic-snn_2019}                            & 2019                           & IZH                                              & 21            & STDP                                   & NN research                     & 110/--                 & 21/20                                                        & 21/20                  & FF-FC          & Spartan-6                         & 22\%/7\%*/0\%                          & 84.1                          \\
            \hline
            Kuang et al.\cite{kuang_digital_2019}                                       & 2019                           & LIF                                              & 50            & --                                     & T-STDP                          & 50/1.88k               & classification task (5x5 binary map)                         & 50/1.88k               & WTA            & Stratix III                       & 23\%/4\%*/58\%                         & --                            \\
            \hline
            Asgari et al.\cite{asgari_low-energy_2020}                                  & 2020                           & LIF                                              & 16            & --                                     & STDP                            & 16/64                  & neuroscience ("decision part of brain")                      & 16/64                  & WTA            & Kintex-7                          & 35\%/3\%*/43\%                         & 143                           \\
            \hline
            Liang et al.\cite{liang_113ujclassification_2021}                           & 2021                           & IF                                               & 512           & w: 8b                                  & --                              & 512/400k               & neuroscience                                                 & 512/400k               & 512/400k       & Virtex-7                          & 4\%/2\%/--                             & 50                            \\
            \hline
            Deng et al.\cite{deng_reconstruction_2021}                                  & 2021                           & LIF                                              & 66            & --                                     & --                              & 66/155                 & neuroscience (auditory SNN)                                  & 66/155                 & 66/155         & Cyclone IV                        & 44\%/17\%*/0\%                         & 65.03                         \\
            \hline
            Heittmann et al.\cite{heittmann_simulating_2022}\textsuperscript{$\dagger$} & 2022                           & LIF/IZH                                          & 77.2k         & float                                  & --                              & 111k/--                & neuroscience (cortical microcircuit)                         & 77.2k/300M             & BIO            & Zynq-7045\textsuperscript{4}      & 49\%/74\%/--                           & 150                           \\
            \hline
            Hu et al.\cite{hu_binarized_2022}                                           & 2022                           & IF                                               & 4.1k          & w: 1b                                  & --                              & 4.1k/--                & classification task (N-MNIST, 16x16)                         & 4.1k/--                & SCNN           & Stratix V                         & --                                     & 100                           \\
            \hline
            \textbf{Carpegna et al.\cite{carpegna_spiker_2024}}                         & \multirow{2}{*}{\textbf{2024}} & \multirow{2}{*}{\textbf{LIF\textsuperscript{5}}} & 138           & \multirow{2}{*}{\textbf{m: 4b, w: 5b}} & \multirow{2}{*}{\textbf{--}}    & \textbf{1.22k/--}      & \textbf{classification task (MNIST)}                         & \textbf{138/102k}      & \textbf{FF-FC} & \multirow{2}{*}{\textbf{Artix-7}} & \multirow{2}{*}{\textbf{9\%/13\%/0\%}} & \multirow{2}{*}{\textbf{100}} \\
                                                                                        &                                &                                                  & \textbf{220}  &                                        &                                 & \textbf{550/--}        &                                                              & \textbf{220/144k}      & \textbf{FF-R}  &                                   &                                        &                               \\
            \hline
            Karakchi\cite{karakchi_scratchpad_2024}                                     & 2024                           & LIF                                              & 2.3k          & w: 8b                                  & --                              & 4K/--                  & --                                                           & 2.3k/--                & --             & Zynq UltraScale                   & 32\%/1\%/--                            & 350                           \\
            \hline
            Saulquin et al.\cite{saulquin_modnef_2025}                                  & 2025                           & LIF                                              & 598           & m: 16b, w: 8b\textsuperscript{6}       & --                              & 1.55k/1.33M            & classification task (MNIST)                                  & 598/386k               & FF-FC          & Zynq-7020                         & 39\%/82\%/2\%                          & 125                           \\
            \hline
            Azmine et al.\cite{azmine_spikespec_2025}                                   & 2025                           & LIF                                              & 18            & w: 16b                                 & TSTDP/RSTDP                     & 18/288                 & spectrum sensing                                             & 18/288                 & LSM            & Virtex-7                          & 1\%/1\%*/--                            & 200                           \\
            \bottomrule
            \multicolumn{13}{l}{\textsuperscript{1}Structural plasticity through cutting unwanted synapses.}                                                                                                                                                                                                                                                                                                                                                                                                          \\
            \multicolumn{13}{l}{\textsuperscript{2}Oscillatory Dynamic Link Matcher (OLDM)\cite{pichevar2003oscillatory}.}                                                                                                                                                                                                                                                                                                                                                                                            \\
            \multicolumn{13}{l}{\textsuperscript{3}Specific type of tutored learning via Teaching Unit -- weights modified according to the tutor signal to activate specific outputs when desired.}                                                                                                                                                                                                                                                                                                                  \\
            \multicolumn{13}{l}{\textsuperscript{4}System used 27 Zynq-7045 devices, the utilization shown is per device.}                                                                                                                                                                                                                                                                                                                                                                                            \\
            \multicolumn{13}{l}{\textsuperscript{5}Two different LIF implementations with varying levels of simplification.}                                                                                                                                                                                                                                                                                                                                                                                          \\
            \multicolumn{13}{l}{\textsuperscript{6}Weights 3-8b were tested, but the authors selected 8b.}
        \end{tabular}
    }
\end{table}

FPP/CCM is one of three classes that support two additional Traits as defined in Section \ref{sec:taxonomy}, which result in specific synergies and possible downsides stemming from this pair-wise support. Those classes merge fully parallel operation with the co-location of memory and computation, bringing them closer to truly brain-like operation than the classes mentioned before. Due to the fact that all necessary data is stored on-chip, close to the processing logic and that every neuron has its own bespoke datapath, those systems have the potential of updating the entire network the fastest out of the clocked-driven subcategory, but they are also extremely dependent on the amount of on-chip resources in terms of the actual size of the SNN that they can support. \textbf{Landmark architectures for this class include four systems, belonging to one of three main organizational schemes}, namely: \textbf{(i)}multiple PEs connected to a common control

\noindent
unit, \textbf{(ii)}direct representations of the implemented SNNs, and \textbf{(iii)}NoC-like structures. Thus, as \textit{landmarks} of this class, we selected \textbf{Spiker+ presented by Carpegna et al.\cite{carpegna_spiker_2024}}, the system presented by \textbf{Wang et al.}\cite{wang_general-purpose_2015}, and \textbf{HEENS presented by Vallejo-Mancero et al.\cite{vallejo-mancero_real-time_2024}}, belonging to the three aforementioned groups, respectively. \textbf{Spiker+} was a complete Python-to-FPGA framework for creating SNNs for image inference and other classification tasks, which was a direct continuation of \textbf{Spiker}, referenced in Section \ref{sec:overview_class1}. Architecture-wise, the generated system comprised three layers of hierarchical units - network, layer and neuron and supported six different versions of the LIF model. \textbf{Wang et al.}\cite{wang_general-purpose_2015} presented an NMA for general-purpose LSM with a reconfigurable reservoir and pre-trained, task-specific fully-connected output layers. The authors implemented the reconfigurability of the reservoir's neurons via power gating (turning off the unused parts of the FPGA). The plasticity was implemented with probabilistic weight updates and an induced teaching signal. The number of output neurons depended on the output of the most demanding task the system was designed to perform. Finally, \textbf{Hardware Emulator of Evolving Neural Systems (HEENS)} was a NoC-like structure connected utilizing AER-SRT\cite{dorta2016aer} protocol, which theoretically allowed for expansion to up to 127 devices in a ring topology.

\newpage
\subsection{NMA.FPP/AST}
\label{sec:overview_class5}

FPP/AST architectures support two Traits: fully parallel operation and asynchronous network update. Because the necessary data is stored off-chip, they can support large SNNs and potentially achieve high operation speeds, due to fully parallel operation and updating neuron states only when spikes occur. However, those systems can also demand higher complexity due to the need to synchronize many parallel PEs as the simulation of the SNN progresses, which puts a high demand on on-chip logic resources, with alleviated requirement for the on-chip memory. \textbf{Landmark architectures for this class include a single system.} As we found only two examples of such systems in the surveyed group, this list is less representative than we would have hoped for, and we do agree that a more extensive search needs to be done to acquire more data about this kind of implementations. Nevertheless, based on the data we were able to gather, we selected the system presented by \textbf{Ipatov et al.\cite{ipatov_development_2019}} as a good example of this class. The system was organized as a Network-on-Chip (NoC) and relied heavily on the crossbar concept, i.e., inside every one of the M cores, every neuron
\begin{table}[h]
  \caption{NMA.FPP/AST implementations. Hardware implementation comparison with quantity capabilities (* -- Flip-Flops as on-chip memory reported, \textsuperscript{$\dagger$} -- \textit{multi-device-by-design} system).}
  \label{tab:class5}
  \resizebox{\textwidth}{!}{
    \begin{tabular}{p{0.15\linewidth}|c|cccccp{0.15\textwidth}cp{0.08\textwidth}p{0.13\textwidth}p{0.15\textwidth}c}
      \toprule
      \multirow{2}{*}{\textbf{Source}}                              & \multirow{2}{*}{\textbf{Year}} & \textbf{Supp.}  & \textbf{\#PE} & \textbf{Num.}    & \textbf{Online}   & \textbf{Max.}          & \textbf{Use case}                         & \textbf{\#Neu./\#Syn.} & \textbf{Topo.}         & \textbf{Device}                      & \textbf{Utilization}   & \textbf{Freq.} \\
                                                                    &                                & \textbf{models} &               & \textbf{repres.} & \textbf{learning} & \textbf{\#Neu./\#Syn.} &                                           &                        &                        &                                      & \textbf{Logic/Mem/DSP} & (MHz)          \\
      \midrule  \midrule
      \textbf{Ipatov et al.\cite{ipatov_development_2019}}          & \textbf{2019}                  & \textbf{LIF}    & \textbf{131k} & \textbf{--}      & \textbf{--}       & \textbf{131k/67M}      & \textbf{--}                               & \textbf{131k/67M}      & \textbf{A2A}           & \textbf{Kintex-7\textsuperscript{1}} & \textbf{--}            & \textbf{--}    \\
      \hline
      Yang et al.\cite{yang_bicoss_2021}\textsuperscript{$\dagger$} & 2021                           & LIF/IZH/HH      & 4M            & 8b               & T-STDP/P-STDP     & 4M/6.14B               & neuroscience/ classification task (MNIST) & 4M/6.14B               & BFT\textsuperscript{2} & Cyclone IV\textsuperscript{3}        & --                     & --             \\
      \bottomrule
      \multicolumn{13}{l}{\textsuperscript{1}Implementation used 2 Kintex-7 devices, but this architecture can be implemented on a single chip with potential expansion to additional FPGAs.}                                                                                                                                                                                         \\
      \multicolumn{13}{l}{\textsuperscript{2}Butterfly Fat Tree (BFT) topology for SNN.}                                                                                                                                                                                                                                                                                              \\
      \multicolumn{13}{l}{\textsuperscript{3}Implementation used 35 Cyclone IV devices.}
    \end{tabular}
  }
\end{table}

\noindent
PE was interconnected with others in an N x N array fashion, and every connection could be turned \textbf{on} or \textbf{off}, with actual synaptic weights being stored off-chip. The system was designed in such a way that it was easy to connect another device with the same structure to expand the network. As for the second architecture in this group, \textbf{BiCoSS, presented by Yang et al. \cite{yang_bicoss_2021}} (and expanded in \cite{yang_nadol_2024}), utilized a particular connectivity scheme between the units, specifically the Butterfly Fat Tree (BFT). However, the system was multi-device by design, making it tricky to compare it fairly with single-chip systems.

\subsection{NMA.CCM/AST}
\label{sec:overview_class6}
\begin{table}[h]
  \caption{NMA.CCM/AST implementations. Hardware implementation comparison with quantity capabilities (* -- Flip-Flops as on-chip memory reported, \textsuperscript{$\dagger$} -- \textit{multi-device-by-design} system).}
  \label{tab:class6}
  \resizebox{\textwidth}{!}{
    \begin{tabular}{p{0.15\linewidth}|c|cccccp{0.15\textwidth}cp{0.08\textwidth}p{0.13\textwidth}p{0.15\textwidth}c}
      \toprule
      \multirow{2}{*}{\textbf{Source}}                                 & \multirow{2}{*}{\textbf{Year}} & \textbf{Supp.}            & \textbf{\#PE} & \textbf{Num.}                     & \textbf{Online}          & \textbf{Max.}                  & \textbf{Use case}                                  & \textbf{\#Neu./\#Syn.} & \textbf{Topo.}        & \textbf{Device}                & \textbf{Utilization}   & \textbf{Freq.} \\
                                                                       &                                & \textbf{models}           &               & \textbf{repres.}                  & \textbf{learning}        & \textbf{\#Neu./\#Syn.}         &                                                    &                        &                       &                                & \textbf{Logic/Mem/DSP} & (MHz)          \\
      \midrule  \midrule
      Pearson et al.\cite{pearson_real-time_2005}                      & 2005                           & LIF                       & 10            & 16b                               & --                       & 144k/1.09M                     & neuroscience (basal ganglia)                       & 1.2k/9.12k             & BIO                   & Virtex-II                      & 100\%/100\%/--         & 50             \\
      \hline
      Cheung et al.\cite{cheung_scalable_2006}                         & 2006                           & LIF                       & 1             & m: 16b                            & --                       & --                             & image processing                                   & 1k/--                  & SCNN                  & Virtex-II                      & 32\%/--/--             & 120            \\
      \hline
      Cheung et al.\cite{cheung_parallel_2009}                         & 2009                           & IZH                       & 32            & m: 18b, w: 9b                     & --                       & 848/--                         & NN research                                        & 800/640k               & A2A                   & Virtex-5                       & 36\%/98\%/100\%        & 110.47         \\
      \hline
      Ang et al.\cite{ang_spiking_2011}                                & 2011                           & IF                        & 1             & --                                & yes\textsuperscript{1}   & 24/--                          & auto-associative memory                            & 4/16                   & FF                    & Cyclone II                     & 13\%/1\%/--            & 50             \\
      \hline
      Luo et al.\cite{luo_real-time_2014}                              & 2014                           & LIF                       & 48            & --                                & --                       & 97k/96k                        & neuroscience (8k granular cells)                   & 8k/--                  & BIO                   & Virtex-7                       & 88\%/93\%/82\%         & --             \\
      \hline
      Holanda et al.\cite{holanda_dhyana_2016}                         & 2016                           & IZH                       & 16            & --                                & --                       & 256/65.5k                      & neuroscience                                       & 208/100                & 3D array              & Stratix IV                     & 99\%/--/--             & 56             \\
      \hline
      Luo et al.\cite{luo_real-time_2016}                              & 2016                           & LIF                       & 48            & 40b                               & --                       & 97k/96k                        & neuroscience (Golgi-granular cells [960:9600])     & 97k/96k                & BIO                   & Virtex-7                       & 88\%/93\%/82\%         & 121.945        \\
      \hline
      Lin et al.\cite{lin_digital_2017}                                & 2017                           & LIF/IZH/HH                & 8             & --                                & --                       & 2.8k/7.84M                     & NN research                                        & 2.8k/7.84M             & A2A                   & Virtex-7                       & 89\%/38\%/22\%         & --             \\
      \hline
      Yang et al.\cite{yang_real-time_2019}\textsuperscript{$\dagger$} & 2019                           & HH                        & 36            & --                                & STDP                     & 1.04M/60M                      & neuroscience (thalamocortical/basal ganglia)       & 173k/10M               & BIO                   & Stratix III\textsuperscript{2} & --                     & 100            \\
      \hline
      Fang et al.\cite{fang_event-driven_2019, fang_encoding_2020}     & 2019                           & LIF                       & 1             & w: 16b                            & --                       & 610/476k                       & classification task (MNIST)                        & 610/476k               & FF-FC                 & Cyclone V                      & --                     & 75             \\
      \hline
      Zhang et al.\cite{zhang_asynchronous_2019}                       & 2019                           & LIF                       & 16            & m: 13b, w: 8b                     & --                       & 1K/1M                          & classification task (MNIST)                        & 906/406k               & FF-FC                 & Virtex-7                       & --                     & --             \\
      \hline
      Mitchell et al.\cite{mitchell_small_2020}                        & 2020                           & LIF                       & 1             & m: 16b, w: 8b                     & --                       & 256/4.1k                       & NN research                                        & 60/3.6k                & A2A                   & iCE40                          & --                     & --             \\
      \hline
      Nambiar et al.\cite{nambiar_scalable_2020}                       & 2020                           & LIF                       & 2             & w: 8b                             & --                       & 590k/--                        & classification task (MNIST)                        & 266/269k               & FF-FC                 & Kintex-7                       & 28\%/12\%1\%           & 100            \\
      \hline
      Sakellariou \& Paliouras\cite{sakellariou_fpga_2021}             & 2021                           & LIF                       & 16            & w: 12b                            & --                       & 4.1k/524k                      & classification task (MNIST)                        & 310/238k               & FF-FC                 & Zynq UltraScale+               & 75\%/67\%/0\%          & 100            \\
      \hline
      \textbf{Wang et al.\cite{wang_triplebrain_2022}}                 & \textbf{2022}                  & \textbf{LIF}              & \textbf{4}    & \textbf{m: 24b, w: 16b}           & \textbf{SOM-STDP/R-STDP} & \textbf{--\textsuperscript{3}} & \textbf{neuroscience/ classification task (MNIST)} & \textbf{256/201k}      & \textbf{FF-FC}        & \textbf{Zynq-7045}             & \textbf{5\%/25\%/4\%}  & \textbf{250}   \\
      \hline
      Zhao et al.\cite{zhao_099--438_2023}                             & 2023                           & LIF                       & 16            & --                                & --                       & 68/7744                        & classification task (biosignals)                   & 68/7.74k               & FF-FC                 & --                             & --                     & 0.3            \\
      \hline
      \textbf{Wang et al.\cite{wang_marmotini_2024}}                   & \textbf{2024}                  & \textbf{LIF}              & \textbf{16}   & \textbf{w: 8b}                    & \textbf{--}              & \textbf{--}                    & \textbf{classification task (MNIST)}               & \textbf{2.56k/528k}    & \textbf{SCNN}         & \textbf{Kintex UltraScale}     & \textbf{93\%/18\%/--}  & \textbf{150}   \\
      \hline
      Vallejo-Mancero et al.\cite{vallejo-mancero_real-time_2024}      & 2024                           & LIF\textsuperscript{7}    & 160           & w: 16b                            & STDP                     & 1.28k/41k                      & classification task (MNIST)                        & 354/25.4k              & WTA                   & Zynq-7045                      & --/61\%/--             & 125            \\
      \hline
      Karthikeyan and Subbulakshmi\cite{karthikeyan_3d_2025}           & 2025                           & LIF                       & 256           & m: 32b, w:16b                     & --                       & 32.8k/4.19M                    & NN research                                        & 16.4k/65.5M            & FF-FC                 & Virtex-4                       & --                     & --             \\
      \hline
      Zhong et al.\cite{zhong_morphbungee-lite_2025}                   & 2025                           & IF                        & 16            & --                                & EO-DeepTempo             & 1k/262k                        & classification task (MNIST)                        & 928/238k               & FF-FC                 & Zynq-7045                      & 7\%/21\%/0\%           & 100            \\
      \hline
      \textbf{Yang et al.\cite{yang_crosscut_2025}}                    & \textbf{2025}                  & \textbf{IF/LIF/IZH}       & \textbf{64}   & \textbf{w: 8b\textsuperscript{4}} & \textbf{--}              & \textbf{262k/288M}             & \textbf{classification task (MNIST)}               & \textbf{4.1k/590k}     & \textbf{SCNN}         & \textbf{Virtex UltraScale+}    & \textbf{40\%/48\%/--}  & \textbf{150}   \\
      \hline
      Aliyev et al.\cite{aliyev_exploring_2025}                        & 2025                           & LIF                       & 276           & w: 4b                             & --                       & 14.7k/--                       & classification task (CIFAR-10)                     & 8.19k/7.7M             & SCNN                  & Virtex UltraScale+             & 7\%/31\%/--            & 100            \\
      \hline
      Cheng et al.\cite{cheng_fpga-based_2025}                         & 2025                           & LIF                       & 16            & w: 8b                             & --                       & 4k/4.19M                       & classification task (MNIST)                        & 970/501k               & FF\textsuperscript{5} & Zynq UltraScale                & 36\%/85\%/--           & 250            \\
      \hline
      Zhang et al.\cite{zhang_low-power_2025}                          & 2025                           & LIF/IF\textsuperscript{6} & 5             & w: 4b/16b\textsuperscript{6}      & TSTDP                    & 4.42k/1k                       & classification task (CHB-MIT)                      & 4.42k/1k               & VFA                   & Zynq UltraScale                & 1\%/1\%/--             & 100            \\
      \bottomrule
      \multicolumn{13}{l}{\textsuperscript{1}The architecture supported learning in terms of synaptic delay adjustment.}                                                                                                                                                                                                                                                                                                             \\
      \multicolumn{13}{l}{\textsuperscript{2}Implementation used six Intel/Altera Stratix-III devices.}                                                                                                                                                                                                                                                                                                                              \\
      \multicolumn{13}{l}{\textsuperscript{4} 1-, 2- and 4bit weights also supported.}                                                                                                                                                                                                                                                                                                                                               \\
      \multicolumn{13}{l}{\textsuperscript{3}Authors state that the system can support multiples of 64 neurons with 1024 synapses per neuron.}                                                                                                                                                                                                                                                                                       \\
      \multicolumn{13}{l}{\textsuperscript{5}FF topology with 50\% sparsity of connections between first and second layer.}                                                                                                                                                                                                                                                                                                          \\
      \multicolumn{13}{l}{\textsuperscript{6}The implemented network consisted of a layer of differently modelled neurons and synaptic weights - IF with 4b synapses and LIF with 16b weights.}                                                                                                                                                                                                                                      \\
    \end{tabular}
  }
\end{table}

CCM/AST architectures approach the topic of brain-like design through support for co-location of memory and computing, and asynchronous operation, with the actual physical PEs utilizing some form of TDM to simulate a larger number of neurons. Those systems, as per our definition, are rather close to fully brain-like systems (FULL), i.e.,

\noindent
%
these architectures will act similarly to FULL for a particular subset of the implemented neurons when they are active and/or receiving spikes. With those architectures, we expect support for rather large networks (if an appropriately powerful platform is used). Moreover, due to the nature of operation of those systems, usually a NoC-like structure is employed (mostly mesh, but parallel PEs with a common bus are also present) that allows for larger systems to efficiently communicate between their subcomponents. \textbf{Landmark architectures for this class include three systems.} From the aforementioned NoC-like structures we can select a few \textit{landmarks}, namely\textbf{ CROSSCUT presented by Yang et al.\cite{yang_crosscut_2025}, Marmoniti presented by Wang et al.\cite{wang_marmotini_2024} and TripleBrain presented by \textbf{Wang et al.}\cite{wang_triplebrain_2022}}. \textbf{CROSSCUT} was an architecture arranged in a "hybrid Tree-Mesh NoC", meaning the PEs (or cores) were arranged in groups of four -- dubbed Quadtrees by the authors -- connected to a common router within the mesh. The authors argued, based on their observations, that this organization reduced the average distance between any two cores of the system, increasing the connectivity performance in comparison to regular NoCs. Apart from that, due to this system targeting SCNNs primarily, they also introduced the concept of Synapse Compress Mechanism, which allowed for processing only the synapses related to the specific inter-layer connectivity, which they argued allowed for reduction in computation and utilized resources. \textbf{Marmotini} was a 2D NoC-like architecture that primarily targets SCNNs. The PEs/cores in the network were of two types: DENSE, responsible for convolutional and fully-connected layer operations with dense synaptic connections, and SPARSE, responsible for fully-connected layers only with sparse connectivity. Finally, \textbf{TripleBrain} was an NMA consisting of Neural Processing Tiles (NPTs) that shared a common bus, supporting 64 neurons and 1024 synapses per NPT, with a focus on classification tasks. The architecture realized learning through self-organizing maps, STDP, the merge of the two (SOM-STDP), and reinforced STDP (R-STDP).

\subsection{NMA.FULL}
\label{sec:overview_class7}

FULL architectures support all Traits listed in Section \ref{sec:bg_snn}, and thus their operation can be deemed the most \textit{brain-like}, with fully on-chip processing and memory, fully parallel operation, and asynchronous update of the SNN's neurons. This high level of inspiration from biology has its cost, however, as those systems are the most bound by the on-chip resources, as the entire system, with asynchronous-operation-enabling logic circuitry, network parameters, and bespoke PEs for every neuron in the network, must fit on a single device. This leads to those systems potentially supporting the smallest networks, which we observed while surveying the architectures - those systems rarely cross the barrier of ten thousand implemented neurons, and if so, it is done at the cost of possible synaptic connectivity. \textbf{Landmark architectures for this class include four systems, belonging to three main organizational schemes,} namely: \textbf{(i)}NoC-like structures, \textbf{(ii)}parallel PEs with a common spike arbiter, \textbf{(iii)}direct implementations of the simulated network in hardware. This last category in particular is the most literal \textit{implementation of an SNN in digital hardware}, however, because of this fact there are not many design choices that can be taken during construction of such systems. For the first group, we selected the \textbf{Dynamic Adaptive Neural Network Array (DANNA) presented by Dean et al.\cite{dean_dynamic_2014}} as well as its successor \textbf{DANNA2\footnote{The successor architecture followed the same primary idea, yet it introduced enough changes to be a distinct system concept.} introduced by Mitchell et al.\cite{mitchell_danna_2018} and Reconfigurable Architecture for Neuromorphic Computing (RANC) introduced by Mack et al.\cite{mack_ranc_2021}} as \textit{landmarks}. \textbf{DANNA/DANNA2} was rather unorthodox, as it consisted of a 2D grid of multi-purpose blocks which could behave as neuron PEs, synaptic fan-outs, or pass-through \textit{elements}. The system supported different connection schemes through the layout of the elements of a selected type in the array. The \textbf{RANC} was a complex ecosystem for realizing different NMAs, utilizing a SW/HW co-design approach. In the paper, the authors used it to implement an FPGA-based version of IBM TrueNorth\cite{akopyan2015truenorth} - NoC of PEs with crossbars for connectivity and support for programmable synaptic delays. Moreover, this system was extended by \textbf{Nguyen-Dinh et al.}\cite{nguyen-dinh_novel_2024} with improvements to router organization, namely by grouping the cores in batches of four and connecting them to a common router\footnote{The selected organization scheme resulted in 81x greater throughput than an equivalent implementation with RANC, but with larger resource utilization.}. An example from the second group was presented by \textbf{Bonabi et al.}\cite{bonabi_fpga_2012}, where the architecture consisted of an array of PEs connected to an AER management system, from which specific spike events were propagated to be processed by appropriate PEs. The direct implementations of the simulated SNNs can be seen, e.g., in the article by \textbf{Windhager et al.}\cite{windhager_spiking_2025} (FF-FC).

\begin{table}[h]
    \caption{NMA.FULL implementations. Hardware implementation comparison with quantity capabilities (* -- Flip-Flops as on-chip memory reported, \textsuperscript{$\dagger$} -- \textit{multi-device-by-design} system).}
    \label{tab:class7}
    \resizebox{\textwidth}{!}{
        \begin{tabular}{p{0.15\linewidth}|c|cccccp{0.15\textwidth}cp{0.08\textwidth}p{0.13\textwidth}p{0.15\textwidth}c}
            \toprule
            \multirow{2}{*}{\textbf{Source}}                         & \multirow{2}{*}{\textbf{Year}} & \textbf{Supp.}  & \textbf{\#PE}  & \textbf{Num.}            & \textbf{Online}        & \textbf{Max.}                       & \textbf{Use case}                           & \textbf{\#Neu./\#Syn.} & \textbf{Topo.} & \textbf{Device}     & \textbf{Utilization}   & \textbf{Freq.}             \\
                                                                     &                                & \textbf{models} &                & \textbf{repres.}         & \textbf{learning}      & \textbf{\#Neu./\#Syn.}              &                                             &                        &                &                     & \textbf{Logic/Mem/DSP} & (MHz)                      \\
            \midrule  \midrule
            Allen et al.\cite{allen_plasticity_2005}                 & 2005                           & IF              & 8              & w: 1b                    & STDP                   & 8/320                               & classification task (olfaction)             & 8/320                  & FF-FC          & Virtex-II           & 26\%/18\%*/--          & 0.002                      \\
            \hline
            Girau \& Torres-Huitzil\cite{girau_fpga_2006}            & 2006                           & LIF             & 256            & 12b                      & --                     & 256/480                             & image processing (segmentation)             & 256/480                & 2D array       & Virtex-II           & 83\%/35\%*/--          & 50                         \\
            \hline
            Cassidy et al.\cite{cassidy_fpga_2007}                   & 2007                           & LIF             & 32             & m: 16b, w: 8b            & STDP                   & 64/4.1k                             & robotics                                    & 32/4.1k                & A2A            & Spartan-3           & 44\%/34\%/--           & 50                         \\
            \hline
            Bonabi et al.\cite{bonabi_fpga_2012}                     & 2012                           & HH              & 16             & 32b                      & --                     & 16/256                              & neuroscience (neuronal pool)                & 16/256                 & A2A            & Virtex-7            & 13\%/32\%*/84\%        & 91                         \\
            \hline
            \textbf{Dean et al.\cite{dean_dynamic_2014}}             & \textbf{2014}                  & \textbf{LIF}    & \textbf{2.5k}  & \textbf{9b}              & \textbf{LTP/LTD}       & \textbf{10k/10k\textsuperscript{1}} & \textbf{NN research}                        & \textbf{2.5k/--}       & \textbf{RAND}  & \textbf{Virtex-7}   & \textbf{93\%/--/--}    & \textbf{8}                 \\
            \hline
            Nanami \& Kohno\cite{nanami_fpga-based_2016}             & 2016                           & DSSN            & 16             & 18b                      & --                     & 16/4.1k                             & neuroscience (spiking behaviour)            & 16/4.1k                & RAND           & --                  & --                     & 100                        \\
            \hline
            Wilson et al.\cite{wilson_reconfigurable_2017}           & 2017                           & MBED            & 100            & 8b                       & yes\textsuperscript{3} & 100/200                             & neuroscience (C. Elegans locomotory system) & 100/200                & BIO            & Virtex-5            & 68\%/76\%*/--          & 9.884                      \\
            \hline
            Farsa et al.\cite{farsa_low-cost_2019}                   & 2019                           & LIF             & 6              & --                       & --                     & 6/130                               & classification task (5x5 binary map)        & 6/130                  & FF-FC          & Virtex-6            & 7\%/1\%/--             & 189.071\textsuperscript{2} \\
            \hline
            Gupta et al.\cite{gupta_fpga_2020}                       & 2020                           & LIF             & 16             & w: 24b                   & STDP                   & 800/12.5k                           & classification task (MNIST)                 & 16/12.5k               & WTA            & Virtex-6            & 38\%/4\%/8\%           & 100                        \\
            \hline
            \textbf{Mack et al.\cite{mack_ranc_2021}}                & \textbf{2020}                  & \textbf{LIF}    &                & \textbf{w: 8b}           & \textbf{--}            & \textbf{141k/--}                    & \textbf{classification task (CIFAR-10)}     & \textbf{5.89k/--}      & \textbf{SCNN}  & \textbf{Alveo U250} & \textbf{--}            & \textbf{--}                \\
            \hline
            Prashanth \& Ahmed\cite{prashanth_fpga_2021}             & 2021                           & --              & 4              & --                       & --                     & 4/6                                 & neuroscience (spiking behaviour)            & 4/6                    & A2A            & Cyclone V           & 10\%/8\%/--            & --                         \\
            \hline
            \textbf{Truong-Tuan et al.\cite{truong-tuan_fpga_2021}}  & \textbf{2021}                  & \textbf{LIF}    & \textbf{1.27k} & \textbf{w: 8b}           & \textbf{--}            & \textbf{1.28k/328k}                 & \textbf{classification task (MNIST)}        & \textbf{1.27k/326k}    & \textbf{FF-FC} & \textbf{Kintex-7}   & \textbf{74\%/6\%/--}   & \textbf{100}               \\
            \hline
            \textbf{Nguyen-Dinh et al.\cite{nguyen-dinh_novel_2024}} & \textbf{2024}                  & \textbf{LIF}    & \textbf{1.27k} & \textbf{w: 8b}           & \textbf{--}            & \textbf{1.28k/328k}                 & \textbf{classification task (MNIST)}        & \textbf{1.27k/326k}    & \textbf{FF-FC} & \textbf{Alveo U250} & \textbf{20\%/9\%*/--}  & \textbf{77}                \\
            \hline
            Windhager et al.\cite{windhager_spiking_2025}            & 2025                           & LIF             & 1.58k          & w: 2b\textsuperscript{4} & --                     & 1.58k/863k                          & classification task (MNIST)                 & 1.58k/863k             & FF-FC          & Zynq UltraScale+    & 21\%/10\%/0\%          & 300                        \\
            \bottomrule
            \multicolumn{12}{l}{\textsuperscript{1}The system can support at maximum 10k elements, where every element is a neuron, synapse or a pass-through.}                                                                                                                                                                                                                                        \\
            \multicolumn{12}{l}{\textsuperscript{2}The authors reported 412.371MHz for a single neuron.}                                                                                                                                                                                                                                                                                               \\
            \multicolumn{12}{l}{\textsuperscript{3}The learning can be performed by manually adjusting the weights on-the-fly.}                                                                                                                                                                                                                                                                        \\
            \multicolumn{12}{l}{\textsuperscript{4}Only -1, 0 and 1 values supported for weights.}
        \end{tabular}
    }
\end{table}

\subsection{Non-NMA structures related to SNN research}
\label{sec:overview_non_nma}

In this section, we present systems that could not be classified as NMAs, presented interesting optimizations of NMAs sub-units or were single neuron implementations. There is a number of structures that we were not able to fairly compare with the implementations listed in sections from \ref{sec:overview_class0} to \ref{sec:overview_class7}, due to the lack of distinctive features of NMAs, as we defined them in Section \ref{sec:taxonomy}. There are many such implementations reported in the literature, e.g., \parhl{Palumbo et al. (2017) \cite{palumbo_feasibility_2017}} presented an architecture for simulating Swarm Intelligence (SI) using SNNs on a soft-core processor and systolic array for for neuronal dynamics, \parhl{Shahsavari et al. (2021)\cite{shahsavari_neuromorphic_2021}} presented \textbf{Partially Ordered Event-Triggered System (POETS)} architecture - a general-purpose parallel architecture of a number of RISC-V cores arranged in a NoC on multiple FPGAs, and recently \textbf{IzhiRISC-V}\cite{szczerek2025izhirisc} was presented, a RISC-V core with designated custom instructions for IZH neurons -- those kind of Application Specific Instruction set Processors (ASIPs) have been getting increasing attention in the recent years. In many papers, the authors were focusing on optimizing parts of NMAs. As an interested reader may find information on optimizing specific their NMA designs extremely useful, we list such papers, grouped by the sub-units being optimized: spike encoding - \cite{gerlinghoff_resource-efficient_2022}, exponential function unit - \cite{kim_hardware-efficient_2021}, STDP-based learning - \cite{gomar_digital_2018, nouri_digital_2018, pedroni_forward_2016, belhadj_fpga-based_2008, wu_efficient_2025, le_digital_2025}, ion channel dynamics unit - \cite{jokar_digital_2017, mak_field_2005}, soma unit - \cite{pourhaj_fpga_2010}, multipliers for SNN processing - \cite{nidesh_kanna_reconfigurable_2025}, alternative to SNN-based processing (P systems) - \cite{pena_efficient_2019}. While surveying the neuromorphic architectures, we realized that many researchers were focusing on implementing single neurons on FPGAs with specific use cases (like PWM generation\cite{jalilian_pulse_2017}), hardware optimizations, accuracy of simulation and biological plausibility of the results in mind. Examples of such works, grouped by the neuron model the authors aimed at implementing, are as follows: HH - \cite{levi_digital_2018, bonabi_fpga_2014}, LIF - \cite{grassia_digital_2016, s_m_mishra_fpga-based_2025}, AdEx - \cite{heidarpour_cordic_2016, gomar_digital_2014}, HR - \cite{heidarpur_digital_2017, hayati_digital_2016, kazemi_digital_2014}, ML - \cite{hayati_digital_2015}, IZH - \cite{pu_low-cost_2021, haghiri_multiplierless_2018, zhang_digital_2025, sruthi_modeling_2025}, WIL - \cite{imani_digital_2018}, Pinsky-Rinzel (PR) - \cite{rahimian_digital_2018}, DSSN - \cite{miao_fpga_2025}, FHN - \cite{ghanbarpour_hardware-efficient_2025}  multiple - \cite{soleimani_efficient_2018}, other - \cite{zhang_neural_2022}.

\newpage
\section{Trends, Tendencies, and Discussion}
\label{sec:trends}
We surveyed 135 articles describing NMAs as defined in Section \ref{sec:taxonomy} that provided enough information to apply the Taxonomy rules, derive further metrics, and draw observed trends and potential predictions about the future of the field. An additional 36 papers consisted of non-NMA structures and NMA-related research, as described and listed 

\begin{wrapfigure}[11]{o}{0.4\textwidth}
  \centering
  \includegraphics[width=\linewidth, trim={10pt 30pt 10pt 45pt}]{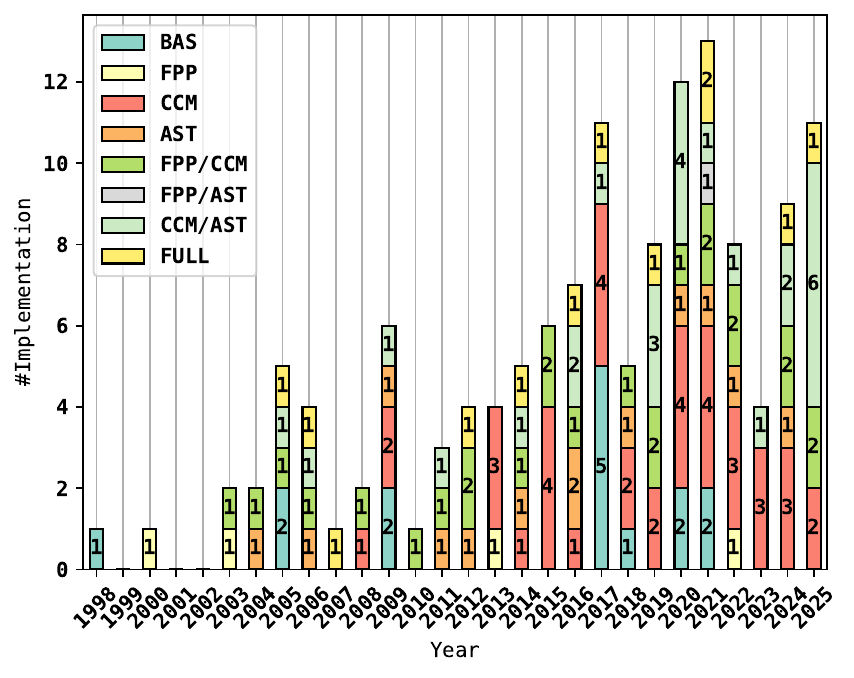}
  \caption{Class population structure through the years.}
  \label{fig:classes}
\end{wrapfigure}
\noindent
in Section \ref{sec:overview_non_nma}, provide valuable insights into implementation details of particular sub-parts of the NMA systems, as well as different views on what may be considered \textit{neuromorphic} for the wider audience. We focused on surveying the most interesting examples of NMAs that fit into our devised Taxonomy from approximately the last 25 years of digital neuromorphic system research and on analyzing potential research paths. We began our analysis by focusing on the class distribution over the years. As Figure \ref{fig:classes} suggests, the surveyed NMAs on FPGAs have been appearing in considerable numbers starting from the years 2014-2015, yet we managed to find interesting representatives from the early 2000s (\cite{waldemark_pulse_1998} is a bit of a special case,

\begin{wrapfigure}[11]{o}{0.4\textwidth}
  \centering
  \includegraphics[width=\linewidth, trim={10pt 30pt 10pt 45pt}]{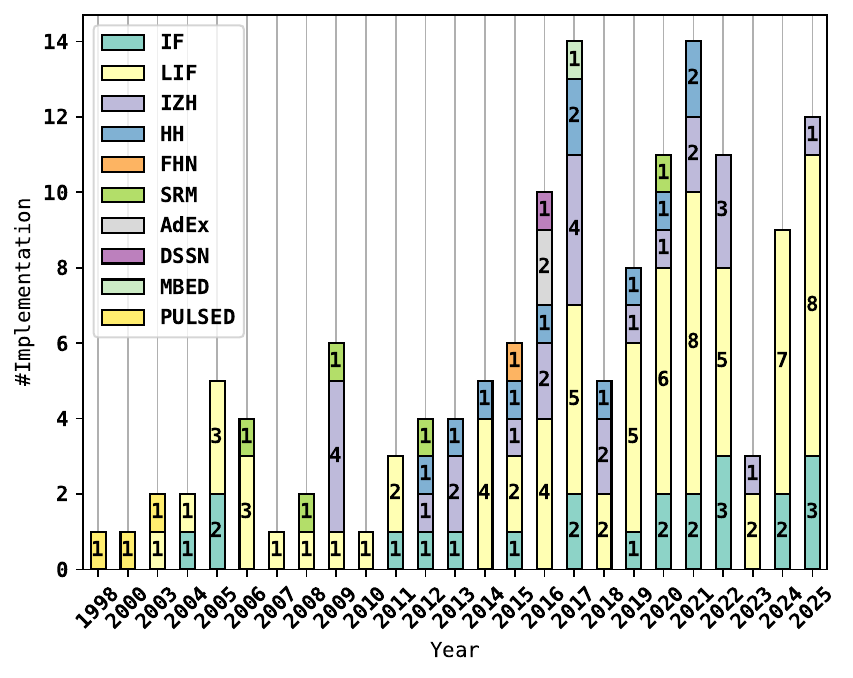}
  \caption{Spiking neuron models through the years.}
  \label{fig:models}
\end{wrapfigure}
\noindent
as only one representative was found from before the year 2000). Class CCM is the most populous with 39 examples (29\%), followed by CCM/AST with 26 examples (20\%) and FPP/CCM with 25 examples (18\%). As the CCM Trait is the easiest to implement -- the entire system needs to fit \textit{on-chip} -- we expected this outcome. We interpret this outcome as a tendency to implement the NMAs fully on-chip, possibly to increase the performance by avoiding costly auxiliary memory transfers. However, the co-location of computation and memory usually prevents the authors from implementing systems supporting a vast number of neurons -- those are usually simulated with systems that can offload the necessary parameters to larger

\begin{wrapfigure}[10]{o}{0.4\textwidth}
  \centering
  \includegraphics[width=\linewidth, trim={10pt 30pt 10pt 45pt}]{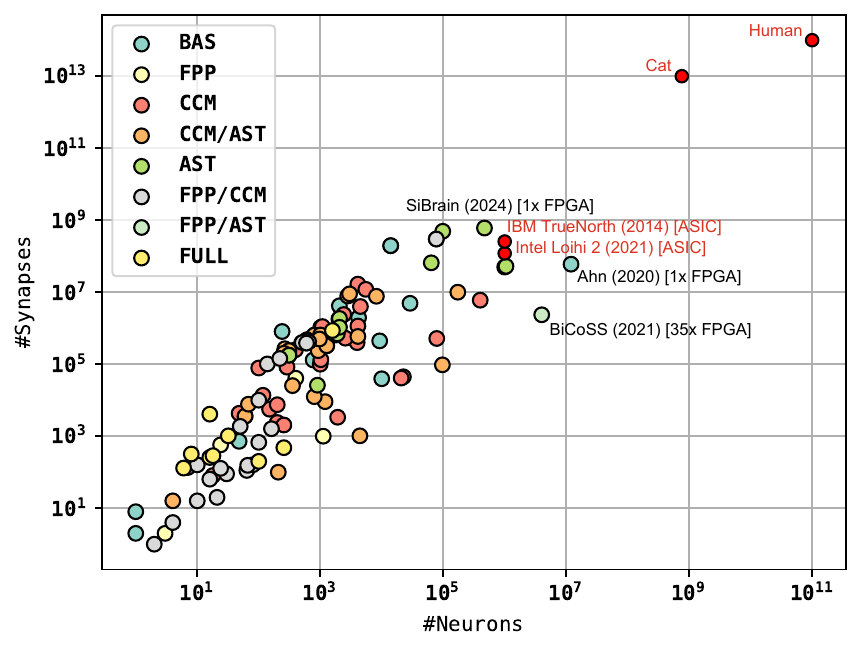}
  \caption{Complexity of the implemented architectures.}
  \label{fig:neu_vs_syn}
\end{wrapfigure}
\noindent
external memory, such as BAS systems. What is interesting is that in 2025, the majority of the surveyed systems belonged to the class CCM/AST, where the on-chip-implemented asynchronous PEs are time-domain multiplexed. We believe this trend will continue, possibly switching to FULL implementations as FPGA platforms continue to develop. We briefly analyzed the share of different neuron models supported by the surveyed architectures over the years. Indisputably, LIF-like neurons are the most common in the surveyed group, with 65\% of systems supporting this model, as shown in Figure \ref{fig:models}. We believe there are two main reasons for that. The surveyed neuroscientific applications focused more on the spiking behavior of the populations of neurons, rather than the faithful replication of the dynamics of individual neurons. In contrast, the computer science applications usually do not require a high level of biological plausibility to achieve satisfactory results. The second most popular choice within the surveyed group was IZH with 18\% of the architectures, and HH was in third place with a share of approximately 9\%. We can deduce that some researchers believe that IZH neurons, which allow for simulating biological behaviors (despite a lower focus on biological plausibility), can also be a time-worthy consideration for NMAs in both neuroscientific and computer science scenarios. Naturally, when a simulation needs to faithfully represent the biological dynamics, the HH model is used.

\begin{wrapfigure}[11]{o}{0.4\textwidth}
  \centering
  \includegraphics[width=\linewidth, trim={10pt 30pt 10pt 40pt}]{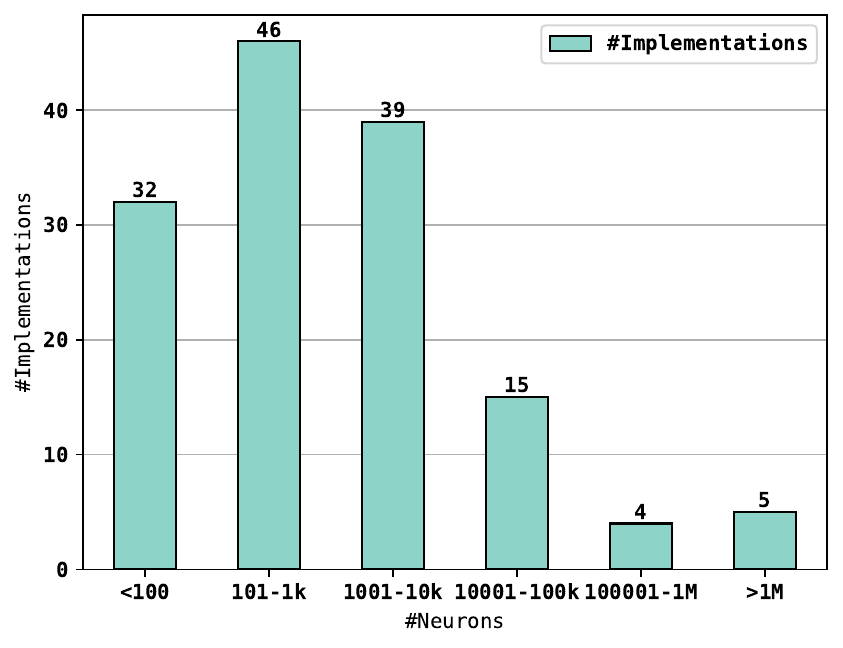}
  \caption{Sizes of implemented SNNs.}
  \label{fig:neuron_sizes}
\end{wrapfigure}
\noindent
We believe it may be beneficial to explore supporting more elaborate spiking models, as they may facilitate the application of modeled discoveries in the neuroscience domain. As a fitting example, we can consider STDP, which is used in many implementations and allows for online unsupervised learning. One of the more important questions related to the NMAs, primarily to those focusing on neuroscientific simulations, is: \textbf{how far are we from replicating the complexity of the human brain in hardware?}. A glance at Figure \ref{fig:neu_vs_syn} shows that, as of now, the single-die digital NMAs are not quite there yet, with numbers being 2-3 decades short of the target size of 10\textsuperscript{11} neurons and 10\textsuperscript{14} synapses. However, we observe a trend in this field toward supporting larger networks, in both ASIC- and FPGA-based solutions.

\begin{wrapfigure}[9]{o}{0.4\textwidth}
  \centering
  \includegraphics[width=\linewidth, trim={10pt 30pt 10pt 75pt}]{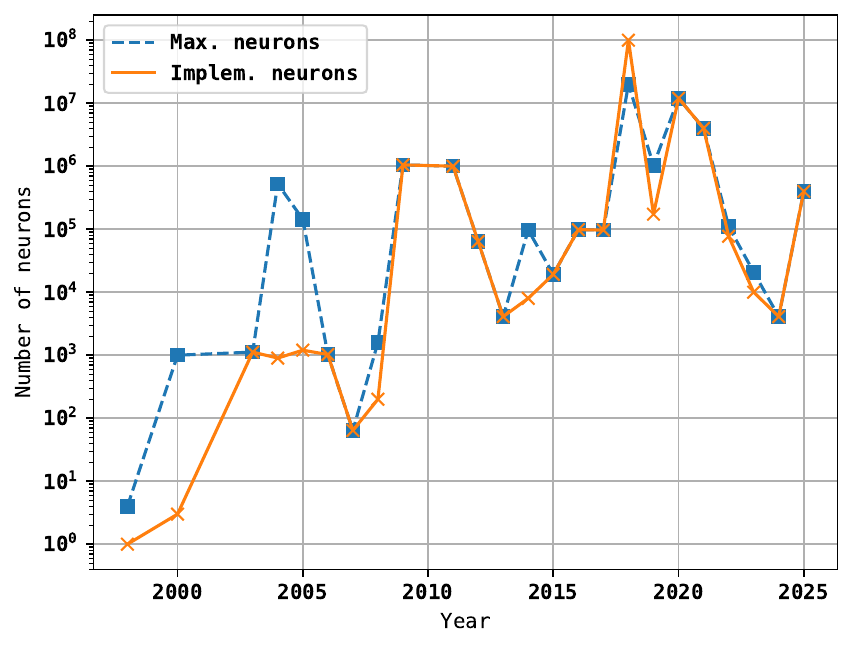}
  \caption{Max. vs the implemented number of neurons.}
  \label{fig:neuron_max}
\end{wrapfigure}
\noindent
What is interesting is that even though ASICs theoretically offer higher performance, due to bespoke silicon for neuromorphic computation, FPGA implementations do not lag far behind Intel Loihi 2 or IBM TrueNorth -- e.g., SiBrain\cite{chen_sibrain_2024} as seen in Figure \ref{fig:neu_vs_syn}. When we consider the continuous development of neuromorphic computing, with continuous efforts to understand, model, and utilize the biological phenomena in hardware, we believe it makes sense to consider reconfigurable computing as a valid paradigm for supporting neuromorphic systems, due to the ease with which novel algorithms and methodologies, such as online learning paradigms, can be added.

\begin{wrapfigure}[10]{o}{0.4\textwidth}
  \centering
  \includegraphics[width=\linewidth, trim={10pt 30pt 10pt 50pt}]{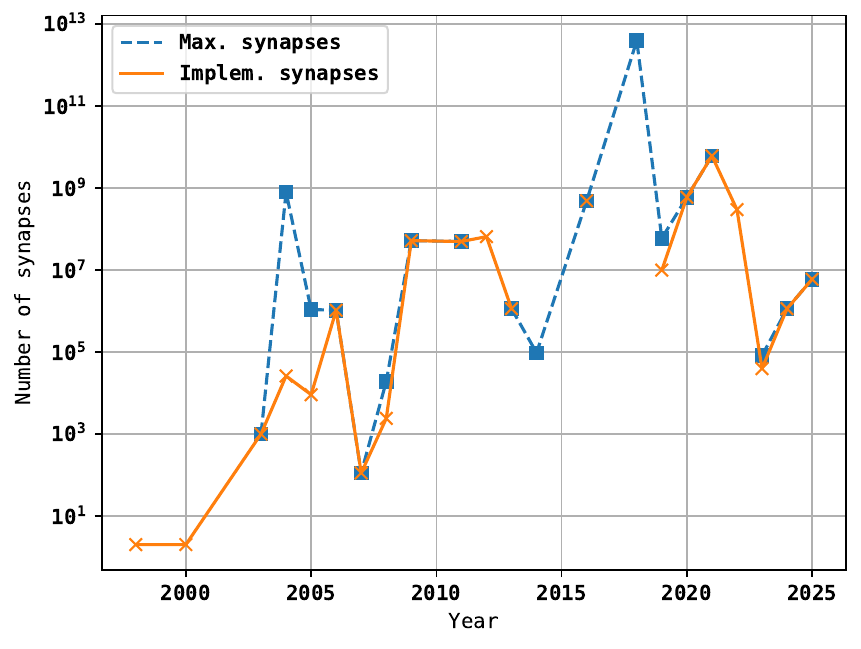}
  \caption{Max. vs the implemented number of synapses.}
  \label{fig:synapse_max}
\end{wrapfigure}
\begin{wrapfigure}[9]{o}{0.4\textwidth}
  \centering
  \includegraphics[width=0.9\linewidth, trim = {30pt 30pt 30pt 55pt}]{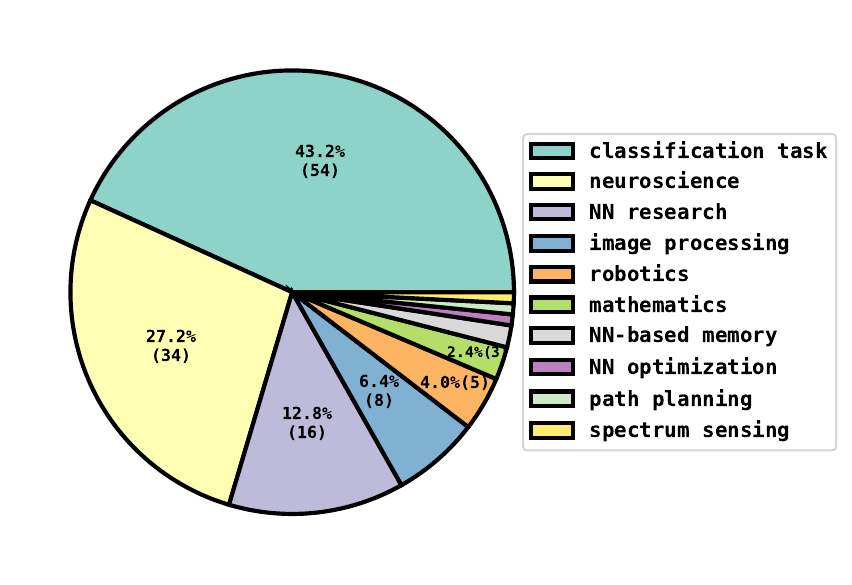}
  \caption{Breakdown of selected use cases.}
  \label{fig:use_case}
\end{wrapfigure}

\noindent
The ASIC implementations provide a finite number of properties, which cannot be easily expanded -- usually requiring fabrication of an entirely new circuit. Reconfigurable-computing-based systems are free from such constraints, allowing for the reconfiguration of existing designs to utilize better the state of knowledge of the human brain in hardware. In Figure \ref{fig:neuron_sizes}, we showed the share of different sizes of implemented SNNs, and a majority of them did not exceed ten thousand neurons. This observation is crucial, as even though a part of the neuromorphic community aims to support the complexity of the human brain in hardware, many test cases selected for the NMAs were classification tasks, as shown in Figure \ref{fig:use_case}. Especially for MNIST-like datasets, the FF-FC and SCNN topologies, often chosen for this task, do not require a vast number of neurons to achieve satisfactory accuracy. Due to this fact, some authors seem not to implement larger networks, relying instead on well-defined and well-known ANN benchmarks, such as classification tasks on image datasets. Though allowing for quick comparison with ANNs (especially CNNs) in terms of accuracy and power consumption, we believe this may not be the correct approach, as digital NMAs and SNNs may excel at different tasks -- as suggested by Davies at el.\cite{davies2021advancing}. Moreover, we analyzed \textit{overprovisioning}, i.e., how the largest surveyed systems from each year declare the maximum number of neurons and synapses and sizes they actually implement -- this is shown in Figures \ref{fig:neuron_max} and \ref{fig:synapse_max}. We can see that the promised sizes are often several decades higher than the sizes implemented in hardware for a specific use case. The anomaly in Figure \ref{fig:neuron_max} for year 2021 refers to the

\begin{wrapfigure}[11]{o}{0.4\textwidth}
  \centering
  \includegraphics[width=0.9\linewidth, trim = {30pt 30pt 30pt 40pt}]{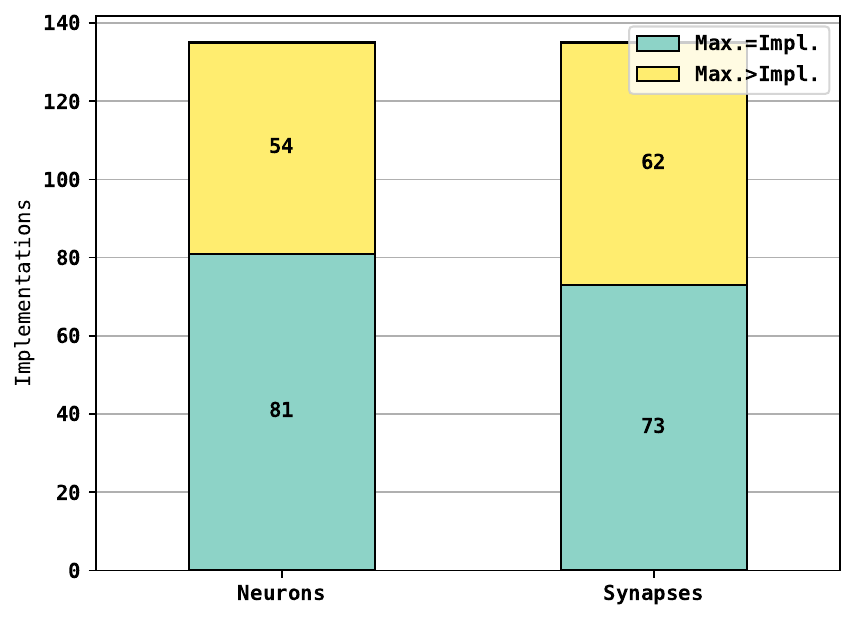}
  \caption{Ratio of systems using declared max. number of synapses and neurons.}
  \label{fig:neu_decl_imp}
\end{wrapfigure}
\noindent
system presented in \cite{wang_fpga-based_2018}, where the most complex theoretical system included 4T synapses with 20M neurons, however the number of implemented neurons for a system with smaller number of synapses was set to 100M. In Figure \ref{fig:neu_decl_imp}, we can see the ratio of systems that implemented the declared maximum number of both synapses and neurons, and even though there are theoretically more systems that actually do that, we have to keep in mind Figure \ref{fig:neuron_sizes} -- the majority of those systems assume a rather small number of neurons as their maximum capacity. Based on this information, we propose that a sizable number of implementations are overprovisioned, without an implemented use case that fully showcases the architecture's capa--

\noindent
--bilities. We also underline the need for unified set of benchmarks that would facilitate this process. Recently one such attempt called \textbf{NeuroBench}\cite{yik2025neurobench} has been proposed, but no such suite was used in the surveyed articles. Classification tasks and neuroscience alone were used as targets for about 70\% of the surveyed systems, highlighting the primary focus of different research groups. However, more unusual use cases are found throughout the neuromorphic research landscape (such as path planning), suggesting the validity of pursuing better use cases for neuromorphic computing. The breakdown of the

\begin{wrapfigure}[9]{o}{0.4\textwidth}
  \centering
  \includegraphics[width=0.9\linewidth, trim = {30pt 30pt 30pt 55pt}]{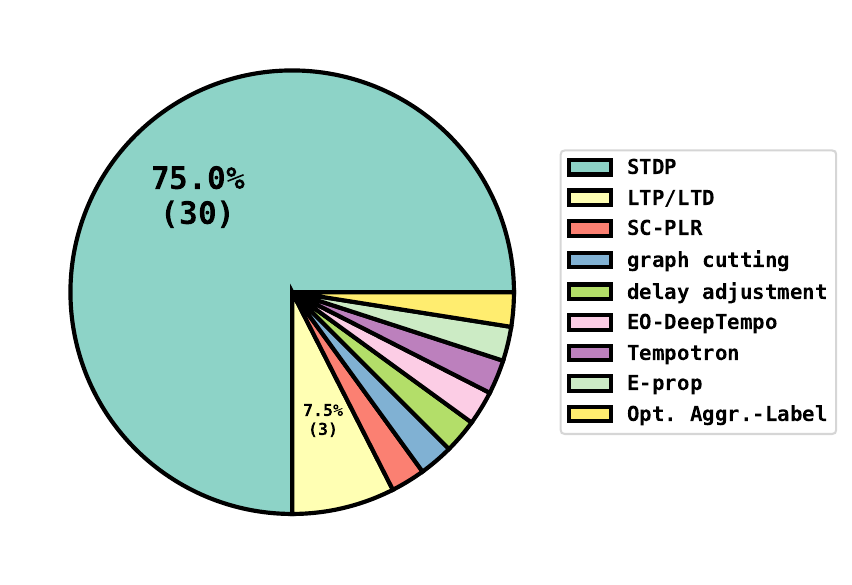}
  \caption{Supported online learning methods.}
  \label{fig:online_learning}
\end{wrapfigure}
\noindent
use cases for NMAs is shown in Figure \ref{fig:use_case}. The architectures supporting some form of online learning accounted for approximately 34\% of the surveyed architectures, showing that including this feature is not common. One can attribute this fact to a significant increase in the complexity of synaptic processing units, which, in addition to standard accumulation of the presynaptic current, must also apply learning rules such as STDP. What is particularly interesting is that the online learning is mainly supported by architectures that were tested with FF-FC, FF, LSM, RAND, and BIO networks, with only a single occurrence for the SCNN network. This may suggest that certain network topologies are better compatible with known online learning methods. Moreover, among the 34\% of the architectures that supported online learning, 75\% supported a form of STDP. Three implementations supported general LTP/LTD processes and there were singular examples of several learning methods: Tempotron\cite{fang_event-driven_2019}, SC-PLR\cite{liu_sc-plr_2024}, or EO-DeepTempo\cite{zhong_morphbungee-lite_2025} -- with the last two being a rather recent addition to this set. Two of the implementations supported a form of structural plasticity: graph cutting\cite{shayani_fpga-based_2008} and delay adjustment (for delay-based pattern memory)\cite{ang_spiking_2011}. Overall, we can summarize that STDP has become a widely used method for performing online learning, with multiple variants available, however due to (most likely) added complexity it is still not widely implemented. We predict that, with continued efforts from the neuroscientific community to better understand the mammalian brain, we will see an increase in the number of systems supporting some form of online learning. This will enable architectures to be more \textit{brain-like} and allow systems to operate with greater autonomy from central data centers performing training, thereby potentially reducing communication overhead and power consumption. Interestingly, the majority of implementations surveyed used AMD/Xilinx platforms, at nearly 84.7\% (116 examples). The rest used Intel/Altera (14.6\%, 20 examples) products, with a single example of Lattice Semiconductor. AMD/Xilinx platforms were thus the most popular, at least in the surveyed group. Expanding on the topic of manufacturers and platforms, only 25 surveyed systems were clocked at a frequency of 200 MHz or higher, suggesting that there is potential for improvement in terms of at least \textit{speed of operation} for NMAs. More on this topic can be found in Appendix \ref{sec:app_ext_freq}.

\begin{wrapfigure}[9]{o}{0.4\textwidth}
  \centering
  \includegraphics[width=\linewidth, trim = {10pt 30pt 10pt 60pt}]{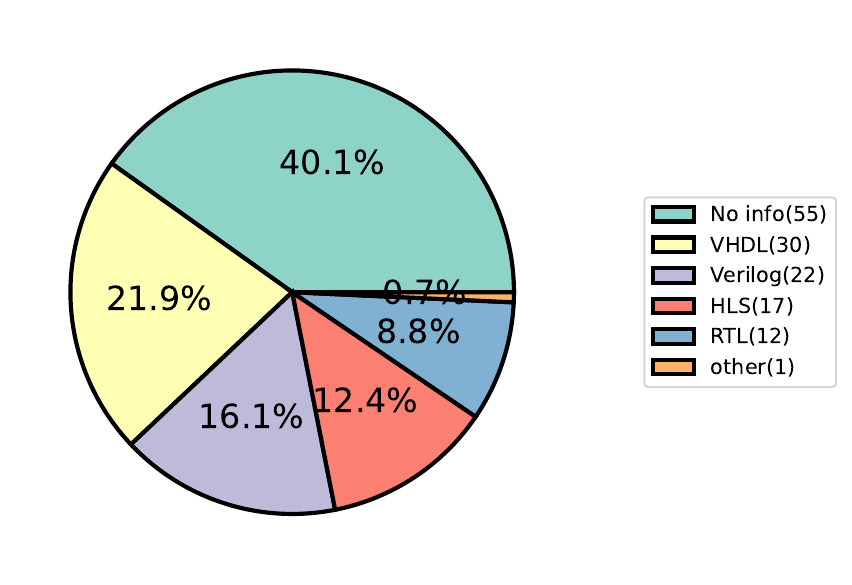}
  \caption{Share of different implementation techniques.}
  \label{fig:implementation-type}
\end{wrapfigure}
We reviewed the surveyed implementations to assess which implementation approach is usually chosen by the authors, be it direct RTL design through Hardware Description Languages (HDLs) or High-Level Synthesis (HLS), as there is a significant trade-off between the two. The former allows for more direct realization of intended functionality in hardware, which brings potentially greater accuracy in operation at the cost of reduced scalability and maintainability (verbose code, low-level description of hardware). On the other hand, HLS usually uses a well-known high-level programming language (such as C++ for Vitis from AMD/Xilinx), which allows for easier implementation, prototyping, and scalability but introduces barriers, as the tools are not capable of implementing certain scenarios\cite{cong_fpga_2022}. What we found interesting is that the authors were not always keen to share this detail, with $40.1\%$ of all articles not providing it (Figure \ref{fig:implementation-type}). From the remaining $\approx 60\%$, nearly $47\%$ used RTL coding through HDLs (VHDL and Verilog being the most popular) and $12.4\%$ used HLS in various forms (from general Xilinx/Intel tools to bespoke custom solutions). The main trends observed in this analysis were that older implementations (e.g., before 2010-2012) often defaulted to VHDL, with HLS appearing more frequently in recent years -- surely due to HLS methodology maturing steadily and achieving decent performance in digital design\cite{cong_fpga_2022}. We also noticed that when authors implemented their own custom high-level synthesis solutions, VHDL was the language targeted in the backend, while Verilog was often chosen for smaller systems. With systems focusing on streaming the data through the processing elements and large-scale systems, we noticed HLS to be often used, most likely due to focus on interconnection, data movements, and optimizations related to those properties, instead of optimizations on the RTL level. Additional analysis can be found in Appendix \ref{sec:app_ext_types}.

At this point, we would like to address perhaps the most important question related to NMAs: \textbf{which class should one use for their application?}. We propose a set of pointers that can aid with this task based on the analysis of \textit{extreme} use cases (i.e., largest SNNs, fastest operation, etc.). We analyzed systems that targeted classification tasks (for the MNIST dataset, as it was the most popular) and neuroscientific simulations, the most populous categories. For neuroscientific simulations selecting a metric that can represent \textit{speed of operation} is not trivial. Different neuron models require different number of operations to advance a simulation timestep and the often reported \textit{speed-up over real-time} should be analyzed considering the timestep selected. Let us assume that a timestep for system A is 0.5s, and 0.01ms for system B. If both of them are reported to achieve speed-up over real time to be $2x$ for the same size of SNN, then system B should be considered \textit{faster}, because it computes orders of magnitude more operations per second. The \textit{largest SNNs} are considered in a \textit{neuron-centric} way, i.e., we consider the number of neurons, disregarding synaptic connections and assuming activity for those neuroscientific experiments to be similar between the implemented use cases. We acknowledge those assumptions to be simplifications, however, a generalized \textit{test suite} for NMAs related to neuroscience was not applied to the systems, and we believe that the number of implemented neurons with assumption of similar neural activity can give a good notion of required computational effort to update an SNN. With this rationale we analyzed the systems for neuroscience, and, after a more thorough analysis, we decided to include NN research-targeting implementations. The reason is that those systems usually target research related to computational neuroscience. We

\begin{wrapfigure}{o}{0.4\textwidth}
  \begin{minipage}{0.4\textwidth}
    \begin{equation}
      \label{eq:rtf}
      RTF = \frac{t_{sym}}{t_{bio}}
    \end{equation}
    \begin{equation}
      \label{eq:nus}
      NUs/s = \frac{N}{hRTF} = \frac{N}{h\frac{t_{sym}}{t_{bio}}}
    \end{equation}
    \begin{equation}
      \label{eq:flops}
      FLOPs/s = NUs/s * CE
    \end{equation}
  \end{minipage}
\end{wrapfigure}
\begin{wrapfigure}[11]{o}{0.4\textwidth}
  \centering
  \includegraphics[width=\linewidth, trim = {10pt 30pt 10pt 65pt}]{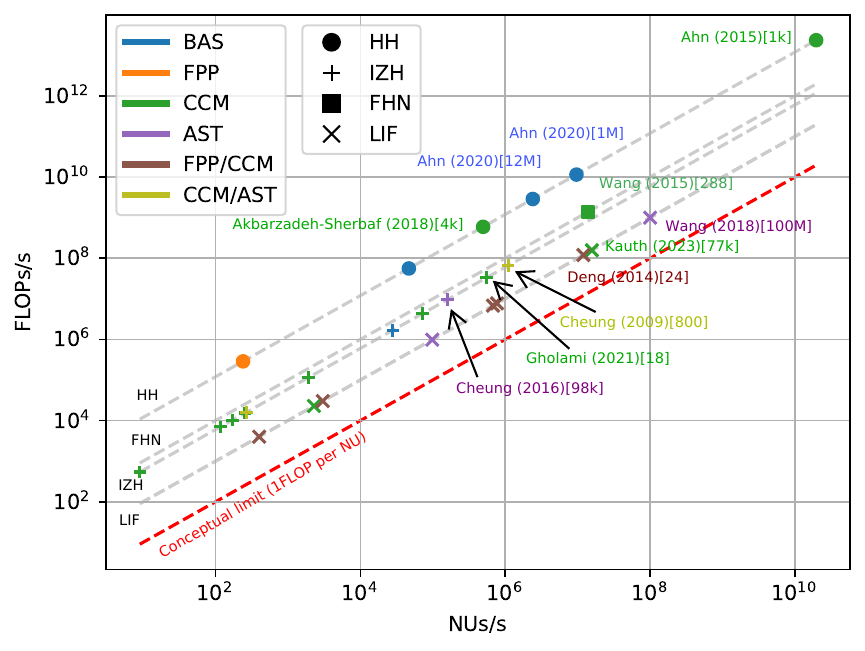}
  \caption{Neuroscience-related systems performance.}
  \label{fig:neuro_speed}
\end{wrapfigure}

\noindent
considered two metrics: \textbf{neural updates per second (NUs/s)} (Eqn. \ref{eq:nus}) and \textbf{FLOPs/s} (Eqn. \ref{eq:flops}), where $RTF$ - real-time factor (Eq. \ref{eq:rtf}), $t_{bio}$ - simulated biological time, $t_{sym}$ - actual wall-clock time of the simulation, $h$ - computational timestep, $N$ - neuron number in SNN, $CE$ - computational effort in FLOPs for the used neuron model (Table \ref{tab:neuron_models}). The results are shown in Figure \ref{fig:neuro_speed}. The $CE$ is simply a coefficient in equation for NUs/s, and as such -- the systems that share a neuron

\noindent
model will be on the same \textit{gray line}. Moreover, $CE >= 1$, as it is not possible to update a neuron with less than 1 operation. We marked the three fastest systems (apart from for the FHN neuron, as there was only one system that used it in the group in question), showing also the number of neurons implemented in square brackets -- systems that are fast but deal with hundreds of neurons are perhaps not better suited for large-scale neuroscientific experiments than those that achieve not as high computation speed but deal with millions of neurons. We can see that CCM Trait is often used -- the fastest system was presented by Ahn\cite{ahn_neuron-like_2015}. However, the systems that follow in order by Ahn\cite{ahn_implementation_2020} was BAS, and Wang et al.\cite{wang_fpga-based_2018} was AST and deal with 

\begin{wrapfigure}[9]{o}{0.4\textwidth}
  \centering
  \includegraphics[width=\linewidth, trim = {10pt 30pt 10pt 75pt}]{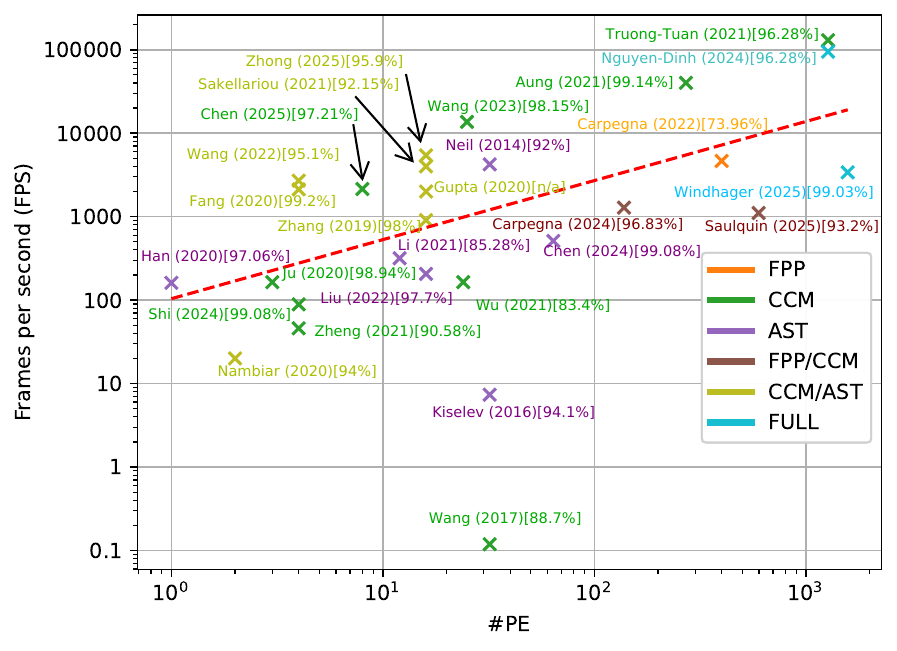}
  \caption{Speed of classification task implementations.}
  \label{fig:mnist_perf}
\end{wrapfigure}
\noindent
a significantly higher number of neurons (orders of magnitude higher in both cases), suggesting that CCM Trait facilitates achieving higher speed, but yields the systems more sensitive to on-chip resource limits. FPP Trait was also used in the surveyed group, especially when LIF neurons were used. For the two articles by Ahn\cite{ahn_neuron-like_2015,ahn_implementation_2020} the architectural concept is similar, yet due to implementation choices we identified those systems as belonging to different classes. However, we can see the trade-off between SNN size and speed of operation clearly. The systems introduced a peculiar memory organization (divided-and-merged) that we believe allowed for overall high speed of operation. System by Wang et al.\cite{wang_fpga-based_2018} simulated the largest network (100M LIF neurons) and the operation speed is high -- inclusion of AST Trait could potentially result in more sparse transfers with external memory. Additionally, the system used minicolumn-based organization of neurons, which could exploit the locality of neuronal activity. We propose that CCM Trait seems beneficial in creating \textit{fast} implementations due to parameters being available on-chip. We expected FPP to be of the highest priority, but it seems not to be always required, suggesting that congestion caused by high number of PEs can have an impact on performance. As for AST, fetching information only during spike events can be beneficial to the speed of the system. For classification tasks, we selected the FPS metric as \textit{speed of operation} coupled with accuracy on selected benchmark. The idea is that, if system A is as fast in inference as system B, accuracy is a factor that plays a great role in differentiating the two implementations. We acknowledge that accuracy is a function of many variables -- including the employed algorithm, which is the simulated SNN topology in this case. However, the surveyed systems often used different SNN topologies, and it is extremely difficult to make the comparison fairly without a unified \textit{test suite}, as mentioned previously. The lack of unified set of benchmarks was raised in the literature\cite{schuman_opportunities_2022} as an important issue, and we agree that introducing such benchmarks would greatly facilitate the comparison between different architectural choices. In Figure \ref{fig:mnist_perf} we present \textit{speed of operation} as a function of \#PE with information about achieved accuracy for MNIST dataset. We can see that the fastest systems were presented by Truong-Tuan et al.\cite{truong-tuan_fpga_2021} and Nguyen-Dinh et al.\cite{nguyen-dinh_novel_2024}, which are improvements to the TrueNorth architecture. Next in line was DeepFire\cite{aung_deepfire_2021} and Wang et al.\cite{wang_resource-efficient_2023}, which both achieved significantly higher accuracy then the fastest systems. Interestingly, three out of four systems belonged to CCM class and Nguyen-Dinh et al. built a FULL system. There was a number of CCM/AST systems that achieved high performance with accuracy $>92\%$. The FPP and FPP/CCM systems achieved comparable FPS and accuracy (with the exception for system by Carpegna \cite{carpegna_spiker_2022}) -- with higher number of PEs, as this stems from the nature of FPP Trait. We propose that CCM Trait is beneficial to performance for classification tasks, as parameters for SNN are held on-chip. However, there are multiple systems that incorporate AST Trait (middle of the graph) and achieve somewhat worse, yet acceptable performance for specific scenarios (1kFPS is still more than enough for real-time processing). Thus, event-driven operation can alleviate communication effort and allow rerouting resources to computation. Finally, FULL systems give good results, which may suggest that contributions from all Traits may be beneficial for operation, albeit challenging to implement. Moreover, the indicated trend line suggests that increased number of \#PEs allows for higher \textit{speed of operation} -- excluding two outliers. This, to an extent, follows Amdahl's law and aligns with intuition that increased number of \textit{computing nodes} should allow for faster computation. Thus, we suggest including CCM and FPP, while considering AST Traits to improve overall system performance.

\begin{wrapfigure}[11]{o}{0.4\textwidth}
  \centering
  \includegraphics[width=\linewidth, trim = {10pt 30pt 10pt 45pt}]{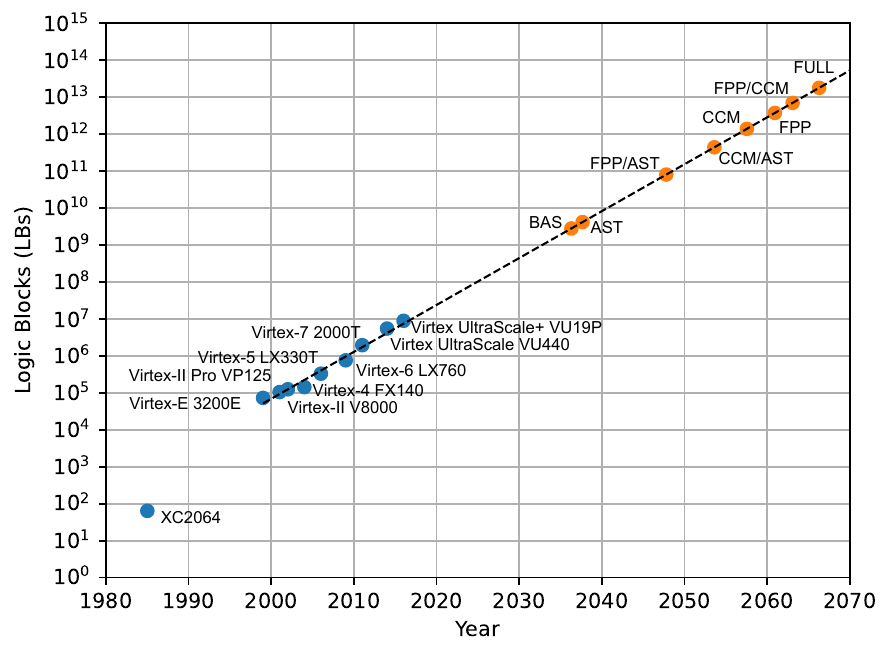}
  \caption{Predicted logic required to implement $10^{11}$ of neurons with a particular class of NMA on FPGAs.}
  \label{fig:pred}
\end{wrapfigure}
Finally, Figure \ref{fig:pred} shows the trend of available logic resources on FPGAs, based on the size of the AMD/Xilinx Virtex FPGA device family (AMD Versal are discarded, because they are CGRAs rather than \textit{pure} FPGAs) through the years and required logic resources for implementing the estimated size of the human brain on a single chip for the most promising architectures from every class. We can see that, if we discard the impact of embedded memory size and improvements in on-chip communication and assume this trend to hold, classes BAS and AST can be expected to achieve human-brain scale around years 2035-2038 -- the rationale can be assumed that they require the least amount of on-chip logic resources and scale the best in single-chip implementations. The rest need more time -- classes CCM/AST, CCM, FPP, FPP/CCM, and FULL would achieve the goal at least in 2055, and class FPP/AST (for this class, the prediction is rather vague here, because it is based on a system that used multiple FPGAs -- BiCoSS) would achieve it after 2045. However, considering that IEEE suggests in their International Roadmap for Devices and Systems (IRDS) from 2022 that Moore's Law will be upheld for an additional 10-15 years\cite{https://doi.org/10.60627/c13z-v363}, only classes BAS and AST seem likely to achieve the aforementioned goal in the predicted years.

\section{Conclusion}
FPGA-based neuromorphic systems can be implemented in various ways, with different advantages and disadvantages, as reflected in the classes of the Taxonomy we provided. Representatives of those classes meet diverse sets of requirements and enable the achievement of various goals, ranging from classification tasks to neuroscientific simulations. We expect the popularity of digital NMAs to continue growing, and that even more brain-like systems will be presented as new discoveries related to the human brain are made.  Moreover, it is apparent, due to the stark differences in design choices between the classes, that creating a unified Design Space Exploration framework that would allow for convenient and automatic search of optimal architectural class for given requirements (e.g., number of neurons, synapses, neuron models and available FPGA platform) is troublesome, but we believe it is necessary in order to simplify the NMA design process and crucial for a fair comparison between the implementations.

\begin{acks}
  This work was supported by Swedish Research Council’s Project Building Digital Brains under Grant 2021-04579. We greatly thank the reviewers for their feedback that helped improve the quality of the survey. No GenAI-related tools were used in creation of this survey, except for tools related strictly to ensuring grammatical correctness (Grammarly).
\end{acks}

\bibliographystyle{ACM-Reference-Format}
\bibliography{references}

\newpage
\appendix
\section{Appendix}
\label{sec:app_ext_analysis}
As hinted in Section \ref{sec:trends}, we provide an extended analysis of the presented data. As some parts of this analysis have a speculative nature -- primarily due to the lack of possibility of fairly assessing all presented architectures on a common platform, with common use cases and so on -- we decided to include it in the Appendix for the interested reader to possibly aid them in the design process of their own NMAs. Additionally, we hope that observations, beliefs, and concerns raised in this Appendix will further convince the reader that the field of digital NMAs (and most likely the neuromorphic computing in general) requires a lot of attention in terms of unifying performance metrics, benchmark suites, and general standardization -- e.g., in a form of Taxonomy, like the one proposed in this article. Moreover, we provide additional clarifications for certain statements from the main text.

\subsection{Impact of targeted SNN topology on the system}
\label{sec:app_ext_topo}
The impact of the selected SNN topology (or topologies) by an architecture on overall performance, adaptability, scalability, and ease of implementation is a topic that deserves its own article -- we refer interested readers to e.g., Motaghian et al.\cite{motaghian_topology-aware_2026}. As this survey focuses on architectural choices and aims to provide a valid categorization scheme for architectural choices during the \textit{implementation} stage rather than the \textit{planning} stage, we did not include this kind of discussion in the main flow of the article. It is, however, evident that many systems are organized in a way that should provide the best performance for a particular topology that the authors were interested in, which is primarily true for the non-neuroscientific accelerators -- e.g., Spiker+\cite{carpegna_spiker_2024} or DeepFire\cite{aung_deepfire_2021}. In those systems, often FF-FC, SCNN, or WTA networks are used, and their structures resemble the original SNN concept -- layers of neurons connected in a way that mimics the original SNN concept. Oftentimes, the authors did not plan to support other topologies and instead focused on improving performance in selected use cases (such as MNIST dataset image inference). We believe that creating bespoke systems for specific use cases can deliver the highest possible performance, but one should always consider the costs of potentially adapting the system for other use cases. If a system is supposed to do only one thing, it is probably enough to stick with one of the architectures presented in this article that target the topologies and use cases of interest. However, as neuromorphic systems continue to be developed and new "appropriate" use cases are suggested\cite{davies2021advancing} (such as CSPs instead of inference tasks, which are usually solved more effectively by regular CNNs), we should consider designing increasingly effective systems with generality in mind.

This topic gets even more peculiar if we consider the impact of the chosen topology on the \textit{actual} number of synapses and neurons implemented. The best example of this is the SCNN, which is built on the concept of \textit{weight sharing}\cite{waibel1988consonant}, where weights are shared throughout the kernel in convolution layers. Thus, the number of actual synapses stored in the system is usually not equal to the number of connections, \textit{per se}. The number of neurons would be equal to the summed sizes of particular kernels plus the output layer. This can, to some degree, explain why SCNNs are often reported to use many more neurons for particular use cases (such as classification tasks) than their FF-FC equivalents. In conclusion, the complexity of the implemented SCNNs, in terms of required synaptic information, can be lower than for FF-FC networks (especially with many hidden layers).

\subsection{Neuroscientific and classification tasks}
\label{sec:app_ext_tasks}
\begin{figure}
    \centering
    \includegraphics[width=0.5\linewidth]{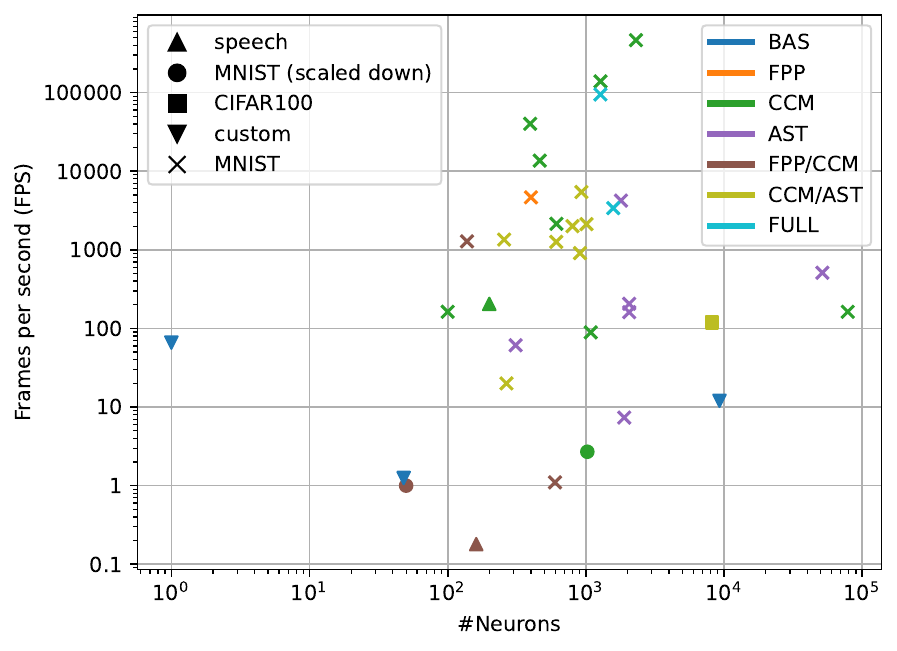}
    \caption{FPS vs number of PEs for MNIST test cases.}
    \label{fig:classification_perf}
\end{figure}
\begin{figure}
    \centering
    \includegraphics[width=0.5\linewidth]{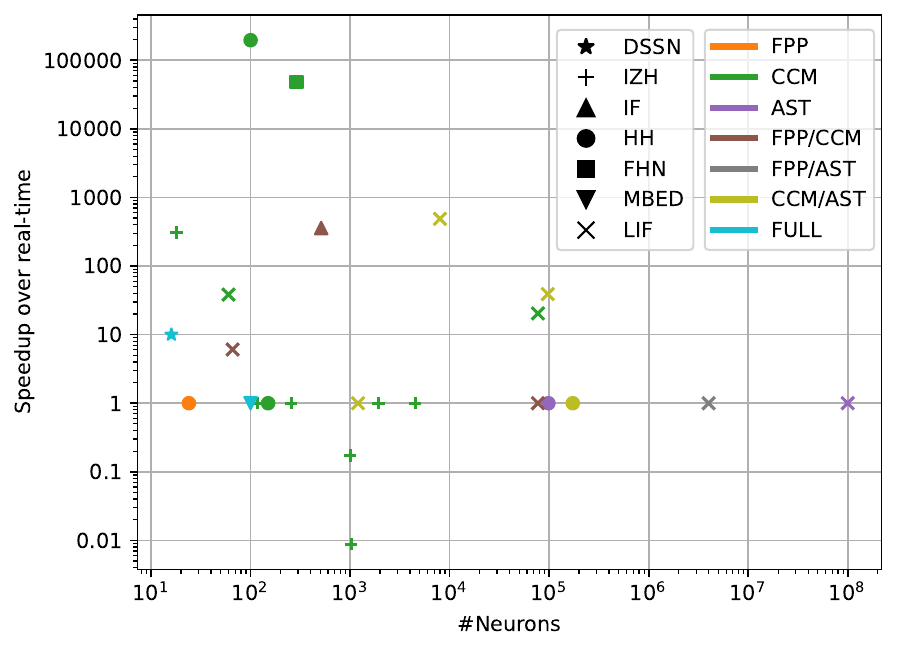}
    \caption{Speed-up over real-time operation for implementations targeting neuroscientific simulations vs \#Neurons.}
    \label{fig:neuroscience_perf}
\end{figure}
\begin{figure}
    \centering
    \includegraphics[width=0.5\linewidth]{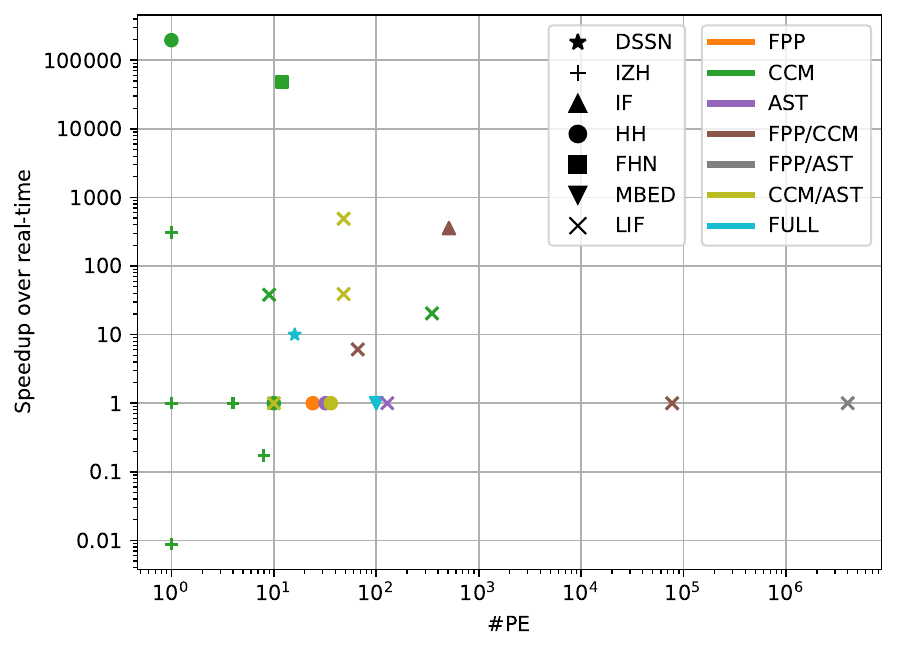}
    \caption{Speed-up over real-time operation for implementations targeting neuroscientific simulations vs \#PEs.}
    \label{fig:neuro_pe}
\end{figure}
In Figures \ref{fig:classification_perf} and \ref{fig:neuroscience_perf}, we took a closer look at possible dependencies between the size of the simulated SNN and the overall speed of operation for both classification tasks and neuroscientific simulations. Generally, we expected the smaller networks to be simulated "faster", mainly due to the reduced communication overhead and processing time, especially with the classes including the FPP and CCM Traits, as those increase the parallelism and reduce memory transfer overhead, respectively. One can argue that for classification tasks, we can observe the opposite - larger networks were simulated faster than smaller ones (especially for CCM and CCM/AST, as the rest of the classes had fewer representatives, and deriving meaningful trends was troublesome). This might be due to larger networks being generally implemented on architectures thought with size in mind that facilitate simulation by employing elaborate communication and processing schemes (neighbour-to-neighbour synchronization, multi-layer broadcast\cite{kauth_neuroaix-framework_2023}). However, this is somewhat vague, as more factors usually come into play, such as the platform used and the type of network implemented (LSMs have fewer, but randomly connected neurons, whereas FF-FC can usually be "streamed through" the accelerator sequentially). Thus, providing a concrete answer to the question about the direct connection between the size of implemented networks and the speed of operation for classification tasks is troublesome. As for the neuroscientific tasks, we can observe a tendency to focus on real-time operation (as pointed out in Section \ref{sec:trends}), but we can also see that the smaller the SNN, the larger the speed-up can generally be achieved, which fits with the initial claim. However, on an in-class basis, one cannot really pinpoint the clear trends. For example, for the CCM class, in Figure \ref{fig:neuroscience_perf}, we can see that regardless of the used neuron model, the results vary without any monotonic trend. In both types of applications, we believe the lack of clear monotony of the results can be due to the aforementioned lack of a proper benchmarking toolbox, which would help the engineers to target "as good as it can be" instead of "good enough for the chosen application". As of now, architectures are being fine-tuned based on the set of arbitrary tests, benchmarks, and cherry-picked applications, which often result in different results for very similar test cases. Providing the researchers with a clear methodology for testing and improving their designs would alleviate this problem. Moreover, in Figure \ref{fig:neuro_pe} we can see that as the number of PEs increases, there is no clear trend that would indicate that the speed of operation increases. On the contrary, we can see that larger systems seem to perform \textit{worse}. However, we have to consider how many systems targeted real-time, often due to the use case they were supposed to support.

\subsection{More on implementation approaches}
\label{sec:app_ext_types}
\begin{figure*}[t!]
    \subfloat[BAS]{%
        \includegraphics[width=.48\linewidth]{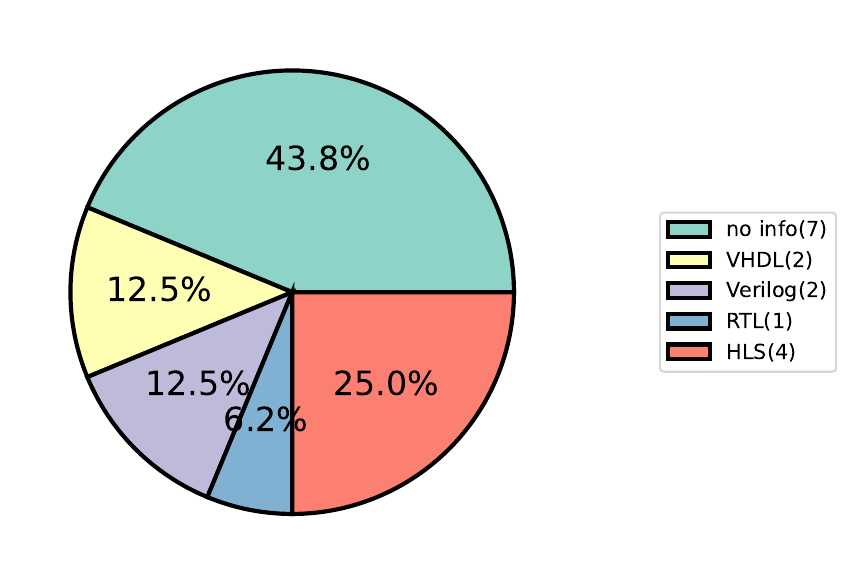}%
        \label{subfig:a}%
    }\hfill
    \subfloat[FPP]{%
        \includegraphics[width=.48\linewidth]{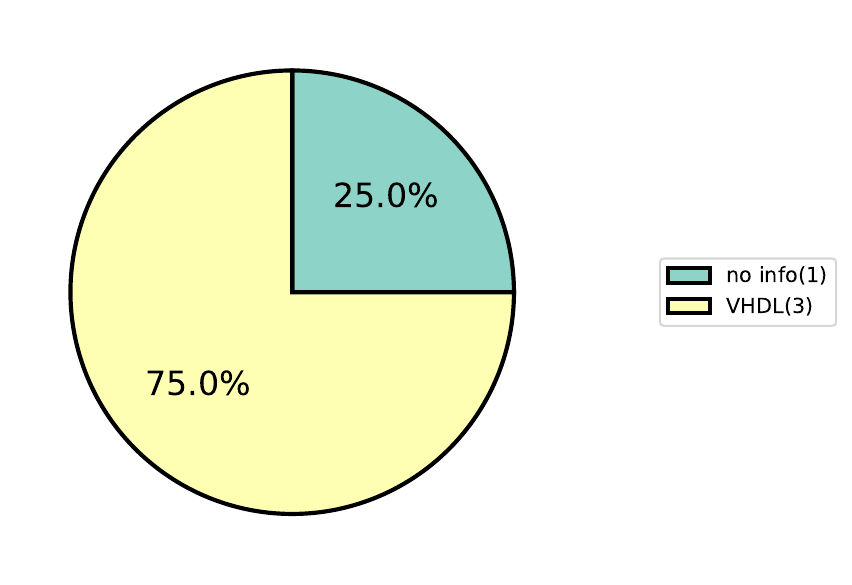}%
        \label{subfig:b}%
    }\\
    \subfloat[CCM]{%
        \includegraphics[width=.48\linewidth]{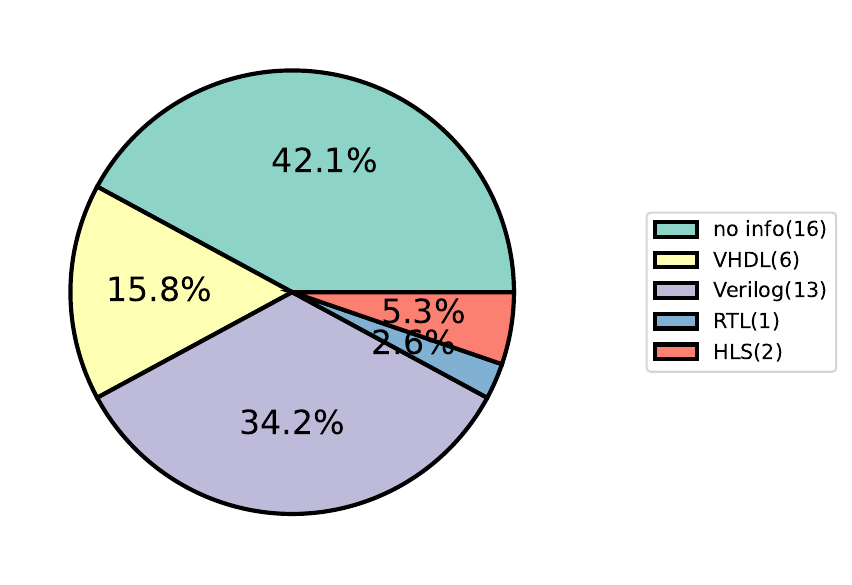}%
        \label{subfig:c}%
    }\hfill
    \subfloat[AST]{%
        \includegraphics[width=.48\linewidth]{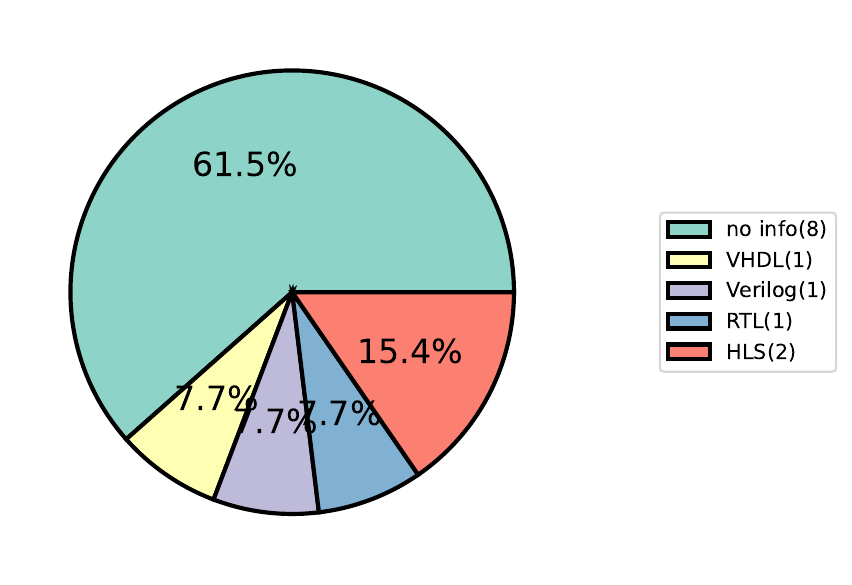}%
        \label{subfig:d}%
    }
    \caption{Share of different implementation types (methods) per Class (Classes BAS -- AST).}
    \label{fig:imp-types-pies-03}
\end{figure*}

\begin{figure*}[t!]
    \subfloat[FPP/CCM]{%
        \includegraphics[width=.48\linewidth]{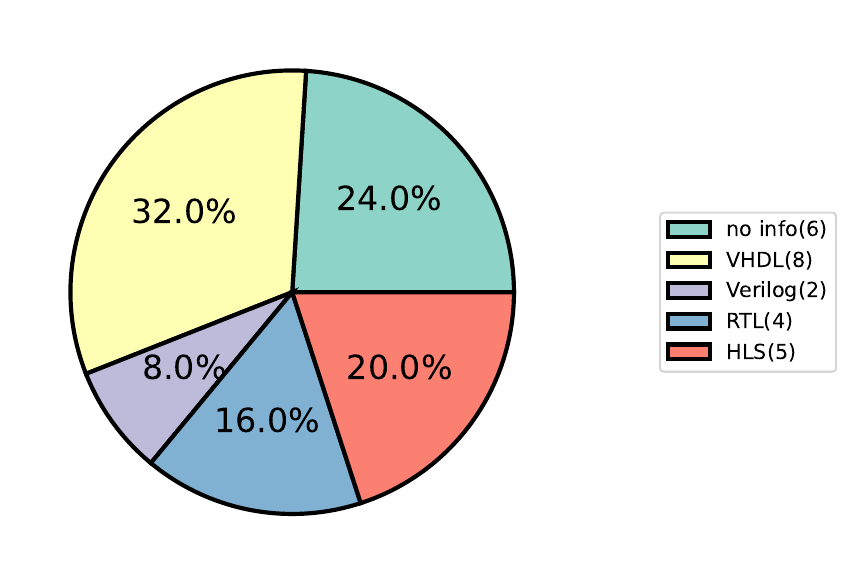}%
        \label{subfig:a}%
    }\hfill
    \subfloat[FPP/AST]{%
        \includegraphics[width=.48\linewidth]{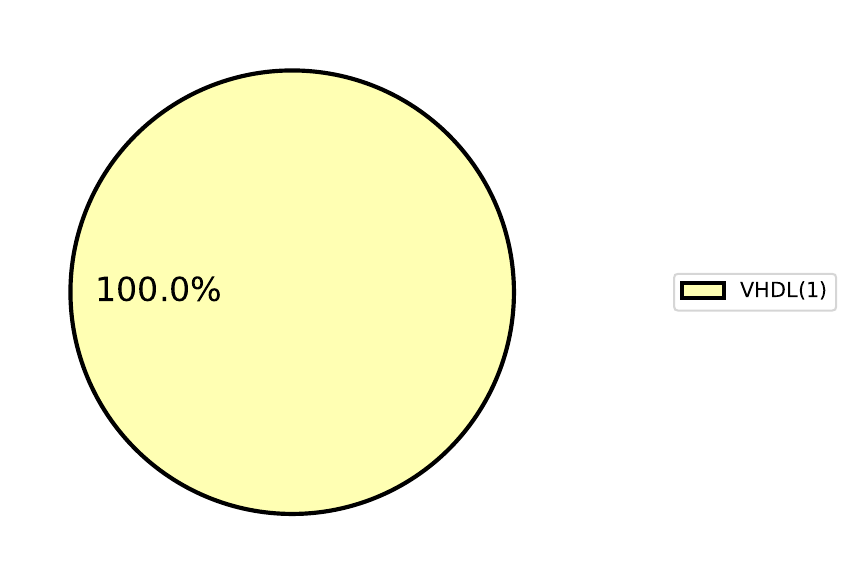}%
        \label{subfig:b}%
    }\\
    \subfloat[CCM/AST]{%
        \includegraphics[width=.48\linewidth]{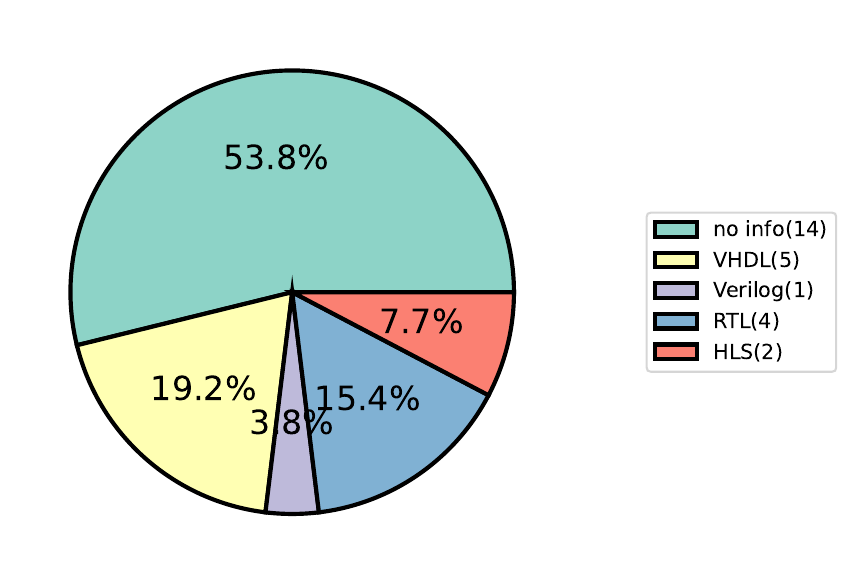}%
        \label{subfig:c}%
    }\hfill
    \subfloat[FULL]{%
        \includegraphics[width=.48\linewidth]{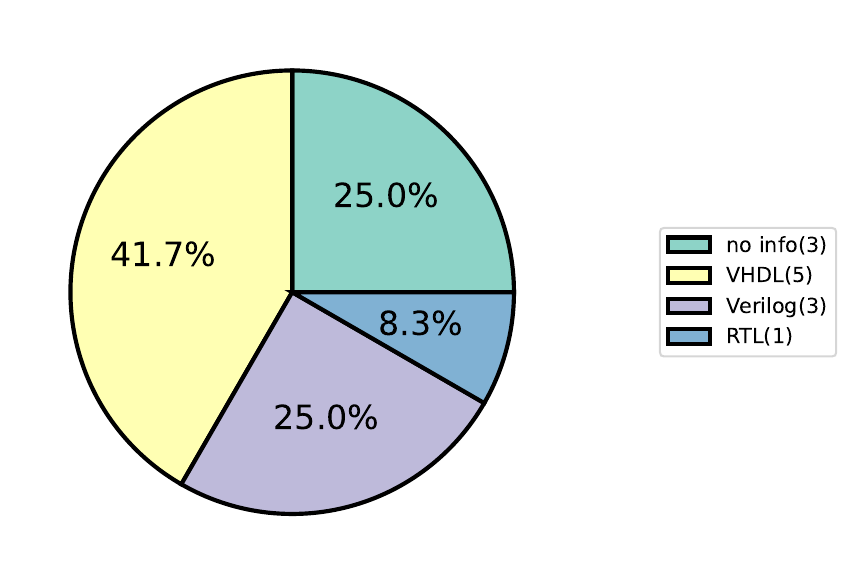}%
        \label{subfig:d}%
    }
    \caption{Share of different implementation types (methods) per Class (Classes FPP/CCM -- FULL).}
    \label{fig:imp-types-pies-47}
\end{figure*}

As discussed in Section \ref{sec:trends}, we observed that not for every implementation did the authors provide the data about the implementation techniques used, i.e., utilization of direct RTL coding, HLS tools, or other possibilities -- the "No info" category in Figure \ref{fig:implementation-type} proves it. In Figures \ref{fig:imp-types-pies-03} and \ref{fig:imp-types-pies-47}, we can see a more detailed breakdown of those choices per proposed Taxonomy class. Based on this information, we aimed at formulating further conclusions and possible trends based on the properties of implemented classes and how easily they are represented with the aforementioned techniques. It is in order to reiterate that the stated trends and observations are based on the observation of a somewhat limited number of systems, and those could potentially be further from the truth should more specimens be surveyed. For the BAS class, we identified five systems implemented with direct RTL coding and four with HLS, which does not favor either approach decisively. As those systems usually use a number of PEs to perform the operations on neural information streamed from external memory (as pointed out in Section \ref{sec:overview_class0} and \ref{sec:trends}), those systems can usually be implemented with both techniques. Apart from directly coding in the streaming interface, one can take advantage of bespoke interfaces like AXI-Stream\cite{arm_axi4_stream} that facilitate the design of this kind of systems. FPP systems included only four representatives, and yet three out of them were implemented using VHDL as the description language. Interestingly, for Spiker\cite{carpegna_spiker_2022}, the decision seems to be connected with the requirement to implement predefined IP blocks, which are later "strapped together" to create a complete system -- the authors seemed to have chosen this approach to have more control over generated hardware ("This stage capitalizes on the modularity of the proposed architecture and utilizes an available library of neuron models. This is the core and the main contribution of the Spiker+ tool. It relieves the user from the need to have specific skills in hardware design: starting from a high-level description of the network, it automatically generates the desired architecture using the VHDL language."\cite{carpegna_spiker_2024}). For class CCM, for which we gathered the most data and categorized the most architectures as members, we can see the tendency towards using RTL coding, with only two examples of HLS tools being used. We believe this can be due to the ease of control in "putting" the entire system on-chip with the RTL coding approach. Moreover, many systems in the CCM class were targeting somewhat smaller SNNs, which reduced the complexity of the overall system (such as Ambroise et al.\cite{ambroise_biorealistic_2013} with 117 neurons and Wu et al.\cite{wu_efficient_2021} with 100 neurons). However, there are also examples of larger systems -- Mohammadhassani et al.\cite{mohammadhassani_digital_2025} with 400k neurons -- and so we cannot say that the size of the simulated network is a direct factor during the choice of implementation approach. Three systems were directly coded in RTL, and two were using HLS for the AST class, which brings even less information than for the BAS class. What is interesting, however, is that the HLS can be used to design dataflow structures (such as NeuroFlow\cite{cheung_neuroflow_2016}) that can deal with the asynchronous operation, depsite HLS using primarily declarative/procedural\footnote{Choosing the more appropriate term to describe C/C++/Java-based HLS we are leaving to the reader.} programming. For class FPP/CCM, the RTL coding seems to be the most popular choice, with nearly three times as many examples in comparison to the HLS-coded solutions. Our rationale in this regard follows the idea of improved control over "putting" the structure on-chip. As for the FPP/AST systems, we are unable to provide meaningful analysis, primarily based on the singular representative who provided implementation approach information. However, we can see how BiCoSS\cite{yang_bicoss_2021} being implemented in VHDL was most likely a good choice, considering a very specific type of multi-device platform used and a particular connectivity scheme (BFT) used -- direct RTL coding usually gives more control over designed hardware, especially considering the verbosity of VHDL. Unfortunately, the class with the most "recent" representatives (CCM/AST) is also the most lacking in terms of information about the implementation methods\footnote{There were a few implementations that we could have deduced that "they must have been implemented in way A or B", but we did not want to conduct guesswork.}. However, it is also worth noting that usually the systems from this class are rather elaborate in terms of communication, memory management and so on, so the authors usually focused on explaining the overall high-level view of the system components instead of how they were coded in. From the data we gathered, we can still observe that HLS is supposedly not that popular, with five times more representatives being directly coded in RTL. We can attribute this to the overall tendency to have a higher level of control when implementing those elaborate systems, however we can also suggest potential HLS benefits here - e.g., generating a mesh-like NoC is conceptually a repetitive process that can potentially be sped up by using a specialized system generator for NoCs with generic PEs, later adjusted for the purposes of SNN simulation. We did not see such a compiler during surveying the systems, but it seems like a valid approach. 
In the FULL class, we did not see a single representative that would be coded in using HLS tools. As this class intuitively requires the most control over how it is implemented, due to the "experimental" nature of the systems, putting every piece of computation, communication, and memory on chip in such a way that ensures correctness and that the design will actually fit on the platform, it is actually rather expected that RTL coding would be the default technique.

\subsection{On clock frequency of the implementations}
\label{sec:app_ext_freq}
\begin{figure*}[t!]
    \subfloat[BAS]{%
        \includegraphics[width=.48\linewidth]{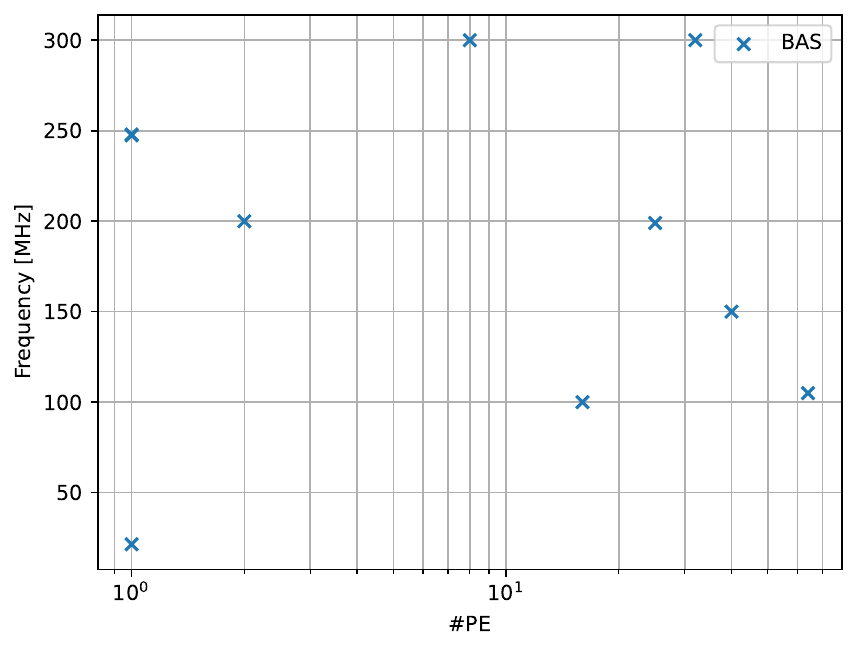}%
        \label{fig:app_freq_03_0}%
    }\hfill
    \subfloat[FPP]{%
        \includegraphics[width=.48\linewidth]{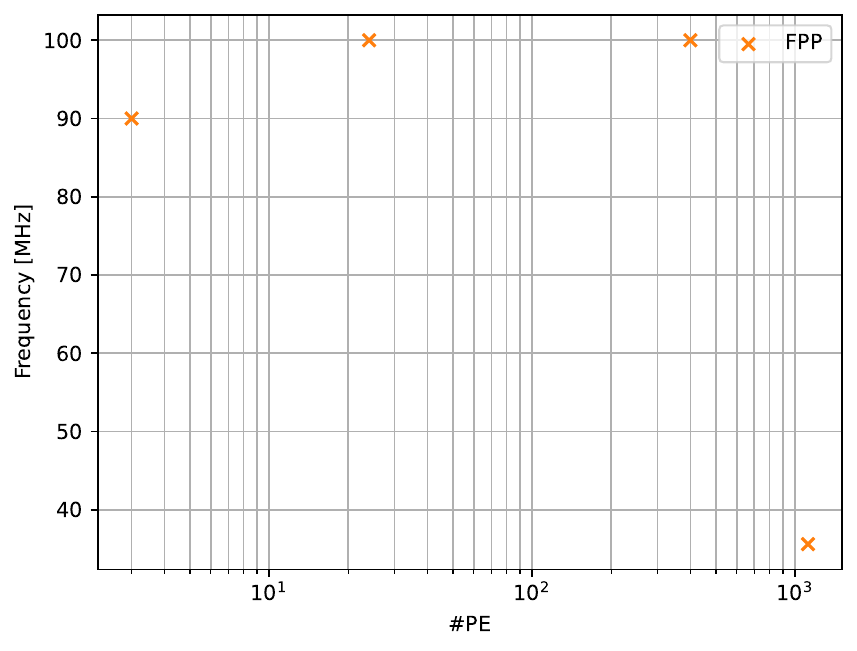}%
        \label{fig:app_freq_03_1}%
    }\\
    \subfloat[CCM]{%
        \includegraphics[width=.48\linewidth]{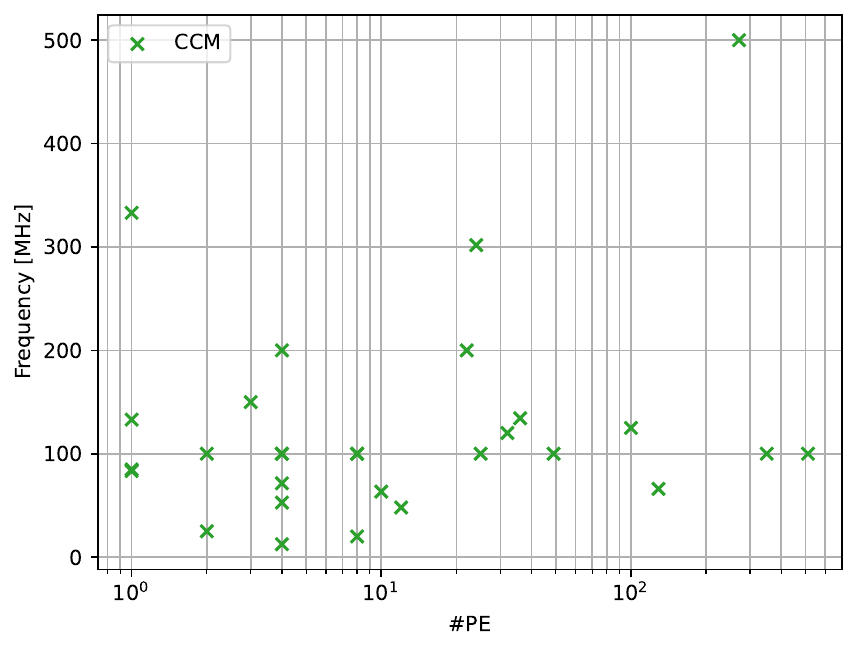}%
        \label{fig:app_freq_03_2}%
    }\hfill
    \subfloat[AST]{%
        \includegraphics[width=.48\linewidth]{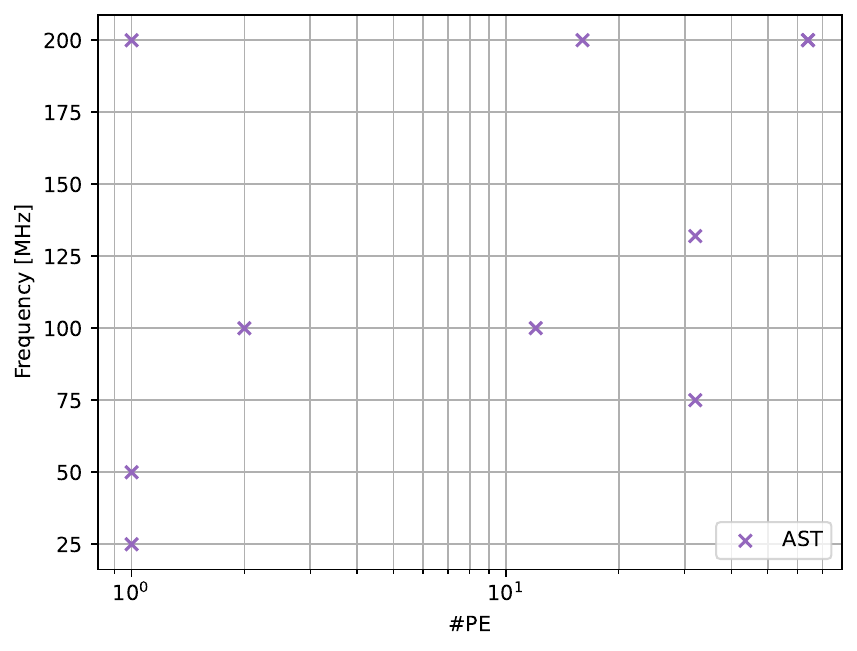}%
        \label{fig:app_freq_03_3}%
    }
    \caption{Clock frequency of the implementations in classes BAS, FPP, CCM and AST.}
    \label{fig:app_freq_03}
\end{figure*}

\begin{figure*}[t!]
    \subfloat[FPP/CCM]{%
        \includegraphics[width=.48\linewidth]{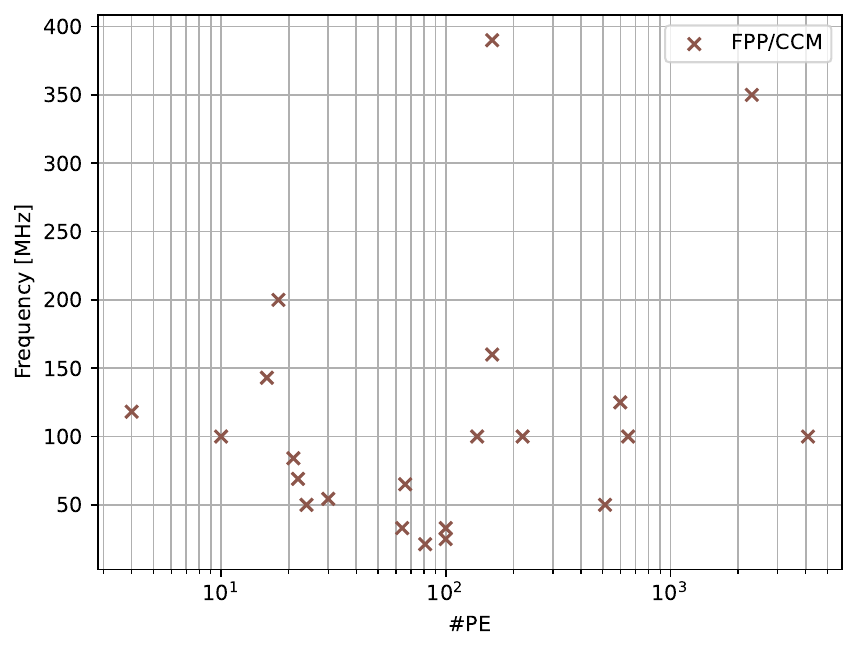}%
        \label{fig:app_freq_47_4}%
    }\hfill
    \subfloat[CCM/AST]{%
        \includegraphics[width=.48\linewidth]{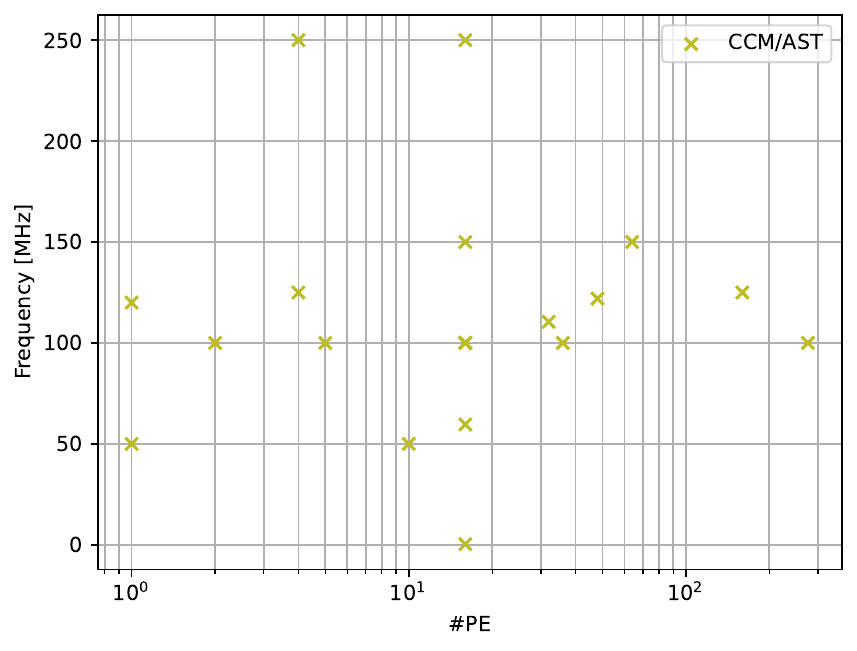}%
        \label{fig:app_freq_47_6}%
    }\\
    \subfloat[FULL]{%
        \includegraphics[width=.48\linewidth]{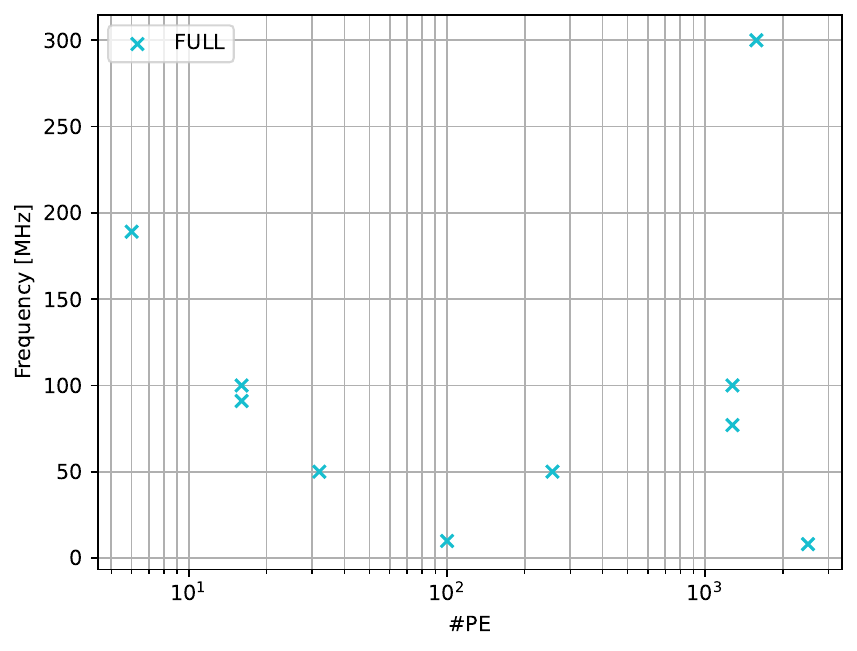}%
        \label{fig:app_freq_47_7}%
    }
    \caption{Clock frequency of the implementations in classes FPP/CCM, FPP/AST, CCM/AST and FULL.}
    \label{fig:app_freq_47}
\end{figure*}

Finally, we tried assessing the potential connection between classes, number of implemented PEs, and the potential maximum frequency the representatives were clocked at -- Figures \ref{fig:app_freq_03} and \ref{fig:app_freq_47} illustrate the gathered frequency data per class (except for FPP/AST as no information was provided for either of the two identified systems). Unfortunately, we did struggle to find any meaningful trends throughout the classes, with only \textit{obvious} observations that maximum frequency was obtained by the implementation from class CCM (DeepFire\cite{aung_deepfire_2021} with 500MHz) and the slowest belonged to class CCM/AST (Zhao et al.\cite{zhao_099--438_2023} with 0.3MHz), where the latter was actually designed to be primarily used taped-out as an ASIC and so the authors found this working conditions to be optimal for their use case -- "The operating point of peak energy efficiency is located at 300 kHz and 0.75 V.". What also can be noticed, is that vast majority of the systems was clocked up to and including 200MHz, with only singular examples reaching higher frequency for every class -- except for FPP, where every representative was clocked at maximum at 100MHz. However, the clock frequency is usually a superposition of the complexity of the system and the platform on which it was implemented. Moreover, the fitting and routing step of synthesis is usually performed with a random starting point, and achieving an optimal implementation is usually done through rerunning the synthesis process multiple times and selecting the best match. As the majority of authors did not state whether this kind of optimization was performed, we cannot assume that the presented systems were implemented with optimal fitting and routing flow for selected hardware. Furthermore, different platforms have different capabilities in terms of clock generation (primarily through varying PLL circuits), and as such, this factor should also be taken into consideration. Concluding, as we were lacking a significant portion of information related to the frequency of operation for the surveyed architectures, we decided to include the information in the Appendix, for completeness, instead of in the main flow of the article.


\subsection{On number representations}
\label{sec:app_ext_repr}
The FPGA-based NMAs usually support the \textit{signed fixed-point arithmetic}, where the parameters like membrane voltage and weights are represented as values with fixed mantissa and integer sizes, as opposed to \textit{floating-point arithmetic} following the IEEE-754 standard. This is because implementing dedicated Floating Point Units (FPUs) for use in ODE solvers multiplication and division requires on-chip DSP blocks, which are often a scarce resource. Fixed-point multiplication and division can often be replaced with a set of bitwise shift and add operations, utilizing the plentiful on-chip logic resources. This, however, introduces a level of error, which only grows as the number of used bits lowers. According to Zhang et al.\cite{zhang_biophysically_2009}, even using the 32-bit fixed-point arithmetic in the context of SNN simulation introduces errors that may cause completely different firing patterns. While this situation is not ideal in neuroscientific simulators, for computer science applications, it is often considered tolerable, as the accurate timings are of lower priority - what usually matters is that, e.g., one of the output neurons becomes more active than the rest, rather than producing a specific spike train.

\end{document}